\documentclass[preprint,12pt,sort&compress]{elsarticle}

\usepackage{amssymb}

\fntext[fn1]{These authors contributed equally to this work.}

\def\fe2{FE$^2$}

\def\bs#1{\boldsymbol{#1}}
\def\bf#1{\mathbf{#1}}

\def\ten#1{\mathbf{#1}}
\def\Ten4#1{\mathbb{#1}}

\def\Real{\mathbb{R}}

\usepackage{float}
\floatstyle{boxed}
\usepackage{graphicx}
\usepackage{subcaption}
\usepackage{natbib}
\usepackage[normalem]{ulem}

\usepackage[T1]{fontenc}    
\usepackage{lmodern}
\usepackage{url}            
\usepackage{booktabs}       
\usepackage{amsfonts}       
\usepackage{nicefrac}       
\usepackage{microtype}      
\usepackage{lipsum}
\usepackage{xcolor}
\usepackage{placeins}
\usepackage{graphicx}
\usepackage{enumerate}
\usepackage{siunitx}
\usepackage{numprint}

\usepackage{natbib} 
\usepackage{footnote}
\usepackage{amssymb,amsmath,amsfonts}
\usepackage{listings}
\usepackage{mathrsfs}
\usepackage{indentfirst}

\usepackage[demo]{adjustbox}
\usepackage{xcolor}

\journal{Computer Methods in Applied Mechanics and Engineering}

\begin{document}

\begin{titlepage}
	\color[rgb]{.4,.4,1}
	\hspace{5mm}

	\bigskip
	
	\hspace{15mm}
	\begin{minipage}{133mm}
		\color{black}
		\sffamily
		\LARGE\bfseries Data-driven methods  for  computational mechanics: A fair comparison between   neural networks and model-free approaches \\[-0.3\baselineskip] 
		
		\vspace{5mm}
		{\large {Preprint of the article published in \\[-0.4\baselineskip] Computer Methods in Applied Mechanics and Engineering (2024) }} 
		
		\vspace{10mm}        
		{\large Martin Zlati\'{c}, Felipe Rocha, Laurent Stainier, Marko \v{C}ana\dj{}ija } 
		
		\large
		
		\vspace{40mm}
		\vspace{5mm}
		
		\small
		\url{https://doi.org/10.1016/j.cma.2024.117289}
		
		\textcircled{c} 2024. This manuscript version is made available under the CC-BY-NC-ND 4.0 license \url{http://creativecommons.org/licenses/by-nc-nd/4.0/}
		\hspace{30mm} 
		\color[rgb]{.4,.4,1} 
		\includegraphics[width=3cm]{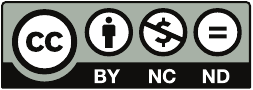}        
	\end{minipage}
\end{titlepage}

\begin{frontmatter}



\title{Data-driven methods for computational mechanics: a fair comparison between neural networks based and model-free approaches}

\author[1,2]{Martin Zlati\'{c}\corref{cor}\fnref{fn1}}
\ead{martin.zlatic@riteh.uniri.hr}
\author[2,3]{Felipe Rocha\fnref{fn1}}
\author[2]{Laurent Stainier}
\author[1]{Marko \v{C}ana\dj ija}

\affiliation[1]{organization={Faculty of Engineering, University of Rijeka},
            addressline={Vukovarska 58},
            city={Rijeka},
            postcode={51000},
            country={Croatia}}
\affiliation[2]{organization={Nantes Université, École Centrale de Nantes, CNRS, GeM, 		UMR 6183},
            addressline={1 Rue de la Noë},
            city={Nantes},
            postcode={44300},
            country={France}}
\affiliation[3]{organization={Univ Paris Est Creteil, Univ Gustave Eiffel, CNRS, UMR 8208, MSME, F-94010, Créteil, France}}

\cortext[cor]{Corresponding author}



\begin{abstract}

We present a comparison between two approaches to modelling hyperelastic material behaviour using data. The first approach is a novel approach based on Data-driven Computational Mechanics (DDCM) that completely bypasses the definition of a material model by using only data from simulations or real-life experiments to perform computations. The second is a neural network (NN) based approach, where a neural network is used as a constitutive model. It is trained on data to learn the underlying material behaviour and is implemented in the same way as conventional models. The DDCM approach has been extended to include strategies for recovering isotropic behaviour and local smoothing of data. These have proven to be critical in certain cases and increase accuracy in most cases. The NN approach contains certain elements to enforce principles such as material symmetry, thermodynamic consistency, and convexity. In order to provide a fair comparison between the approaches, they use the same data and solve the same numerical problems with a selection of problems highlighting the advantages and disadvantages of each approach. Both the DDCM and the NNs have shown acceptable performance. The DDCM performed better when applied to cases similar to those from which the data is gathered from, albeit at the expense of generality, whereas NN models were more advantageous when applied to wider range of applications.

\end{abstract}


\begin{highlights}
\item an in-depth comparison of data-driven and neural network approaches
\item new implementations of the most recent developments in both fields
\item assessment for two relevant examples in finite strain setting
\item synthetic databases and reference cases obtained from the two different isotropic hyperelastic models
\end{highlights}


\end{frontmatter}

\section{Introduction}
In recent years, data-driven techniques have become a popular tool in computational mechanics. In terms of material behaviour, we can distinguish two classes of approaches: i) replacing a constitutive model with a machine learning surrogate model, such as a neural network model, and ii) completely bypassing constitutive models. These strategies differ significantly from  each other, even though the goal is the same, namely to use the increasing abundance of available data to overcome the known drawbacks of classical constitutive modelling. In the context of finite elasticity, constitutive models link some measure of strain to a measure of stress.

Concerning \textit{Machine learning surrogate constitutive models}, these models aim to replace classical models in which material behaviour is described by several material constants obtained from experiments such as uniaxial or biaxial tests. Instead, the experimental data is used to generate a substitute model that is then used in place of the classical models. One possibility is the use of neural networks, which is also explored in this paper. 

Feedforward neural networks have been successfully used in various areas of material modelling. Some examples are the development of surrogate models for modelling biological tissue such as the liver \cite{Mendizabal2020} or the thoracic aorta \cite{Liang2018} in real time, replacing finite element analysis, enhancing multiscale analysis \cite{Rocha2023}, in multiphysics problems \cite{Kalina2024}, in carbon nanotrusses \cite{Canadija2024} and many more.

The application of classical NNs to ordinary hyperelastic behaviour \cite{Weber2021} or adiabatic behaviour \cite{Zlatic2023} by predicting stress from strain shows the ability of NNs to capture more complex behaviour using known basic architectures. However, it has proven beneficial to incorporate certain constraints or modelling strategies, resulting in more accurate models that require smaller data sets. The simplest approach to increase accuracy is to use invariants of deformation tensors as inputs, which significantly reduces the data needed to train the NNs \cite{Shen2004,Zlatic2023}. More recently, input convex neural networks have been presented in \cite{Amos2017} and demonstrated on multi-label prediction, image completion and reinforcement learning problems. This approach was adopted and used on hyperelastic problems by \cite{Thakolkaran2022,Klein2022,Linden2023}. Furthermore, in \cite{Linden2023}, general guidelines are presented for imposing physical constraints such as convexity and objectivity. Another family of NNs, the so-called Constitutive Artificial Neural Networks (CANNs), was proposed by \cite{Linka2021} to integrate well-known methods for modelling hyperelastic behaviour into neural networks and further developed in \cite{Linka2023}, where new activation functions are proposed to replace classical functions such as the hyperbolic tangent or the sigmoid function. Another model based approach for automated discovery of interpretable constitutive models called EUCLID (Efficient Unsupervised Constitutive Law Identification and Discovery) has been developed originally for hyperelastic materials \cite{Flaschel2021}. It has been expanded to the framework of generalised standard materials \cite{Flaschel2023} where the underlying material behaviour is not \textit{a priori} known. The difference between EUCLID and the NN approach investigated in this work is that EUCLID focuses on model discovery and identification from a large candidate pool of existing models whereas NNs are used as material models. It is also important to note that EUCLID does not learn from stress labels, but rather minimizes the force residuals of a FE problem or experiment. In some of the works NNs are trained on stress data \cite{Klein2022,Linden2023,Kalina2024,Canadija2024} which is also done in this work. However, apart from training only on energy or on stress, NNs can be trained on both at the same time as presented in \cite{Vlassis2021}.

On the other hand, a way to \textit{completely bypass constitutive models} is the so-called (model-free) data-driven computational mechanics (DDCM) paradigm \cite{Kirchdoefer2016}, that proposes a generalization at the level of the governing fields equation formulation. The new problem is constrained to satisfy relevant conservation principles (e.g. balance of linear momentum), which is independent of material behaviour, while the interplay with "experimental" material data is ensured in the minimization sense. Recent advances in DDCM has extended its range of application to inelastic materials \cite{Eggersmann2019}, noisy data \cite{Kirchdoerfer2017}, finite strains \cite{Platzer2021}, frequency domain data \cite{Salahshoor2023}, etc. As for applications of DDCM on material identification, the so-called Data-Driven Identification approach (DDI) \cite{Leygue2018,Stainier2019} can learn stress data without postulating underlying constitutive laws, following analogous premises as the aforementioned EUCLID approach. Another related method is based on manifold learning \cite{Ibanez2016} to approximate the local behaviour of experimental data.
Naturally, both classes of techniques present different advantages and hindrances, therefore a skeptical and fair comparison between these two different viewpoints is needed to clarify the domain of applicability for each method.

This paper focuses on the comparison of the accuracy of the two approaches when applied to compressible hyperelastic materials, which is performed using a neural network based approach (representing the class of surrogate models) and the so-called Data-driven Computational Mechanics (representing the model-free approaches). As far as the authors are aware, this is the first work to pursue this goal. In particular, we analyse the performance of these two methods in terms of accuracy on several benchmark problems, which are compared with reference solutions obtained with different hyperelastic models that have also been used to generate synthetic (noisy or not) material data.A comparison of execution times was made between the DDCM approaches and the NNs, but this should be treated with caution. A significant part of generating the NNs is the training itself, which can take a long time, whereas all DDCM approaches are ready to use as soon as the data is available. However, the different execution times of the simulations can be significant if the method is used to run multiple or large simulations. It is important to note that the simulations were performed using the same finite element solver (FEniCs-based implementations in Python) in order to make the comparison as fair as possible, although we are aware of the inherent difficulty of this task (different code optimisation, sophistication of auxiliary libraries, etc.).

\section{Preliminaries}

First, let us introduce the physical problem to be addressed. Consider a domain $\Omega_0 \subset \mathbb{R}^d$ and the partition of its boundary $\partial \Omega_0 = \partial \Omega^0_N \cup \partial \Omega^0_D$, such that a traction vector $\overline{\bf{t}}_0$ and a prescribed displacement $\overline{\bf{u}}$ are known, defined in the $\partial \Omega^N_0$ and $\partial \Omega^D_0$, respectively. Also consider the body forces $\bf{b}_0$ acting on $\Omega_0$. Under the quasi-static assumption, the balance of linear momentum leads to
\begin{subequations}
	\label{eq:equilibrium}	
	\begin{eqnarray} 
		-\text{Div} \ten{F}(\bf{u}) \ten{S} = \bf{b}_0 \quad \text{in } \Omega_0, \\
		\ten{F}(\bf{u}) \ten{S} \bf{n}_0 = \overline{\bf{t}}_0 \quad \text{on } \partial \Omega^N_0, \\
		\bf{u} = \overline{\bf{u}} \quad \text{on } \partial \Omega^D_0.
	\end{eqnarray}
\end{subequations}
where $\bf{u}$, $\ten{F} = \ten{I} + \nabla \bf{u}$, $\ten{S}$, and $\textbf{n}_0$ stand for the displacement field, gradient of deformation, the second Piola-Kirchhoff stress tensor, and the normal vectors on the boundary  $\partial \Omega^N_0$, respectively. In addition, $\ten{S}$ is symmetric as result of the angular momentum balance. Note that we have not carried explicitly the dependence of $\ten{S}$ with respect to displacements $\bf{u}$ on purpose, whose reasons are made clearer in Section \ref{sec:ddcm}. It is also useful  remarking that $\ten{S}$ is the thermodynamic conjugate to Green-Lagrange tensor, given by
\begin{align} \label{eq:compatibility}
	\ten{E} &= \frac{1}{2}(\ten{F}^T \ten{F} - \ten{I}) = \frac{1}{2}(\nabla \bf{u} +  \nabla^T \bf{u} + \nabla \bf{u}^T \nabla \bf{u}).
\end{align}
As already commented, the point of intersection between the two different methodologies is the use of experimental data. Let us assume we have at disposal a finite dataset of $N_d$ points as below
\begin{align}
	D = \left \{ (\hat{\ten{E}}_i, \hat{\ten{S}}_i) \in \Real^{m\times m}_\text{sym} \times \Real^{m\times m}_\text{sym} \text{ for } i= 1, \dots, N_d \right\},
\end{align}
with $\text{m}$ being the dimension of the space ($\text{m}$ = 2,3). In the sequel, we revisit the two alternative methodologies to clarify how each of them process $D$.

\section{Neural Networks for modelling hyperelastic materials}

In this section the neural network model is presented. The neural network model is based on the recent works of \cite{Linka2023, Zlatic2023}, where the basic premise is to replace the strain-energy of a classical model with a neural network. In this paper a new type of feed-forward shallow neural network is proposed which utilises the advantages of mechanically inspired activation functions, but retains the versatility of a neural network. 

Here we employ automatic differentiation of the strain-energy represented by the NN to define the stress and material tangent, such that thermodynamic consistency is ensured. Furthermore, by employing the invariants of the right Cauchy-Green deformation tensor ($\ten{C} = \ten{F}^T\ten{F}$) the NN model is frame indifferent, as well as ensuring symmetry of the stress tensor and material tangent. For the sake of simplicity, only isotropic behaviour is considered, hence only the first three invariants of the right Cauchy-Green deformation tensor are used, defined as usual:

\begin{equation}
	\centering
	I_1 = \text{tr}{(\ten{C})}, \quad I_2 = \frac{1}{2}\left[\text{tr}{(\ten{C})}^2 - \text{tr}{(\ten{C}^2)}\right], \quad I_3 = \det{(\ten{C})},
	\label{eq:invariant_definition}
\end{equation}
and the strain-energy function becomes a neural network with the invariants from Eq.~(\ref{eq:invariant_definition}) as inputs:
\begin{equation}
	\centering
	\psi_\text{NN}(\ten{C}) = f_\text{NN}(\ten{C}) = \tilde{f}_\text{NN}(I_1,I_2,I_3).
	\label{eq:psi_nn}
\end{equation}

The second Piola-Kirchhoff stress tensor can be obtained by deriving the strain-energy w.r.t. the right Cauchy-Green deformation tensor, and by using the chain rule w.r.t. the invariants:

\begin{equation}
	\centering
	\begin{split}
		\ten{S} & = 2\frac{\partial{\psi(I_1,I_2,I_3)}}{\partial{\ten{C}}} = 2\left[	\frac{\partial{\psi}}{\partial{I_1}}\frac{\partial{I_1}}{\partial{\ten{C}}} + 
		\frac{\partial{\psi}}{\partial{I_2}}\frac{\partial{I_2}}{\partial{\ten{C}}} +
		\frac{\partial{\psi}}{\partial{I_3}}\frac{\partial{I_3}}{\partial{\ten{C}}}\right] = \\
		& = 2\left[\left(\frac{\partial{\psi}}{\partial{I_1}} + I_1\frac{\partial{\psi}}{\partial{I_2}}\right)\ten{I} - \frac{\partial{\psi}}{\partial{I_2}}\ten{C} + I_3\frac{\partial{\psi}}{\partial{I_3}}\ten{C}^{-1}\right].
	\end{split}
	\label{eq:2PK_definition}
\end{equation}

Another restriction that should be fulfilled is that for an undeformed state, i.e. $\ten{C} = \ten{I}$, the strain-energy is zero. To fulfil this condition, the specific choice of activation function is introduced and the inputs to the network should be slightly modified. This is discussed in the following section.

\subsection{Architecture of the NN}

\begin{figure}
	\centering
	\includegraphics[scale=0.25]{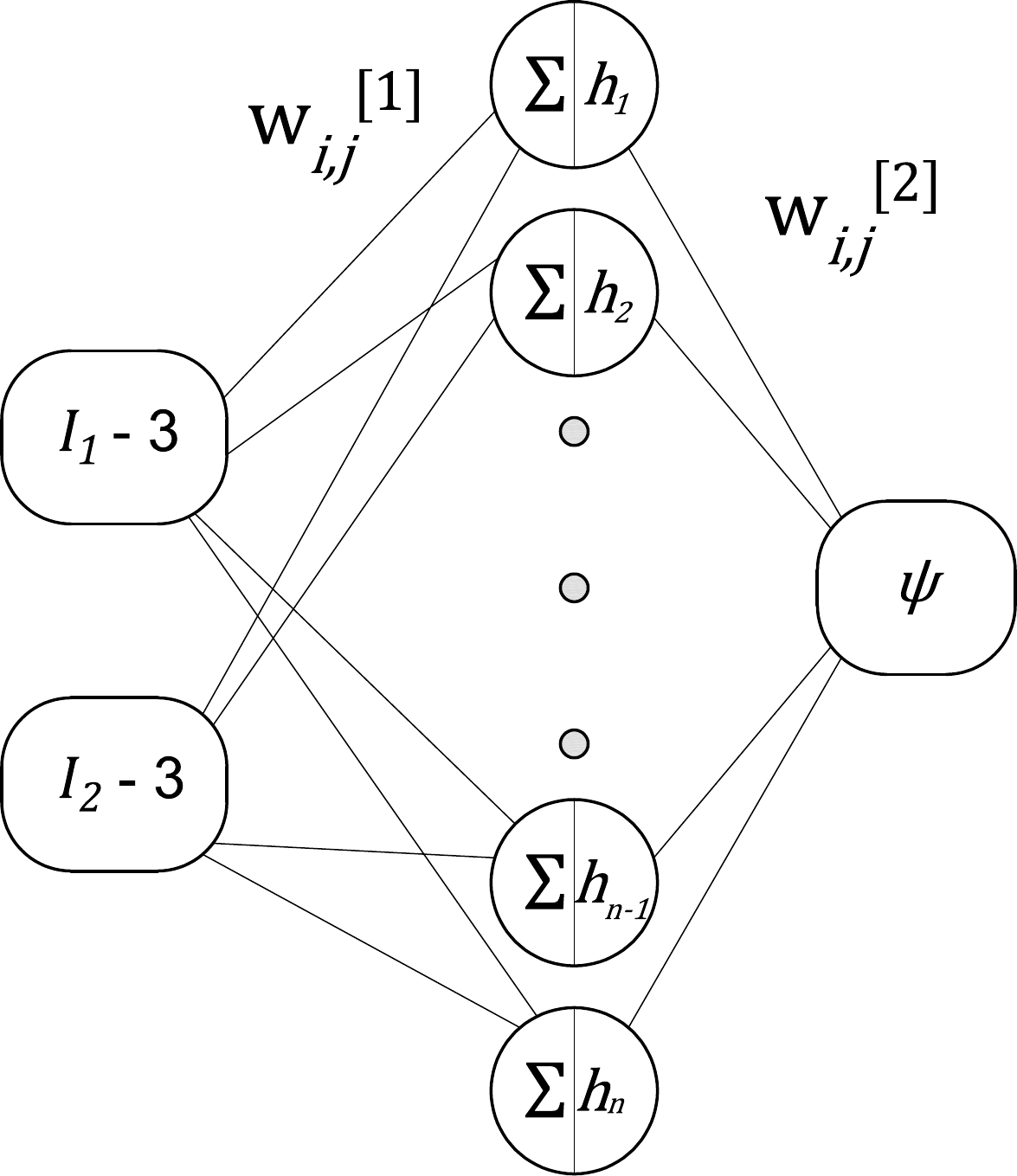}
	\caption{Architecture of the neural network used for modelling compressible hyperelasticity.}
	\label{fig:NN_architecture}
\end{figure}

As mentioned, a shallow feed-forward NN is employed. The NN is developed using the TensorFlow library. The architecture is shown in Fig.~\ref{fig:NN_architecture}, whose choice is motivated in the following. The feed-forward neural network consists of an input, hidden and output layer where information is passed in the forward direction to the neurons in the next layer and passes through an activation function after each neuron in the hidden layer. In Fig.~\ref{fig:NN_architecture} the weights between the neurons are $w_{i,j}^{[l]}$ where $i$ references the neuron from the previous layer and $j$ the neuron in the next layer, and $l$ the number of the next layer. The number of neurons in the hidden layer can vary arbitrarily but in this paper it was constrained to 10 neurons ($n_h = 10$). This size of the hidden layer has proven to give models whose accuracy does not oscillate much from training to training. The changes in quality occur because the weights between the neurons are randomly initialized each time a NN is trained. The size of this NN is comparable to that of \cite{Linden2023} where 1 hidden layer containing 4 or 8 neurons was shown to be viable. The network does not contain biases. When it comes to the choice of activation functions, in \cite{Linka2023} it is proposed to use functions commonly used in hyperelastic modelling. The linear exponential unit was considered in this work:

\begin{equation}
	\centering
	h(x) = \exp{(\alpha x)} - 1,
	\label{eq:linear_exponential_unit}
\end{equation}
where $\alpha$ is a trainable parameter.

\FloatBarrier

Having chosen the activation function, we can construct $\psi_{NN}$ such that the condition normalization condition of the energy $\psi_\text{NN}(\ten{I}) = 0$ can be fulfilled. The NN output can be expressed as 
\begin{equation}
	\centering
	\psi_\text{NN} = \sum_{i=1}^{n_{h}}w_{i,1}^{[2]}h_i.
	\label{eq:output_weight_restriction}
\end{equation}

The activation function becomes equal to zero when its input is zero, i.e. $h(0) = 0$. In an undeformed state the invariants have the following values $I_1 = I_2 = 3, I_3 = 1$ . Thus the inputs from Eq.~(\ref{eq:psi_nn}) need to be redefined by subtracting these values in the undeformed state obtaining the new form:
\begin{equation}
	\centering
	\tilde{f}_\text{NN}(I_1,I_2,I_3) = \bar{f}_\text{NN}(I_1-3, I_2 - 3, I_3 - 1),
	\label{eq:psi_nn_modified_inputs}
\end{equation}
and now for an undeformed state the condition $\psi_\text{NN}(I_1-3, I_2 - 3, I_3 - 1)_{|\mathbf{C}=\mathbf{I}} = 0$ is satisfied. Biases are omitted from the NN to guarantee by construction that this condition is satisfied. Otherwise, the biases would always be present in the activation function violating the normalization condition of the energy. In the case that biases were present an additional normalization term would need to be added in the energy function Eq.~\eqref{eq:psi_nn_modified_inputs}. The expression for the activation now takes the form of
\begin{equation}
	\centering
	h_{j}(I_1-3, I_2 - 3, I_3 - 1) = \exp\big[\alpha_j[w_{1,j}^{[1]}(I_1-3) + w_{2,j}^{[1]}(I_2 - 3) + w_{3,j}^{[1]}(I_3 - 1)]\big] - 1.
	\label{eq:output_one_neuron}
\end{equation}

Also, by choosing the activation function in the form of Eq.~\eqref{eq:linear_exponential_unit} the predicted strain-energy will always remain convex w.r.t. the input invariants, see \ref{sec:convexity_proof}. The possibility of having negative weights $w_{i,j}^{[1]}$ and $\alpha_i$ does not impact the convexity of the NN. By looking at the output $\psi_{NN}$ in Eq.~\eqref{eq:output_weight_restriction} only the weights $w_{i,j}^{[2]}$ need to be non-negative to enforce convexity, since the activation functions are \textit{a priori} convex. In addition, enforcing convexity is desirable because it implies more stable behaviour, which can be advantageous when training the network on noisy data \cite{Asad2022}.

For the  non-negativity of the strain energy, consider the output of the $j$-th neuron in the hidden layer in Eq.~\eqref{eq:output_one_neuron} and the expression for the output of the NN in Eq.~\eqref{eq:output_weight_restriction}. The output of the function $h_j$ will be positive when the value of the exponent of the exponential function is positive. The values of the weights $w_{i,j}^{[1]}$ and $\alpha_i$ can be enforced to be non-negative, however the values $I_1-3, I_2-3$ and $I_3-3$ can not, as $I_1$ and $I_2$ can be take values lower than 3 and $I_3$ can take values lower than 1. This means that the activation functions can not be ensured to be non-decreasing in general, meaning that they could have negative values (with a limit up to -1). Non-negativity of the strain-energy cannot be guaranteed when using the NN architecture with the invariants $I_1$, $I_2$ and $I_3$ described earlier since they can reach values smaller than 3 (in the case of $I_1$ and $I_2$) or 1 (in the case of $I_3$) and thus the activation functions in the form of Eq.~\eqref{eq:linear_exponential_unit} can become decreasing functions, regardless of the constraints imposed on the weights in the NN. In the case of incompressibility, non-negativity could be restricted since the invariants $I_1$ and $I_2$ have a minimum value of 3 and the restriction that the weights $w_{i,j}^{ [1]}$ must be non-negative guarantees that the activation functions are non-decreasing. However, if the NN is trained on its derivatives, i.e. on stress, this would in turn incorporate the non-negativity of energy into training, although not in an exact way.

It should be noted that the condition of the vanishing of the stresses in the state of zero strain, $\mathbf{S}(\mathbf{I})=\mathbf{0}$, is not enforced by the architecture, nor is there a normalisation term for the stress as in \cite{Linden2023}. It is approximated during training, similar to \cite{Klein2022}, i.e. if the NN is correctly learning the material behaviour it should reduce the stress in the undeformed configuration. However, there will always be a very small non-zero stress. 

\subsection{Loss functions and training details}

The neural network that predicts strain-energy can be trained either directly on strain-energy \cite{Thakolkaran2022,Zlatic2023} or on stress \cite{Linka2023,Klein2022}. To implement the two approaches, the following loss functions were used:
\begin{equation}
	\begin{split}
		& 1. \quad L_\psi(\mathbf{\Theta};I_1,I_2,I_3) = \frac{1}{N}\sum_{i=0}^{\text{N}}(\psi_{\text{NN},i}(\mathbf{\Theta}; I_1-3,I_2-3,I_3-1) - \hat{\psi}_i)^2, \\
		& 2. \quad L_\textbf{S}(\mathbf{\Theta};I_1,I_2,I_3,\ten{C}) = \frac{1}{N}\sum_{i=0}^{\text{N}}||\ten{S}_{\text{NN},i}(\mathbf{\Theta}; I_1-3,I_2-3,I_3-1) - \hat{\ten{S}}_i||^2,
	\end{split}
	\label{eq:loss_functions}
\end{equation}
where $\mathbf{\Theta}$ represents the set containing trainable network weights $w_{i,j}^{[l]}$ and the parameters $\alpha_i$, and $N$ is the number of samples. The values $\hat{(\bullet)}$ represent target data, while the values $(\bullet)_\text{NN}$ represent data predicted by the neural network. $L_\psi$ is used as the loss function when training NNs on energy, and $L_\textbf{S}$ is used as the loss function when training the NNs on $2^\text{nd}$ Piola-Kirchhoff stress.
In the second loss function $\ten{S}_\text{NN}$ is obtained by using TensorFlow's autodifferentiation tool to obtain the derivatives of the strain-energy function and passing these values to Eq.~\eqref{eq:2PK_definition}. During training the accompanying values of $\ten{C}$ are also passed. In both cases Adam was used as the optimizer with the default settings (learning rate equal to 0.001). The training was restricted to a maximum of 200 000 epochs and an early-stopping was set to terminate the training if after 5000 epochs the validation loss does not decrease. The data was split in a 75/25 train /test ratio, so 75\% of the data was used to train the NNs, while 25\% was reserved for validation.
The results are presented in Section~\ref{sec_examples}.

Since during training the weights and parameters of the neural networks are initialised randomly, this affects the training. A neural network was trained 10 times on the same dataset and the one with the lowest validation loss was taken as the best trained model for that dataset. This significantly increases the training time required but helps in obtaining the best model. The scaling of the data, which is indeed usually performed in machine learning applications, was found to be unnecessary in this work. No convergence issues were encountered during training.

\section{(Model-free) Data-driven Computational Mechanics in Finite Strains}
\label{sec:ddcm}

As already commented, in the classical setting of solid mechanics, a constitutive law is provided to close set of equations in Eq.~\eqref{eq:equilibrium}. On the other hand, in the (model-free) data-driven computational mechanics setting, as originally proposed by \cite{Kirchdoefer2016}, only a finite sampling of strain-stress pairs is at the disposal. Hence, an appropriate redefinition of the mechanical problem is needed to allow the search for a solution. These ideas are briefly revisited in the rest of this section, particularly following further extensions of the method for the finite strain setting \cite{Nguyen2018}, and also  its variational reformulation \cite{Nguyen2020}, more suited for continuum problems.

First, let us define $\mathbb{Z} = \mathbb{R}^{m\times m}_{sym} \times \mathbb{R}^{m\times m}_{sym}$, the so-called phase-space composed by strain-stress pairs, here composed by Green-Lagrange/Second Piola-Kirchhoff stress tensor couples. Let $\mathcal{Z}$ be the set of all functions with domain in $\Omega_0$ and image $\mathbb{Z}$. In order to distinguish functions to point-wise values, we adopt the following convention $\bf{z} = (\ten{E}, \ten{S}) \in \mathcal{Z}$, $\hat{\bf{z}} = (\hat{\ten{E}}, \hat{\ten{S}})= \bf{z}(\bf{x}) \in \mathbb{Z}$ for some $\bf{x} \in \Omega_0$. Distances in $\mathcal{Z}$ are measured by $d(\bf{z}, \bf{z}^*) = \sqrt{ \int_{\Omega} \| \bf{z}(\bf{x}) - \bf{z}^*(\bf{x}) \|^2_{loc} \, d \Omega }$, such that
$\| (\hat{\ten{E}}, \hat{\ten{S}}) \|_{loc} = \sqrt{ \mathbb{C} \hat{\ten{E}} \cdot \hat{\ten{E}} + \mathbb{C}^{-1} \hat{\ten{S}} \cdot \hat{\ten{S}} }$, where the fourth-order tensor $\mathbb{C}$ is an algorithmic parameter that defines an energy-like norm, balancing contributions of the strain and stress parts.

In practice, an appropriate finite-dimensional subspace of $\mathcal{Z}$ should be considered, for example adopting a finite element discretization. Hence, a given function $\bf{z} \in \mathcal{Z}$ is unequivocally defined by a finite number of control points \cite{Nguyen2020}, usually associated to integration points. Accordingly, the DDCM distance can also be interpretable discretely as a weighted sum of local metrics, where weights are associated to the Gauss quadrature weights and finite element volumes. Finally, the Data-driven problem reads as
\begin{align}
	\min_{\bf{z}' \in \mathcal{Z}_E}  \min_{\bf{z}^* \in \mathcal{Z}_D}  d(\bf{z}',\bf{z}^*),
	\label{eq:dd_problem}
\end{align}
where $\mathcal{Z}_E =  \big\{ \bf{z} \in \mathcal{Z}; \text{\eqref{eq:equilibrium} and  \eqref{eq:compatibility} hold} \big\}$ is the so-called equilibrium manifold, and $
\mathcal{Z}_D = \{ \bf{z} \in \mathcal{Z}; \bf{z}(\bf{x}) \in D; \forall \bf{x} \text{ control point in } \Omega_0 \}$ is the so-called Data-function space \cite{Nguyen2020}. Note that mechanical equilibrium and kinematical compatibility are respected for every function in $\mathcal{Z}_E$, while $\mathcal{Z}_D$ translates the dataset $D$ into discretized functions in $\mathcal{Z}$. 


The standard algorithm used to deal with the double minimisation problem encompasses the fixed-point iteration with the alternated resolution of two subproblems until reaching a certain convergence tolerance. First, given a $\bf{z}^* \in \mathcal{Z}_D$, we seek for the projection onto $\mathcal{Z}_E$. Second, given $\bf{z} \in \mathcal{Z}_E$, we seek for closest point $\mathcal{Z}_D$. The latter sub-problem boils to a nearest neighbour search for each integration point.  The former sub-problem is a continuous constrained minimisation problem, which is then rephrased in an unconstrained format by the incorporation of displacement-like field of Lagrange Multipliers. For convenience, the interested reader can find further details of this procedure in \ref{sec:DDCM_PE} or in more detailed specialized literature \cite{He2020b, Nguyen2020}. 

It is worth mentioning our implementation has been built upon \texttt{ddfenics} \cite{ddfenics}, an opensource library based on FEniCS for DDCM. Therefore, the variational formulations notation adopted in \ref{sec:DDCM_PE}, as opposed to the more wide-spread matrix notation in the same spirit of the seminal DDCM work \cite{Kirchdoefer2016}, allows for a straightforward translation towards FEniCS implementation. Currently, \texttt{ddfenics} interfaces tree-based nearest neighbours algorithms (e.g. \texttt{KDtree}, \texttt{BallTree}) of \texttt{scikit-learn} \cite{scikit-learn}. There are also more advanced data structures to further improve nearest neighbours searches specially coined for DDCM \cite{Eggersmann2021a}, and also more general-purposes high-performance libraries notably \texttt{FLANN} \cite{flann} for fast approximative nearest neighbours searches, and \texttt{kNN-CUDA} \cite{knn-cuda} for GPU-accelerated searches. Although, these implementations can further improve DDCM performance in some situations, they are not currently supported in \texttt{ddfenics}.

Here below, we comment on some slight changes to the original algorithm we consider in manuscript for the sake of comparison, namely: i) accounting for isotropic data and ii) local smoothing of data through a locally convex embedding \cite{He2020a}. Results obtained with the standard strategy will be denoted \texttt{DD}, while \texttt{DDiso}, \texttt{DDLC}, and \texttt{DDLCiso} stand for modification i), ii), and their combination, respectively. Finally, it is worth mentioning that in the nonlinear setting, there are two nested convergence loops, an outer one associated with the alternate minimisation procedure and an inner one associate with the Newton-Raphson method (see \ref{sec:DDCM_PE}).

In this work, the convergence of the outer loop is assessed by the ratio between the data-driven distance (computed between the mechanical and material state) and the same distance definition evaluated for the mechanical states with respect to the null state. Note that such criteria renders comparable errors among different problems and mesh sizes such that a tolerance of $10^{-8}$ was used, with maximum number of iterations of $50$ for all simulations. The metric $\mathbb{C}$ itself is not a hyper-parameter, but it is estimated from the database by taking a low-rank approximation from the SVD decomposition \cite{Eggersmann2021}. In contrast to the NN models, the DDCM approaches do not need to be trained (apart from creating the tree structure for the nearest neighbour search), so there is no variance from run to run and they can be used immediately.

\subsection{Taking into account isotropy}
As the NN model is invariant-based, it natively takes into account information about the isotropy. For a fairer comparison, the standard DDCM (which is agnostic to special underlying hyperelastic model structure) have to be extended to take into account isotropy. The exact way of imposing such structure is considering dataset extended by orbits $\tilde{D} = \{(\ten{Q}^T \hat{\ten{E}} \ten{Q}, \ten{Q}^T \hat{\ten{S}} \ten{Q}); \forall (\hat{\ten{E}}, \hat{\ten{S}}) \in  D;  \forall \ten{Q} \in \mathrm{Orth} \}$. In this work, we perform that approximately by incorporating discretization of such orbits, the entire $\mathrm{Orth}$ being replaced by a certain finite number of rotations as adopted in \cite{Platzer2021} for the ease of implementation. Here we consider two-dimensional problems with rotations parametrized with $N_o$ equally spaced angles.  In 2D, it is easy to see that the rotations are parametrized by just one angle in a half-plane. Therefore, we consider $N_{o} = 100$ equally spaced samples in $[-\pi/2 , \pi/2]$ for all simulations in this paper.

It is worth noticing that the strategy described above does not strongly enforce isotropy in DDCM. Therefore, small deviations to the isotropy are expected for the mechanical states, while the material states are always isotropic since the database only contains isotropic data. As a natural consequence of the spectral representation of isotropic strain energy (in terms of eigenvalues of $\ten{C}$), the second Piola-Kirchhoff stress tensor $\ten{S}$ should be expressed as a linear combination of dyadic products of eigenvectors of $\ten{C}$ (see e.g \cite{Holzapfel2000}). In other words, $\ten{C}$ and $\ten{S}$ should respect collinearity. Therefore, one could not simply straightforwardly use a metric in based exclusively on invariants (or eigenvalues) as it completely disregards deviations on the collinearity. Moreover, the use of Euclidean metric for the phase-space (weighted by constant positive-definite metric tensor) has the advantage of allowing the use of efficient algorithms for nearest neighbours search and gives rise to linear systems with constant left-hand side, which the same factorization can be reused for all iterations.

\subsection{Taking into account noisy and lacking data}
As any surrogate model, NN tends to smooth out the noise present on the data (if not overfitted) and interpolate/extrapolate in zones of lacking. In the standard DDCM the nearest neighbour search prevents interpolation/extrapolation. While this can be desirable property, it turns out to be a hindrance for zones of small density or large noise in the data. The so-called locally convex embedding \cite{He2020a} proposes instead the following modification for Data-functions set:
\begin{align}
	\hat{\bf{z}}^* &= \sum w^*_i \hat{\bf{z}}_i, \\
	\bf{w}^*_i(\bf{z}) &= \text{arg min}_{\bf{w}} \|\bf{z} - \sum_{i\in \mathcal{N}_k(\bf{z})} w_i \hat{\bf{z}}^i \|^2_M, \\
	\text{subject to}: & \sum_{i\in \mathcal{N}_k(\bf{z})} w_i = 1, w_i\geq 0, \forall i \in \mathcal{N}_k(\bf{z}),
\end{align}
with $\mathcal{N}_k(\bf{z})$ being the set of $k$ nearest neighbours indexes. Here we consider $k = 20$ for all simulations where this method is used.
The equality constraint is relaxed and regularised with the recommended penalty terms as in \cite{He2020a}, and the non-negativity of weights is ensured by the use of the Scipy's non-negative leasts-squares function \texttt{scipy.optimize.nnls}.

\section{Numerical examples}\label{sec_examples}

\subsection{Construction of reference solution and datasets}{\label{sec:ref_data}}

To test the solution of this problem with a data-driven solver utilizing FEniCs, we use the Ciarlet and Hartman-Neff compressible hyperelastic models on the Cook membrane as presented in \cite{GammBenchmarks2021}. The problem is restricted to plane strain. Constitutive models used to generate the datasets are:

\begin{align}
	\Psi_{\text{Ciarlet}} &= \frac{\mu}{2} (I_1(\ten{C}) - 3) + \frac{\lambda}{4} (J^{2} - 1) - (\frac{\lambda}{2} + \mu) \log{J} , \label{eq:psi_ciarlet} \\
	\Psi_{\text{HN}} &= \underbrace{a (\bar{I}_1(\ten{C})^3 - 27) + c_{10}(\bar{I}_1(\ten{C}) - 3) + c_{01}(\bar{I}_2^{3/2} - 3\sqrt{3})}_{:=W(\bar{I}_1,\bar{I}_2)} + 
	\underbrace{\frac{k}{50}(J^5 + J^{-5} - 2)}_{:=U(J)} \label{eq:psi_hn}
\end{align}
with $\mu = \SI{185.185}{MPa}$, $\lambda =\SI{432.099}{MPa}$, $a= 3.67\cdot10^{-3}\space \text{MPa}$, $c_{10}=\SI{0.1788}{MPa}$, $c_{01} = \SI{0.1958}{MPa}$, $k=\SI{80}{MPa}$, $\bar{I}_1 = \text{tr}(\bar{\ten{C}}), \bar{I}_2 = \text{tr}(\text{cof} \;\bar{\ten{C}}), J =\det{(\ten{C})}^{1/2} $ and $\bar{\ten{C}} = J^{-1/3}\ten{C}$.

\begin{figure}
	\centering
	\includegraphics[scale=0.4]{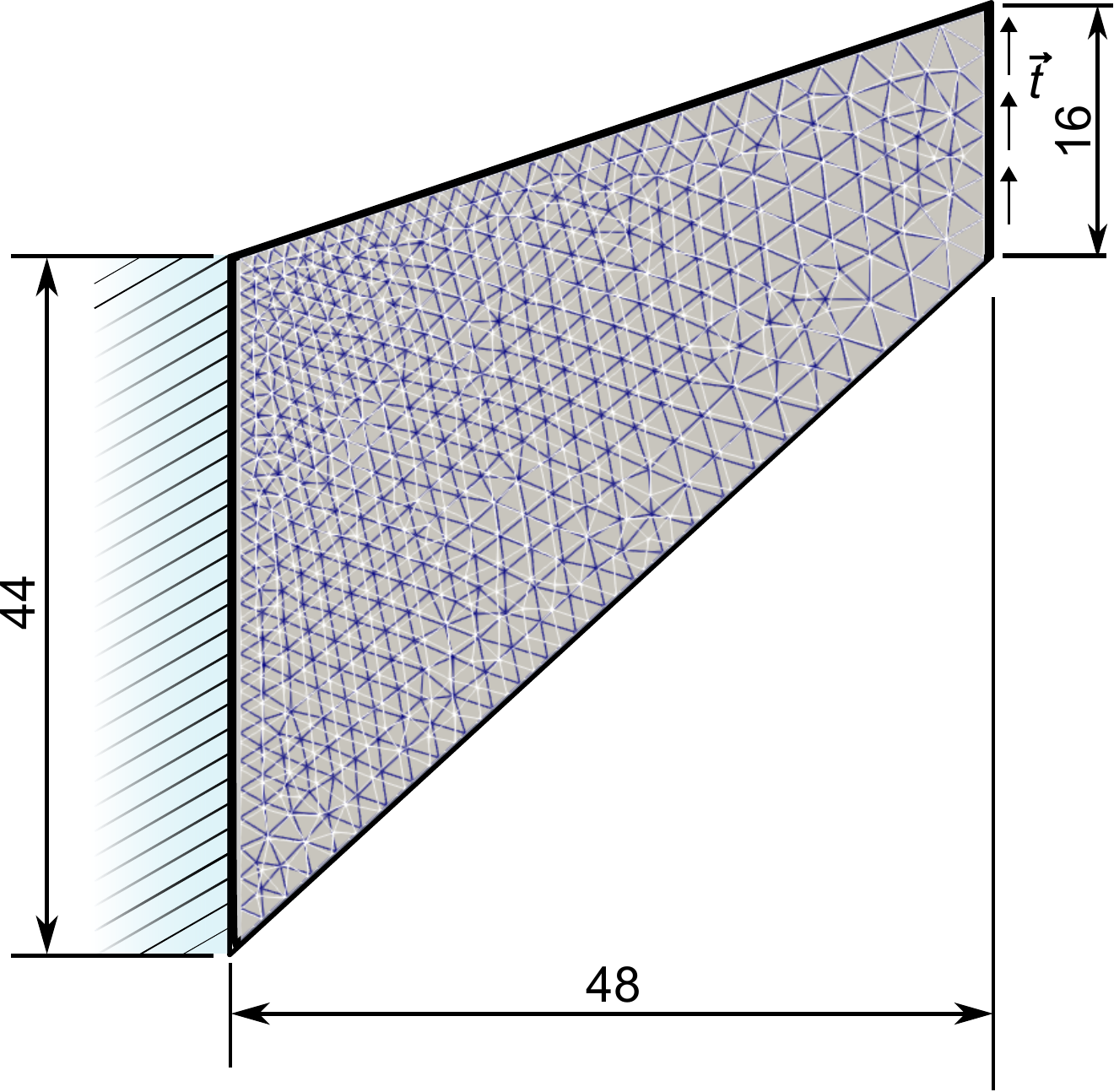}
	\caption{Geometry, boundary and loading conditions of the Cook membrane. The mesh with an overlay of the source and changed mesh is shown as well. Source mesh is solid blue, changed mesh is solid white.}
	\label{fig:geometry_and_mesh_cook}
\end{figure}

For the sake of completeness, the expressions for the second Piola-Kirchhoff tensor are provided below
\begin{gather}
	\ten{S}_{\text{Ciarlet}} = 2 \frac{\partial{\Psi}_{\text{Ciarlet}}}{\partial{\ten{C}}} = \frac{\lambda}{2} (J^{2} - 1) \ten{C}^{-1} + \mu (\ten{I} - \ten{C}^{-1}),\\
	\ten{S}_{\text{HN}} = JU^{'}(J)\ten{C}^{-1} + 2J^{-2/3}\big((W_{,1} + W_{,2}\bar{I}_1)\ten{I} - W_{,2}\bar{\ten{C}}- \frac{1}{3}(W_{,1}\bar{I}_1+2W_{,2}\bar{I}_2)\bar{\ten{C}}\big),
\end{gather}
with $W_{,1}$ and $W_{,2}$ as the partial derivatives of $W(\bar{I}_1,\bar{I}_2)$ with respect to $\bar{I}_1$ and $\bar{I}_2$.
To gather the data necessary for the data-driven calculation a simulation with 4 iterations is done from which the strain and stress data are taken from the centre points of the elements as a source dataset. This dataset was used both for the data-driven simulations and for training the neural networks. In the case of neural networks trained on strain-energy, the strain-energy was calculated using \eqref{eq:psi_ciarlet}~and~\eqref{eq:psi_hn}. Afterwards the mesh is changed so that it no longer corresponds to the source mesh. The source mesh is shown in Fig.~\ref{fig:geometry_and_mesh_cook} with solid blue lines whilst the changed mesh is represented by solid white lines.

The source dataset consisting of 3944 samples is further randomly split into smaller datasets of 100, 500, 1000 and 2000 samples to compare the accuracy of the various DD techniques and the NN on varying availability of data. Afterwards, multiplicative noise is added to each dataset to further explore the sensitivity of the DD and NN approaches. The datasets were generated once before training the NN models or performing the DD calculation and were not resampled from run to run. In all the simulations, pure displacement linear Lagrange triangle elements were used.

\FloatBarrier

\subsection{Cook Membrane with Ciarlet law}{\label{sec:cook_ciarlet}}

\begin{figure}[h!]
	\centering
	\begin{subfigure}{0.5\textwidth}
		\centering
		\includegraphics[width=\textwidth]{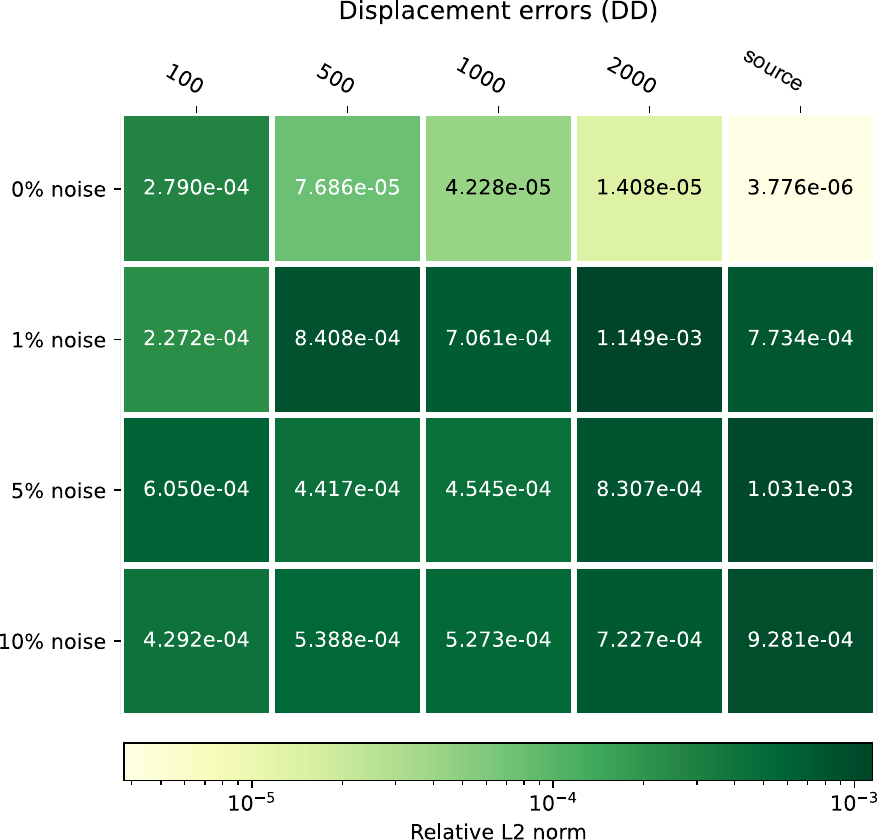}
		\caption{Standard search, original database.}
		\label{fig:errors_disp_C_DD}
	\end{subfigure}%
	\hfill
	\begin{subfigure}{0.5\textwidth}
		\centering
		\includegraphics[scale=0.462]{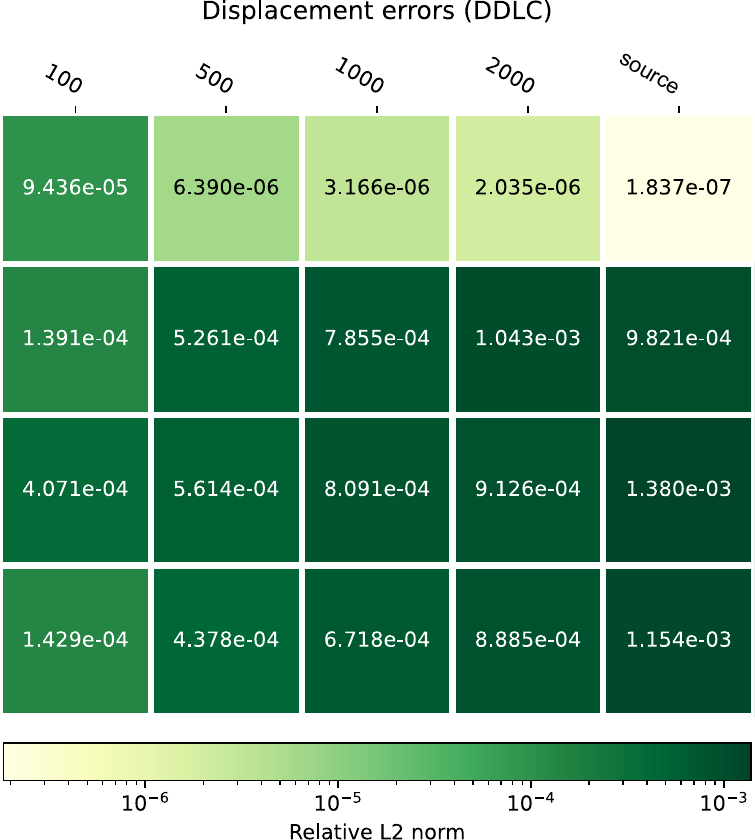}
		\caption{Locally convex search, original database.}
		\label{fig:errors_disp_C_DDLC}
	\end{subfigure}
	\hfill
	\begin{subfigure}{0.5\textwidth}
		\centering
		\includegraphics[width=\textwidth]{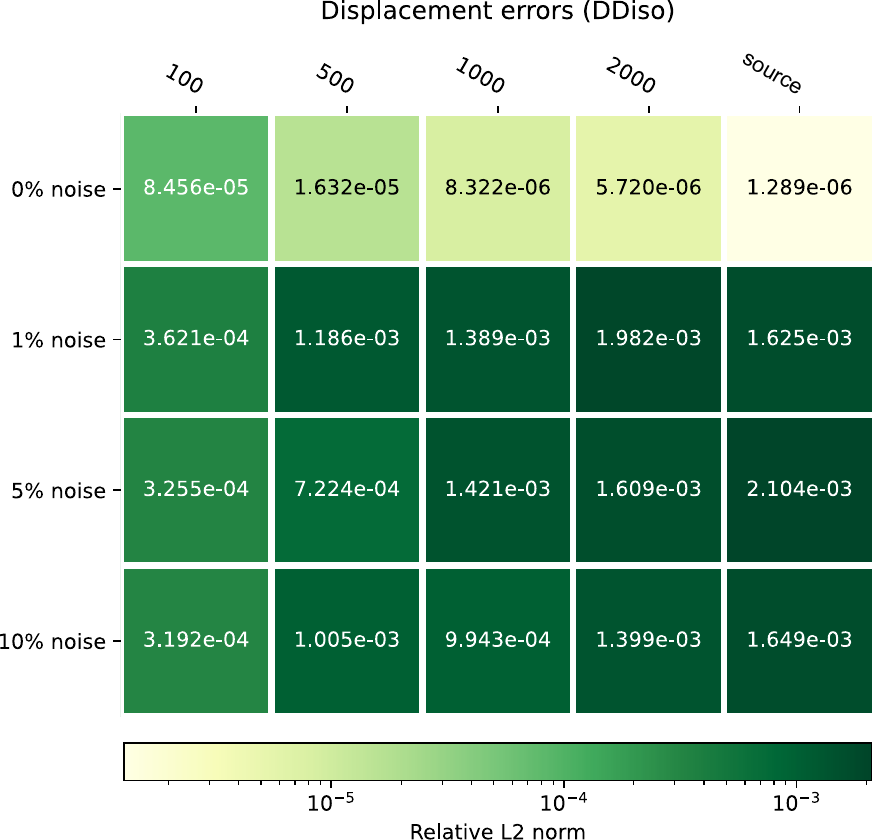}
		\caption{Standard search, enriched database with discretised orbits.}
		\label{fig:errors_disp_C_DDiso}
	\end{subfigure}%
	\begin{subfigure}{0.5\textwidth}
		\centering
		\includegraphics[scale=0.463, trim = {59 0 0 0}, clip=true]{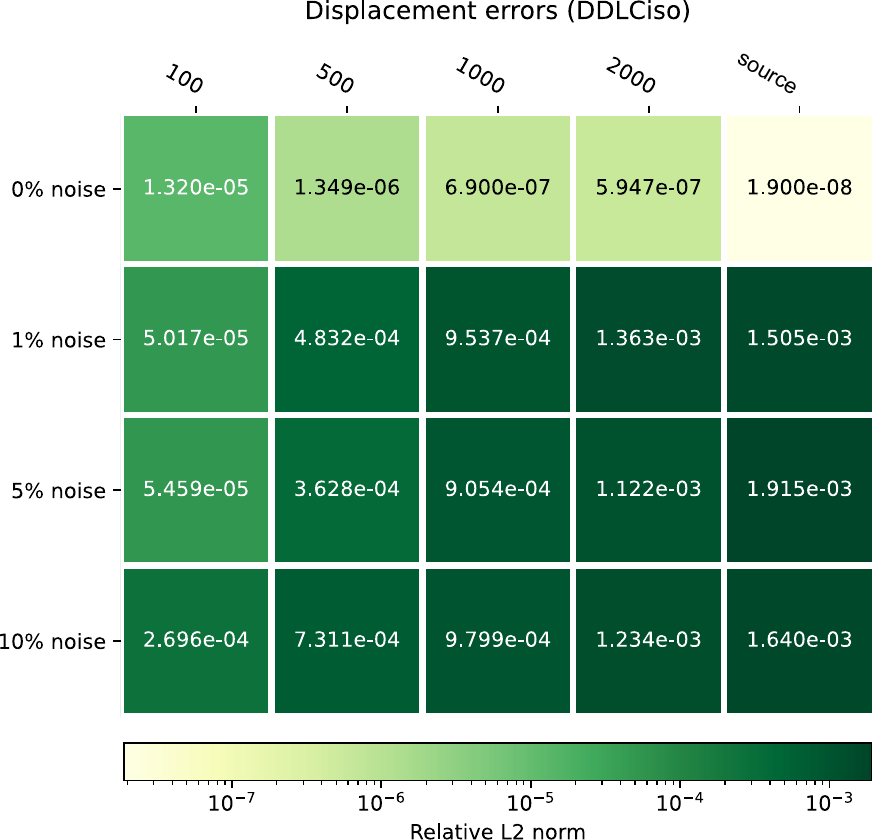}
		\caption{Locally convex, enriched database with discretised orbits.}
		\label{fig:errors_disp_C_DDLCiso}
	\end{subfigure}
	\caption{Relative $L^2$ norm of displacements for Cook membrane with Ciarlet law for DDCM procedures. The labels on top represent the size of the dataset used in a simulation.}
	\label{fig:errors_disps_C_data_driven}
\end{figure}

\begin{figure}
	\centering
	\begin{subfigure}{0.5\textwidth}
		\centering
		\includegraphics[width=\textwidth]{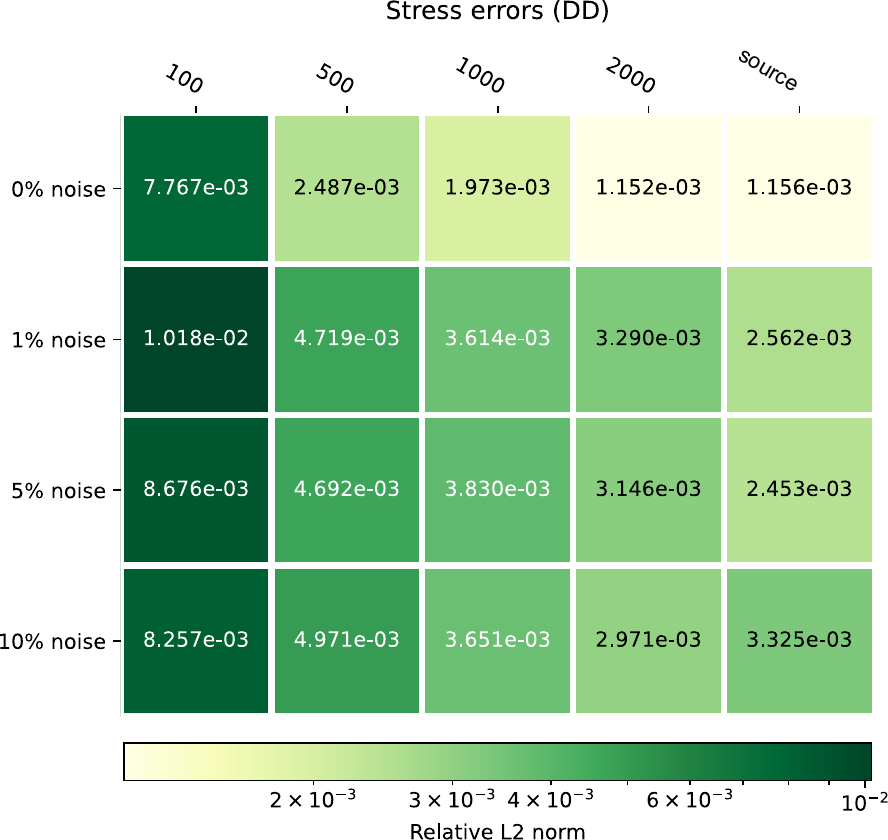}
		\caption{Standard search, original database.}
		\label{fig:errors_stress_C_DD}
	\end{subfigure}%
	\hfill
	\begin{subfigure}{0.5\textwidth}
		\centering
		\includegraphics[scale=0.462, trim={59 0 0 0},clip=true ]{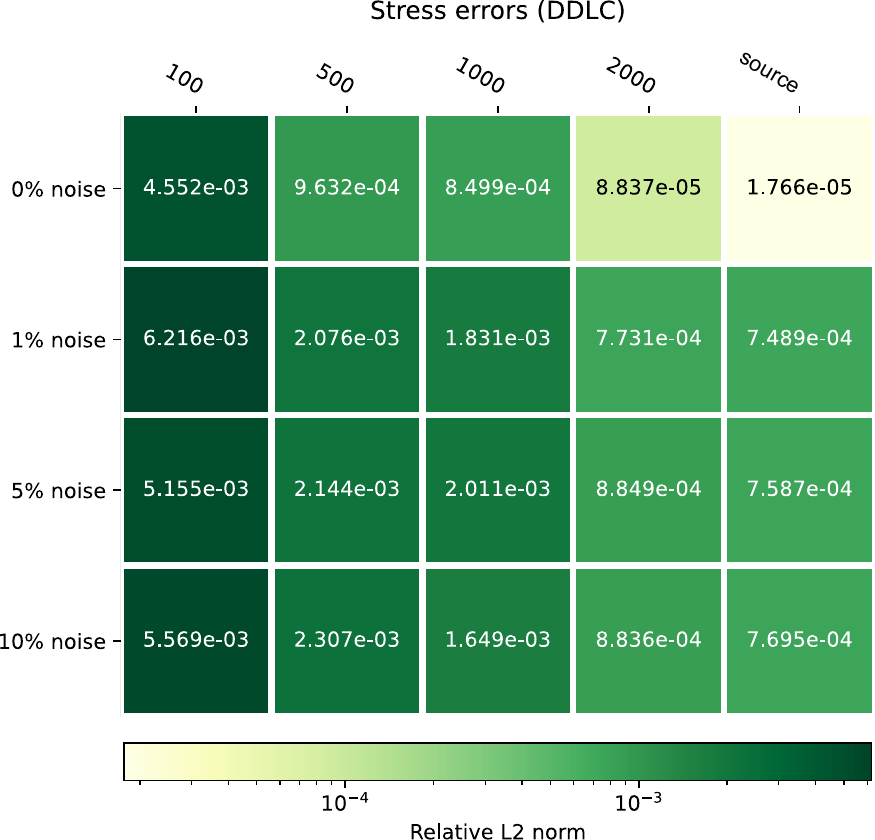}
		\caption{Locally convex search, original database.}
		\label{fig:errors_stress_C_DDLC}
	\end{subfigure}
	
	\begin{subfigure}{0.5\textwidth}
		\centering
		\includegraphics[width=\textwidth]{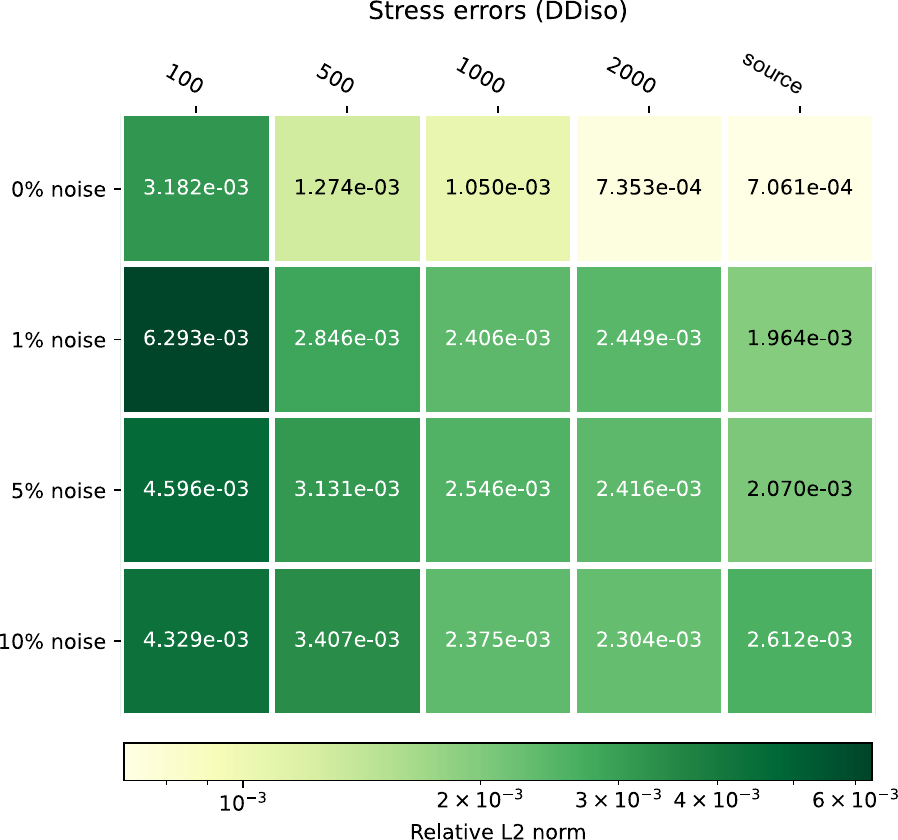}
		\caption{Standard search, enriched database with discretised orbits.}
		\label{fig:errors_stress_C_DDiso}
	\end{subfigure}%
	\hfill
	\begin{subfigure}{0.5\textwidth}
		\centering
		\includegraphics[scale=0.462, trim={59 0 0 0},clip=true ]{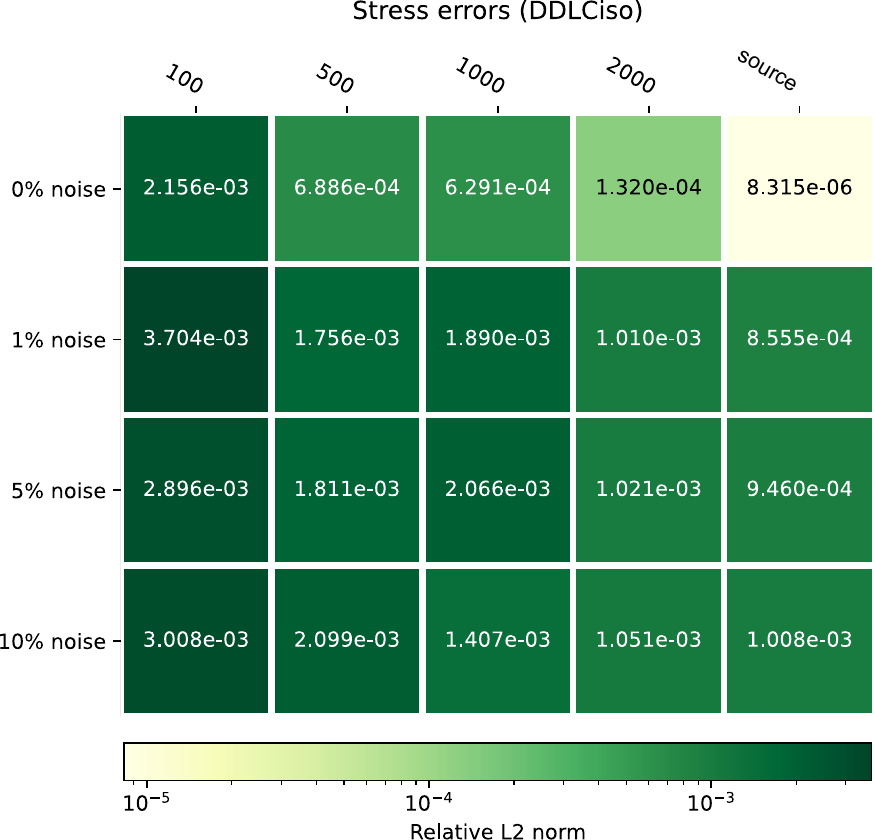}
		\caption{Locally convex, enriched database with discretised orbits.}
		\label{fig:errors_stress_C_DDLCiso}
	\end{subfigure}
	\caption{Relative $L^2$ norm of stress for Cook membrane with Ciarlet law for DDCM procedures.}
	\label{fig:errors_stress_C_data_driven}
\end{figure}

\begin{figure}
	\centering
	\begin{subfigure}{0.5\textwidth}
		\centering
		\includegraphics[width=\textwidth]{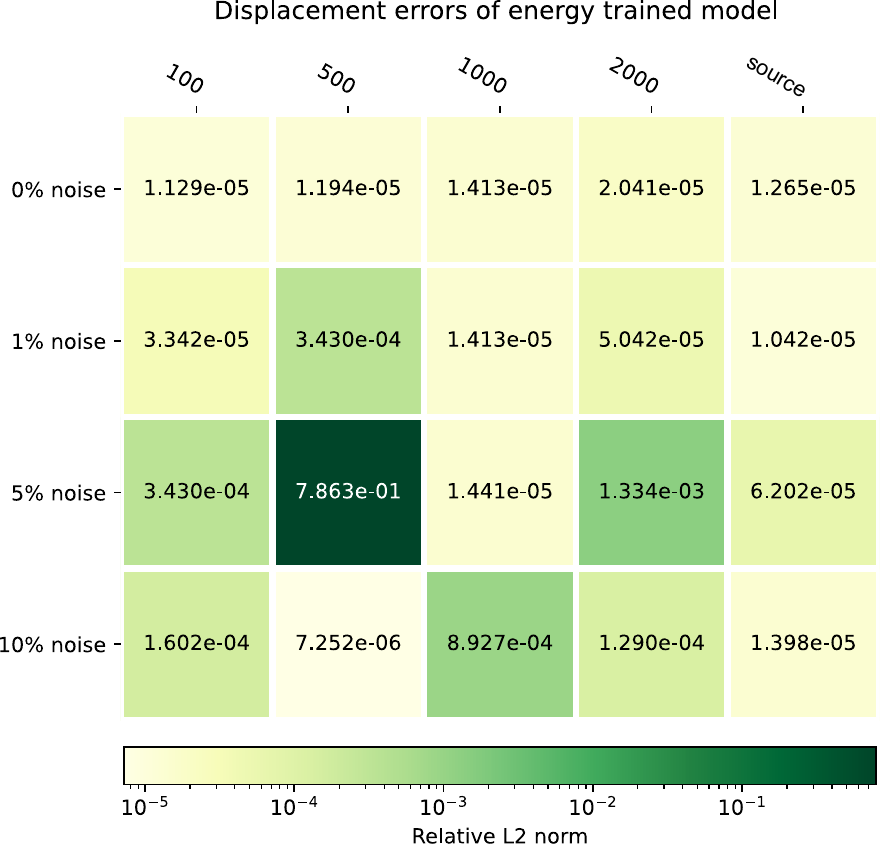}
		\caption{Displacement errors of the best energy trained model .}
		\label{fig:errors_disp_energy_Ciarlet}
	\end{subfigure}%
	\hfill
	\begin{subfigure}{0.5\textwidth}
		\centering
		\includegraphics[scale=0.462, trim={59 0 0 0}, clip=true]{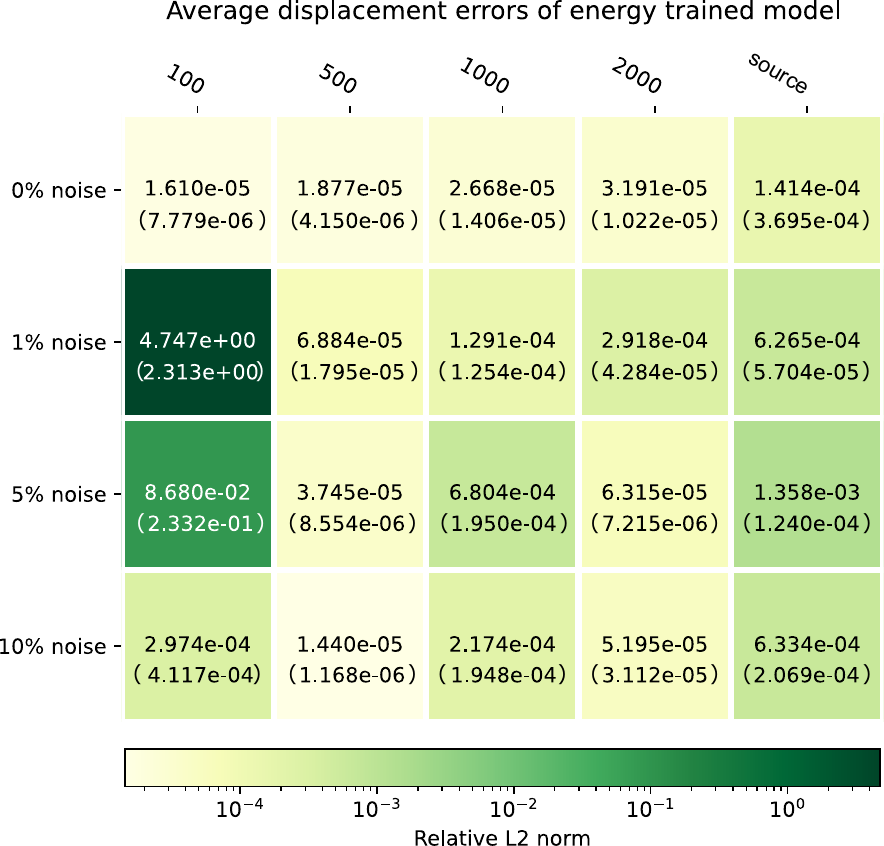}
		\caption{Average displacement errors of energy trained models.}
		\label{fig:errors_disp_energy_Ciarlet_avg}
	\end{subfigure}
	\caption{Relative $L^2$ norm of displacements for neural networks trained on energy, results for Cook membrane with Ciarlet law. The results shown for the average displacement errors are averaged for all training runs of the neural networks. Standard deviations of the averaged errors are shown in parentheses.}
	\label{fig:errors_diff_NN_energy}
\end{figure}

\begin{figure}
	\centering
	\begin{subfigure}{0.5\textwidth}
		\centering
		\includegraphics[width=\textwidth]{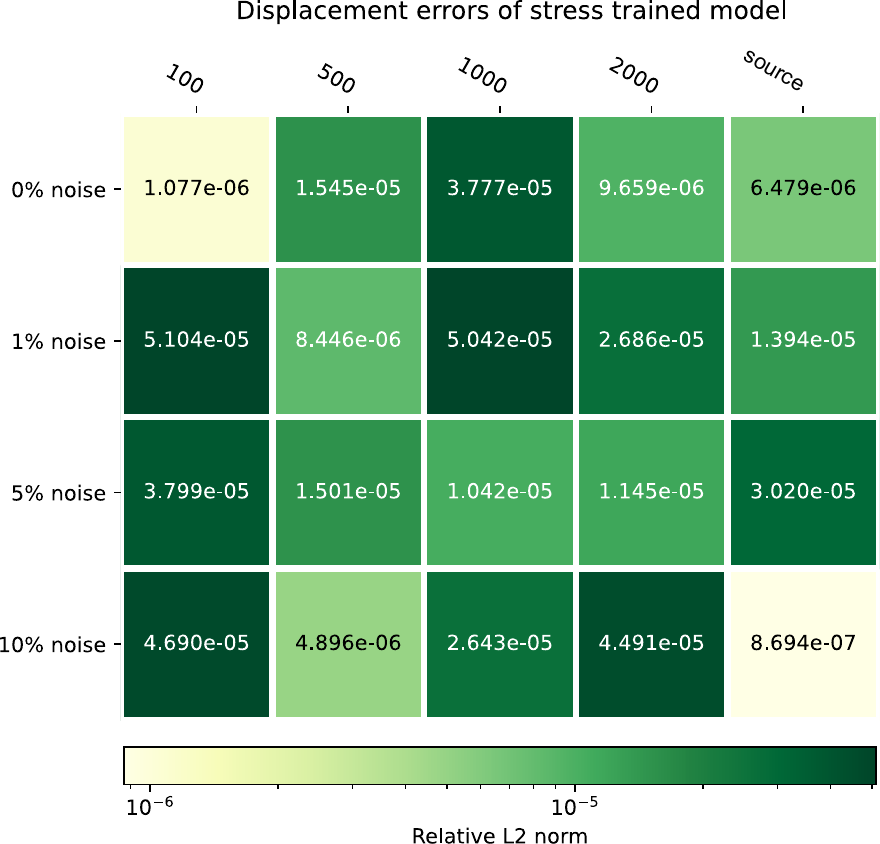}
		\caption{Displacement errors of the best stress trained model.}
		\label{fig:errors_disp_stress_Ciarlet}
	\end{subfigure}%
	\hfill
	\begin{subfigure}{0.5\textwidth}
		\centering
		\includegraphics[scale=0.462, trim={59 0 0 0}, clip=true]{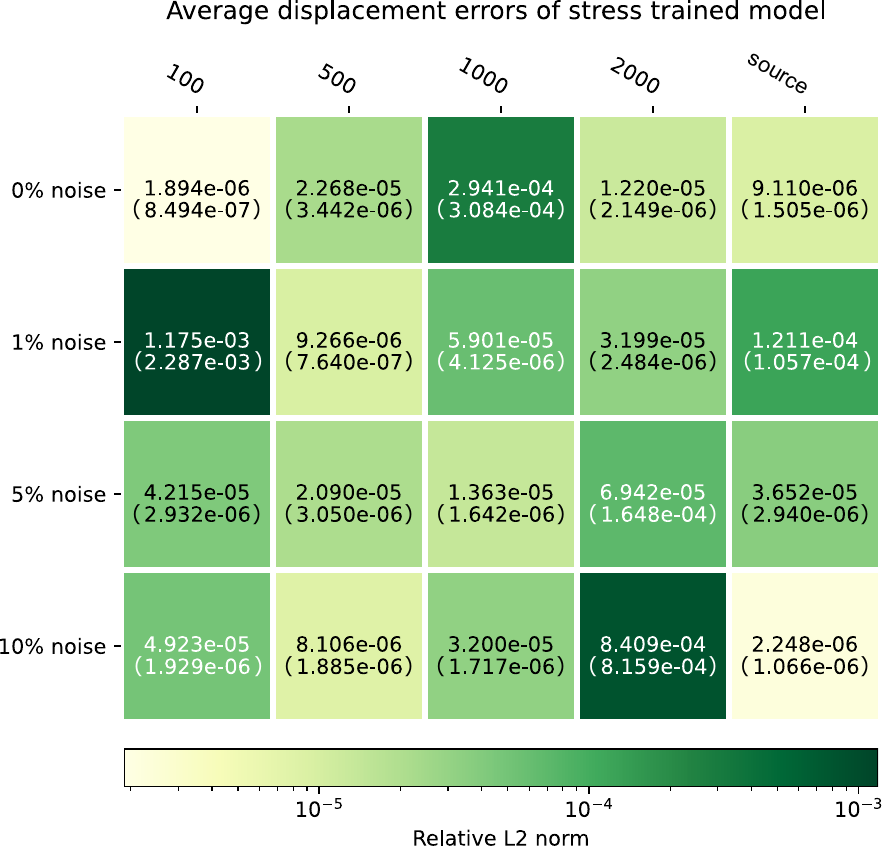}
		\caption{Average displacement errors of stress trained models.}
		\label{fig:errors_disp_stress_Ciarlet_avg}
	\end{subfigure}
	\caption{Relative $L^2$ norm of displacement for neural networks trained on stress, results for Cook membrane with Ciarlet law. The results shown for the average stress errors are averaged for all training runs of neural networks. Standard deviations of the averaged errors are shown in parentheses.}
	\label{fig:errors_diff_NN_stress}
\end{figure}

\begin{figure}[h!]
	\centering
	\begin{subfigure}{0.5\textwidth}
		\centering
		\includegraphics[width=\textwidth]{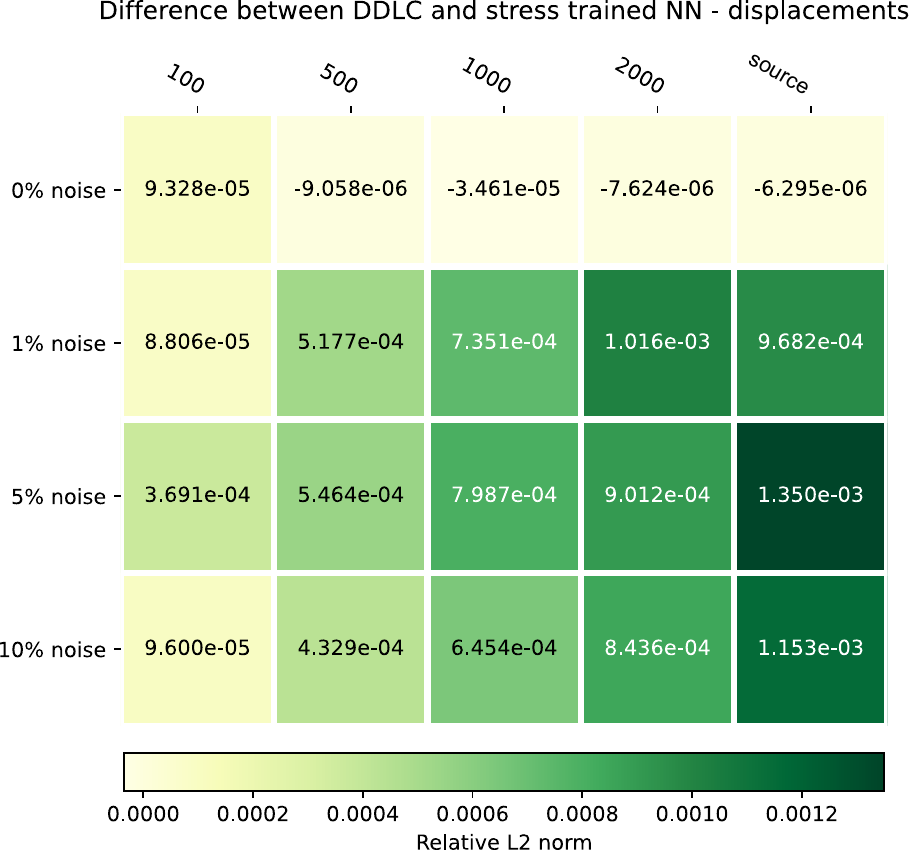}
		\caption{Absolute error difference, displacements.}
		\label{fig:error_diff_DDLC_C_disp}
	\end{subfigure}%
	\hfill
	\begin{subfigure}{0.5\textwidth}
		\centering
		\includegraphics[width=\textwidth]{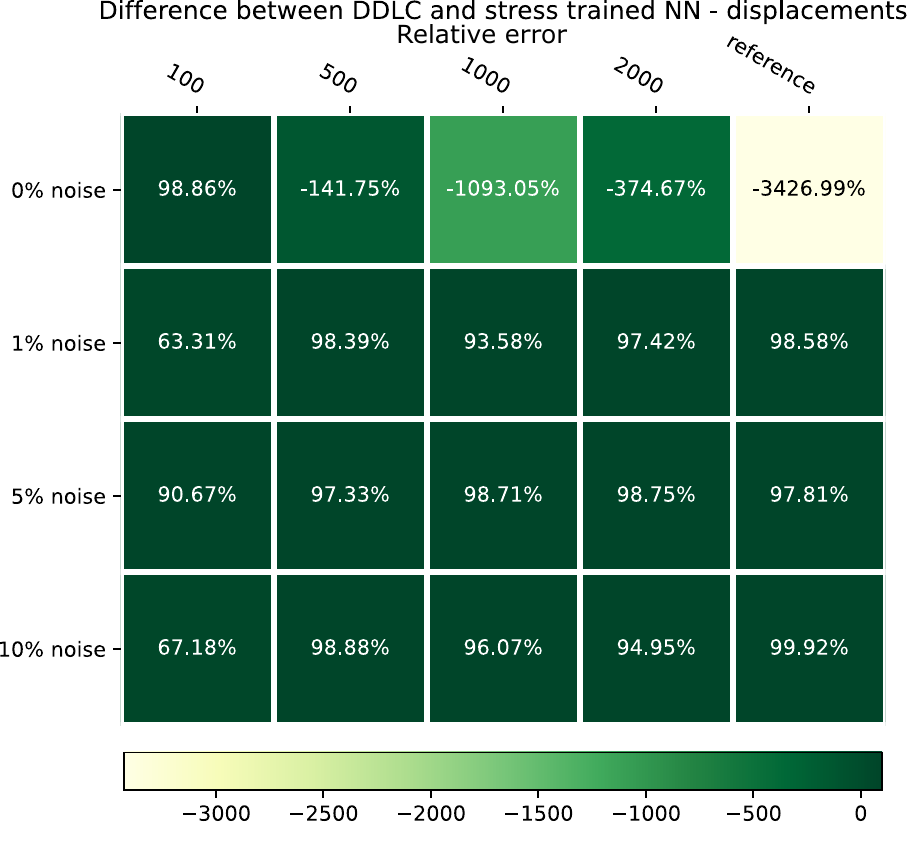}
		\caption{Relative error difference, displacements.}
		\label{fig:rel_error_diff_DDLC_C_disp}
	\end{subfigure}
	\caption{Absolute error difference of the DDLC and NN approach on Cook's membrane with Ciarlet law is shown in the left figure, while the relative difference is shown in the right figure.}
	\label{fig:error_diff_DDLC_C_disp_comparison}
\end{figure}

In this section the DD and NN results are presented. The geometry, boundary conditions and position of the load are shown in Fig.~\ref{fig:geometry_and_mesh_cook}. The mesh used for comparing the DDCM and NN approaches in this section and in Section~\ref{sec:cook_hn} consists of 888 elements. The traction $t$ on the right edge is prescribed to be 20 N/mm. The results are shown for the 5 different datasets with increasing noise levels. The errors presented in Figs. \ref{fig:errors_disps_C_data_driven}-\ref{fig:error_diff_DDLC_C_disp_comparison} are the relative $L_2$ norm of the displacements, strain or stress with respect to the reference FE solution:

\begin{equation}
	\textrm{error}(\bullet) = \frac{\lVert (\bullet) - (\bullet)_{\textrm{ref}} \rVert_2}{\lVert(\bullet)_{\textrm{ref}}\rVert_2},
\end{equation}
and are averaged across the entire domain. The data-driven errors are presented first, followed by the errors for the energy and stress trained neural network models. When presenting the errors for the NN models, figures showing the average error of all the models trained for a case are shown together with the standard deviation in parentheses. In Fig.~\ref{fig:error_diff_DDLC_C_disp_comparison}, the comparison of the relative $L^2$ errors of the two approaches is given. Due to the large number of results available for comparison, many are placed in \ref{sec:additional_figs} for a clearer preview of the results. 
\FloatBarrier

\FloatBarrier

Of the four data-driven approaches compared, the approaches with locally convex search produce the best results, see Fig.~\ref{fig:errors_disps_C_data_driven}. When comparing the DDLC and DDLCiso approaches, DDLCiso performance is worse on noisy data. It is also clear to see that although both the neural networks trained with energy and the neural networks trained with stress perform well on the given example, the errors of the NNs trained with stress are lower, for comparison see Figs.~\ref{fig:errors_disp_energy_Ciarlet}~and~\ref{fig:errors_disp_stress_Ciarlet}. This can be attributed to the fact that when training on stress, the NN is trained on its derivatives resulting in an overall smoother model. In addition, if the performance of the best trained model is compared to the average error of all trained models for a certain case, see Figs.~\ref{fig:errors_diff_NN_energy}~and~\ref{fig:errors_diff_NN_stress}, the difference between the best model and average of all models is much lower when NNs are trained on stress. Therefore, the results of the NNs trained with stress and the DDLC approach is used to compare the NN and model-free approaches. The errors of the DDLC approach and the best stress trained NN results are subtracted, $(\bullet)_{DD} - (\bullet)_{NN}$, so that negative values represent the DD approach is better in the given case whereas positive values represent that the NN approach is better. Additionally, in order to compare the relative performance of the DD and NN approaches, the relative error of the NN model w.r.t. the DDLC model was also calculated by dividing the former error difference with the DD error $\left((\bullet)_{DD} - (\bullet)_{NN}\right)/(\bullet)_{DD}$. With this metric the methods can be compared more directly. In this way both the actual error of the approaches can be seen in addition to their relative performance. In Fig.~\ref{fig:error_diff_DDLC_C_disp} the DDLC  approach was better in most of the non noisy examples, however the NN models outperformed it in all other cases. It should be noted that when the smallest dataset of 100 samples is used  the NN approach outperformed the DDCM approach with no noise included, demonstrating better performance on a small dataset.


\FloatBarrier

\begin{figure}[h!]
	\centering
	\begin{subfigure}{\textwidth}
		\centering
		\includegraphics[width=\textwidth]{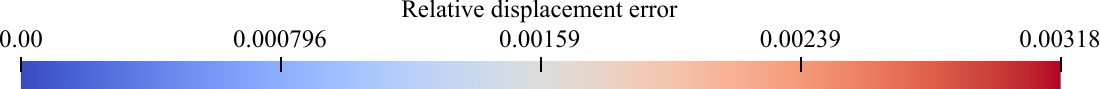}
	\end{subfigure}
	\begin{subfigure}{0.5\textwidth}
		\centering
		\includegraphics[width=0.95\textwidth]{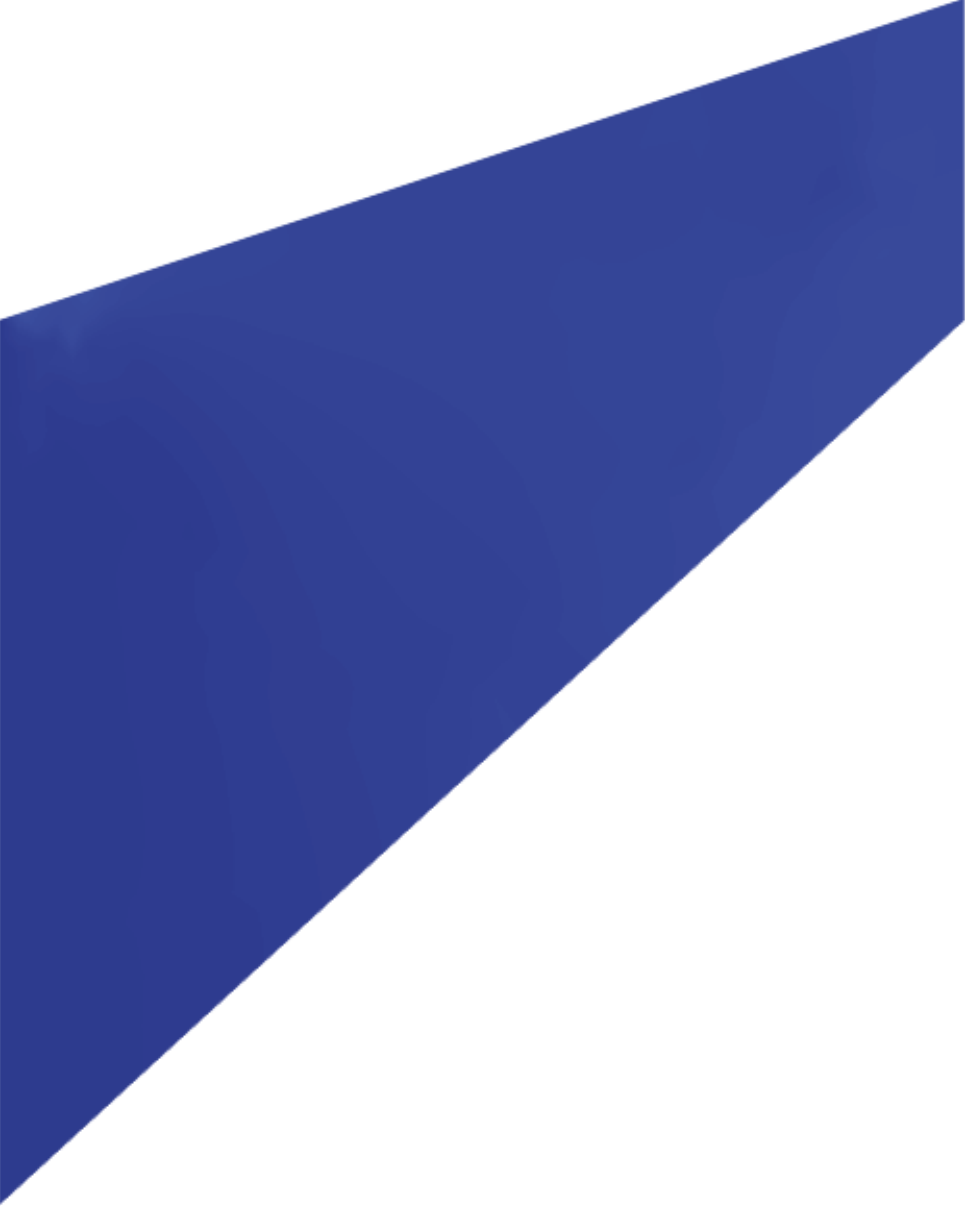}
		\caption{DDLCiso relative displacement error, largest error is $0.00078\%$.}
		\label{fig:error_DDLCiso_disp_contour_cook_C_source}
	\end{subfigure}%
	\hfill
	\begin{subfigure}{0.5\textwidth}
		\centering
		\includegraphics[width=0.95\textwidth]{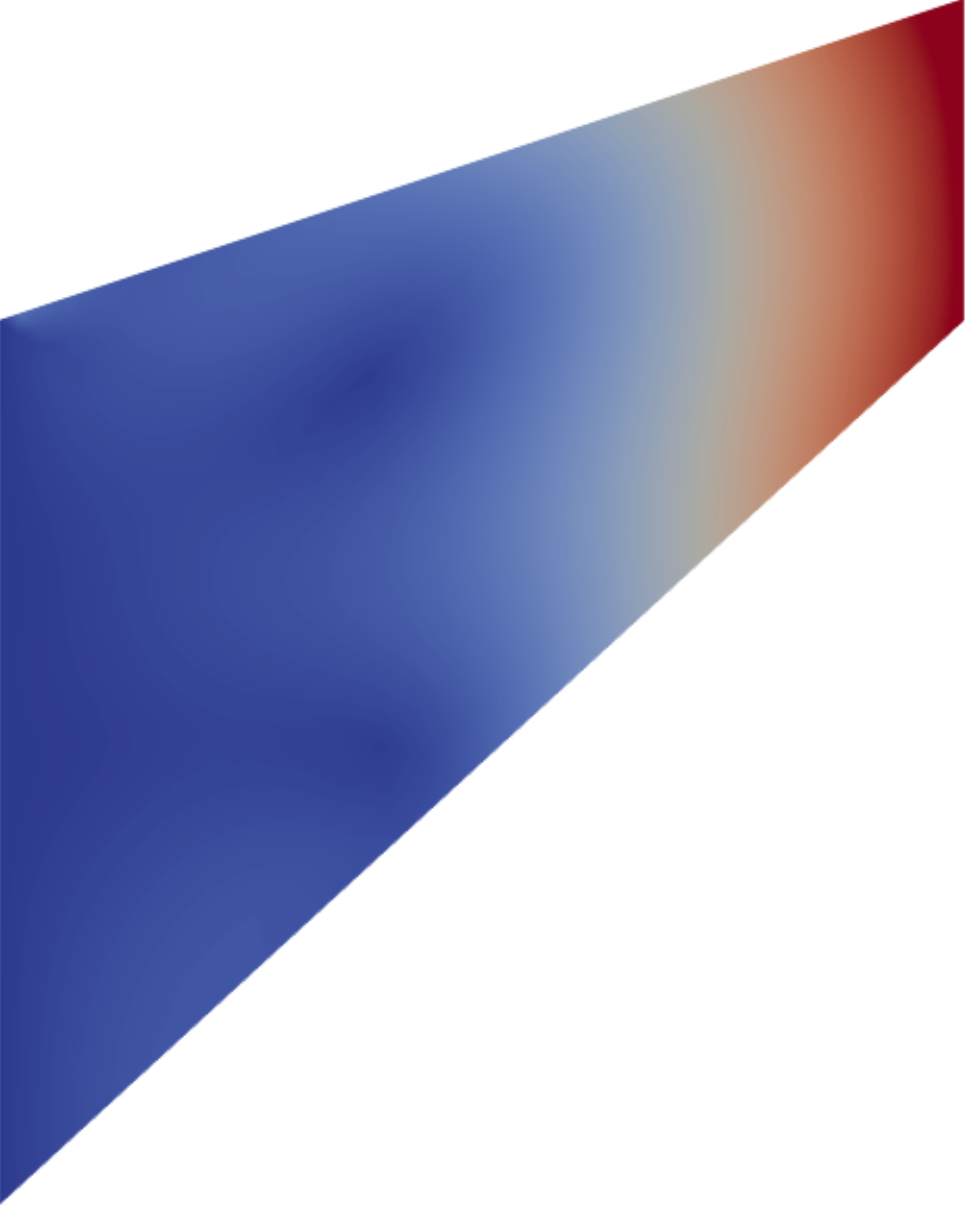}
		\caption{Stress trained NN relative displacement error, largest error is $0.32\%$.}
		\label{fig:error_NN_disp_contour_cook_C_source}
	\end{subfigure}
	\begin{subfigure}{\textwidth}
		\centering
		\includegraphics[width=\textwidth]{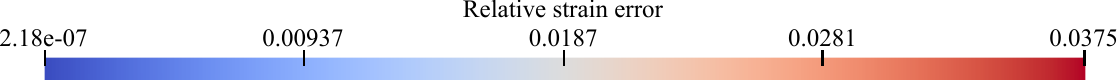}
	\end{subfigure}
	\begin{subfigure}{0.5\textwidth}
		\centering
		\includegraphics[width=0.95\textwidth]{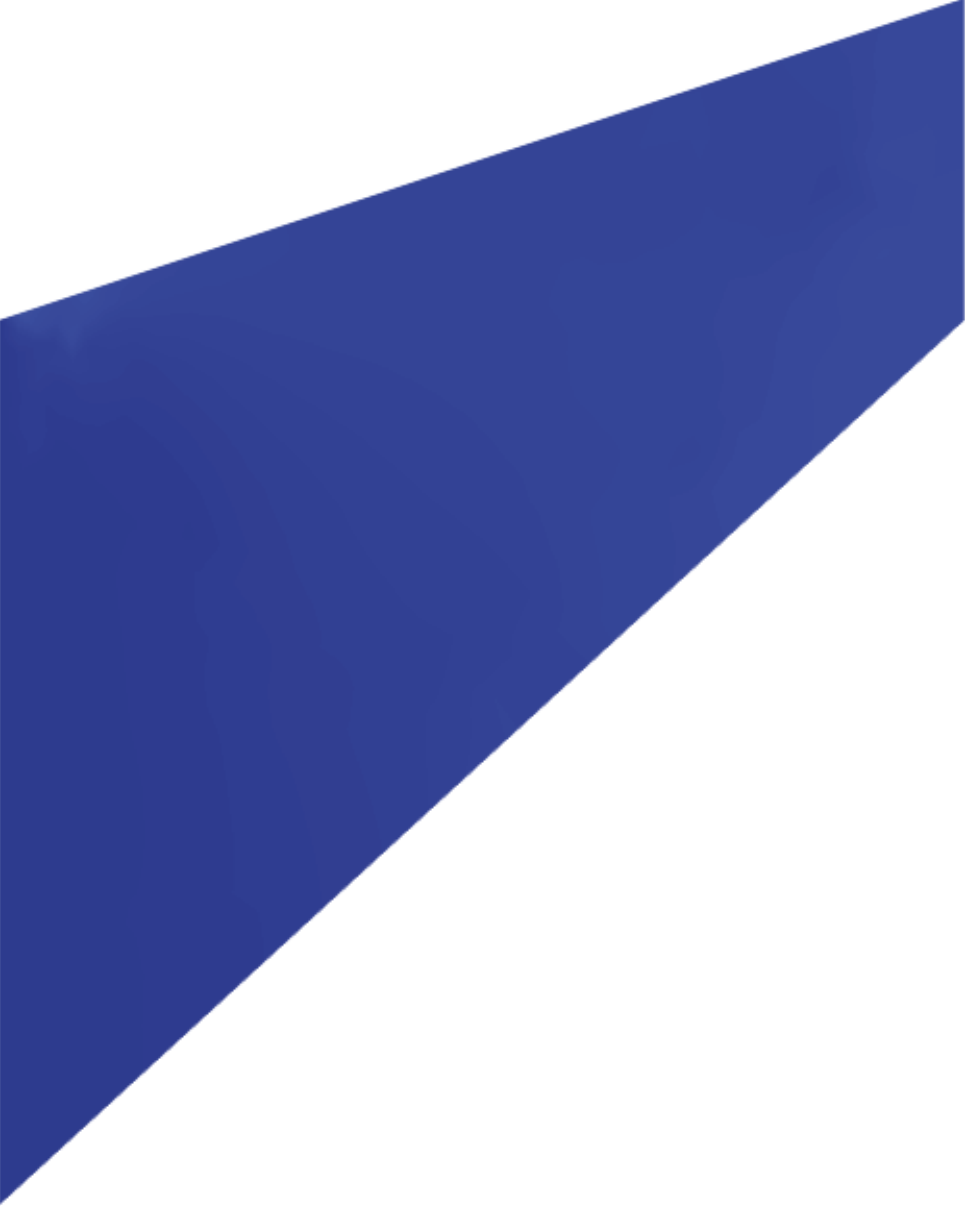}
		\caption{DDLCiso relative strain errors contour plot, largest error is $0.053\%$.}
		\label{fig:error_DDLCiso_strain_contour_cook_C_source}
	\end{subfigure}%
	\hfill
	\begin{subfigure}{0.5\textwidth}
		\centering
		\includegraphics[width=0.95\textwidth]{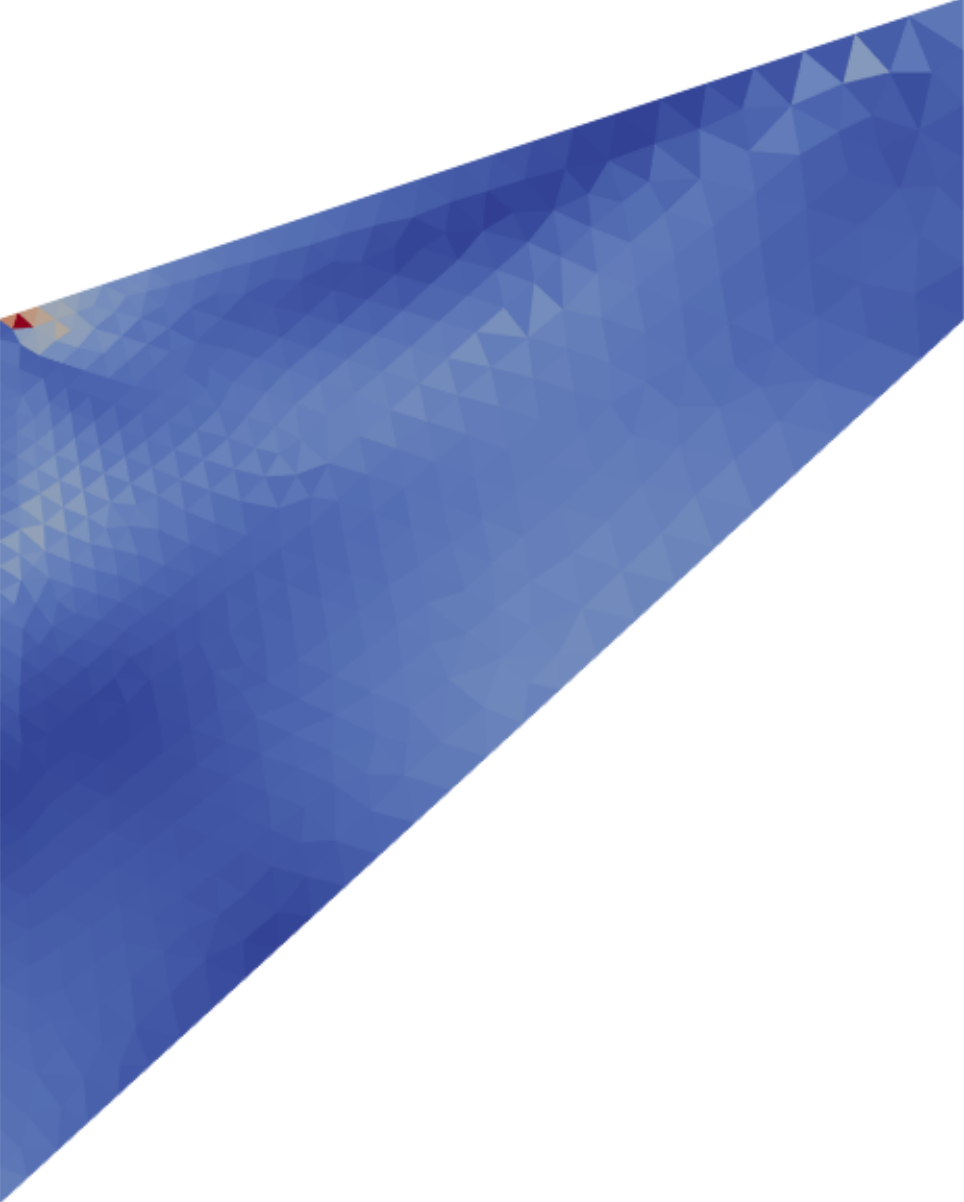}
		\caption{Stress trained NN relative strain errors contour plot, largest error is $3.75\%$.}
		\label{fig:error_NN_strain_contour_cook_C_source}
	\end{subfigure}
\end{figure}
~
\begin{figure}\ContinuedFloat
	\begin{subfigure}{\textwidth}
		\centering
		\includegraphics[width=\textwidth]{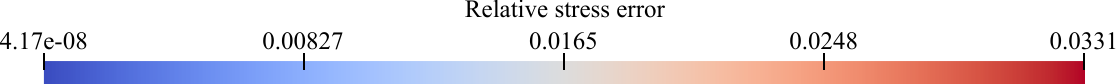}
	\end{subfigure}
	\begin{subfigure}{0.5\textwidth}
		\centering
		\includegraphics[width=0.95\textwidth]{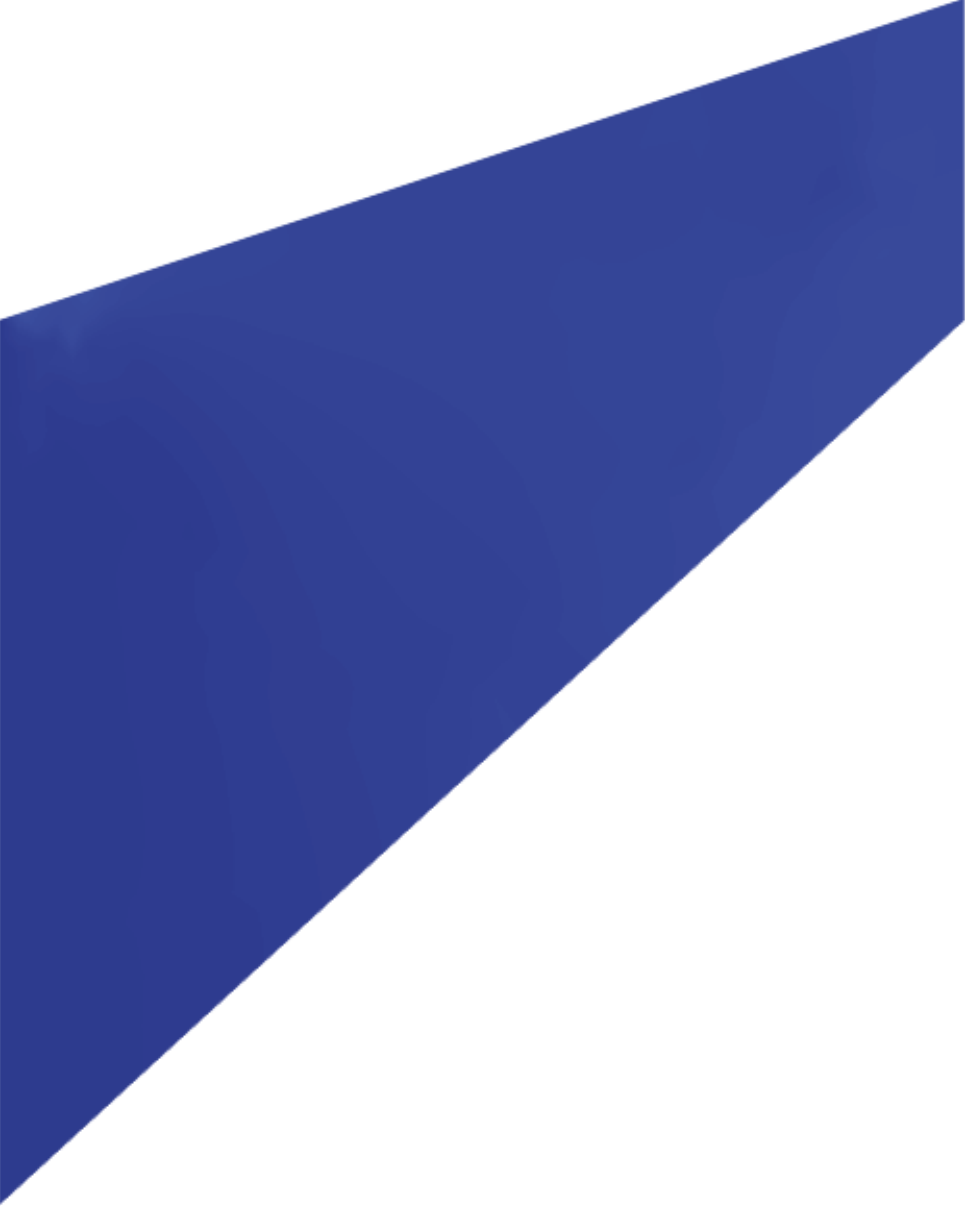}
		\caption{DDLCiso relative stress errors contour plot, largest error is $0.056\%$.}
		\label{fig:error_DDLCiso_stress_contour_cook_C_source}
	\end{subfigure}%
	\hfill
	\begin{subfigure}{0.5\textwidth}
		\centering
		\includegraphics[width=0.95\textwidth]{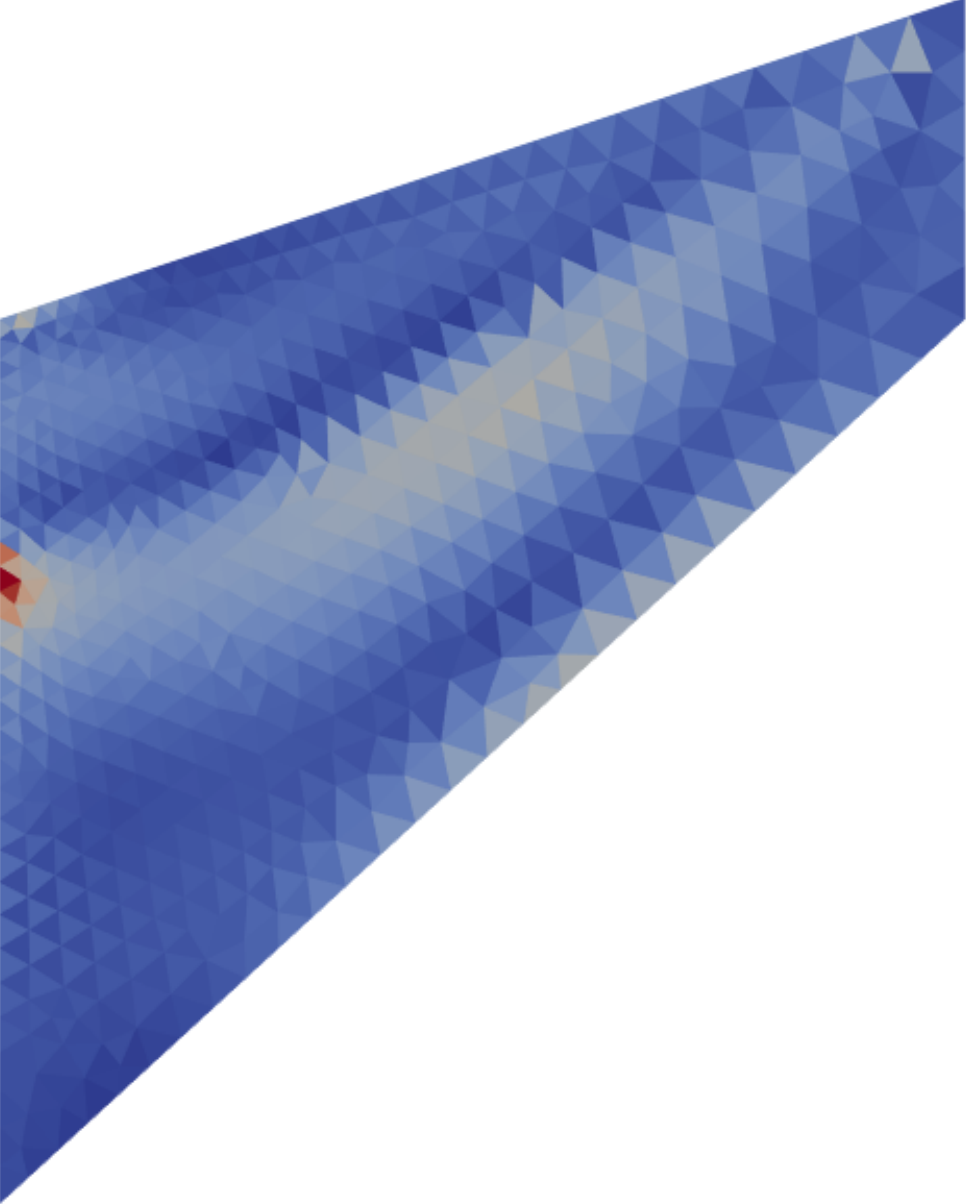}
		\caption{Stress trained NN relative stress errors contour plot, largest error is $3.31\%$.}
		\label{fig:error_NN_stress_contour_cook_C_source}
	\end{subfigure}
	\caption{Contour plots of the displacement, strain and stress errors of the DDLCiso and stress trained NN solutions on the source mesh from which the data was sampled. The results were obtained useing the full source dataset without noise.}
	\label{fig:error_DDLCiso_NN_contour_cook_C_source}
\end{figure}

\FloatBarrier

\begin{figure}[h!]
	\centering
	\begin{subfigure}{\textwidth}
		\centering
		\includegraphics{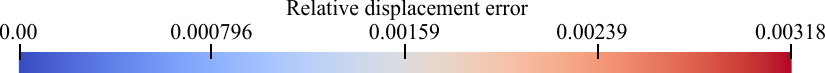}
	\end{subfigure}
	\begin{subfigure}{0.5\textwidth}
		\centering
		\includegraphics[width=0.95\textwidth]{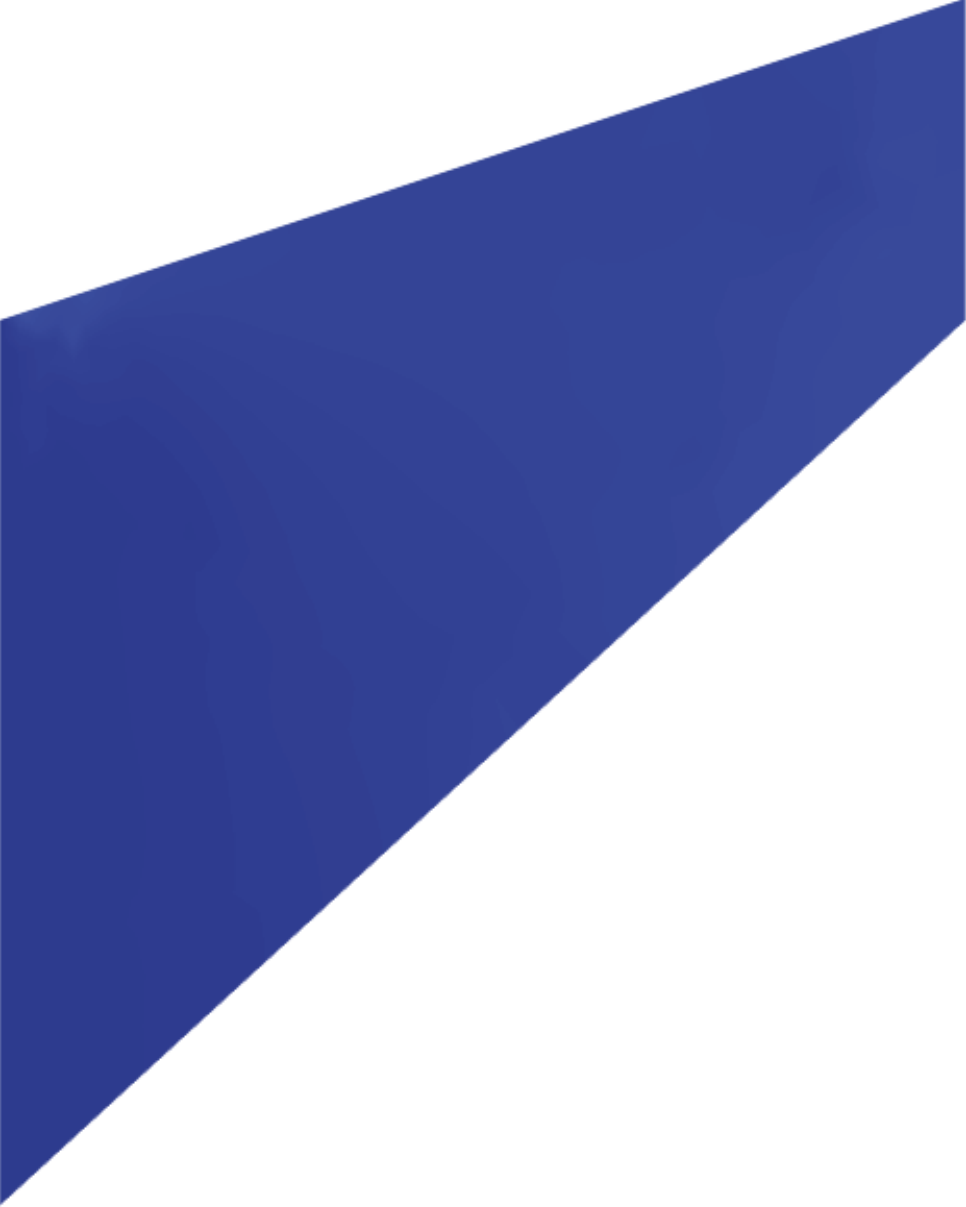}
		\caption{DDLCiso relative displacement error, largest error is $0.019\%$.}
		\label{fig:error_DDLCiso_disp_contour_cook_C}
	\end{subfigure}%
	\hfill
	\begin{subfigure}{0.5\textwidth}
		\centering
		\includegraphics[width=0.95\textwidth]{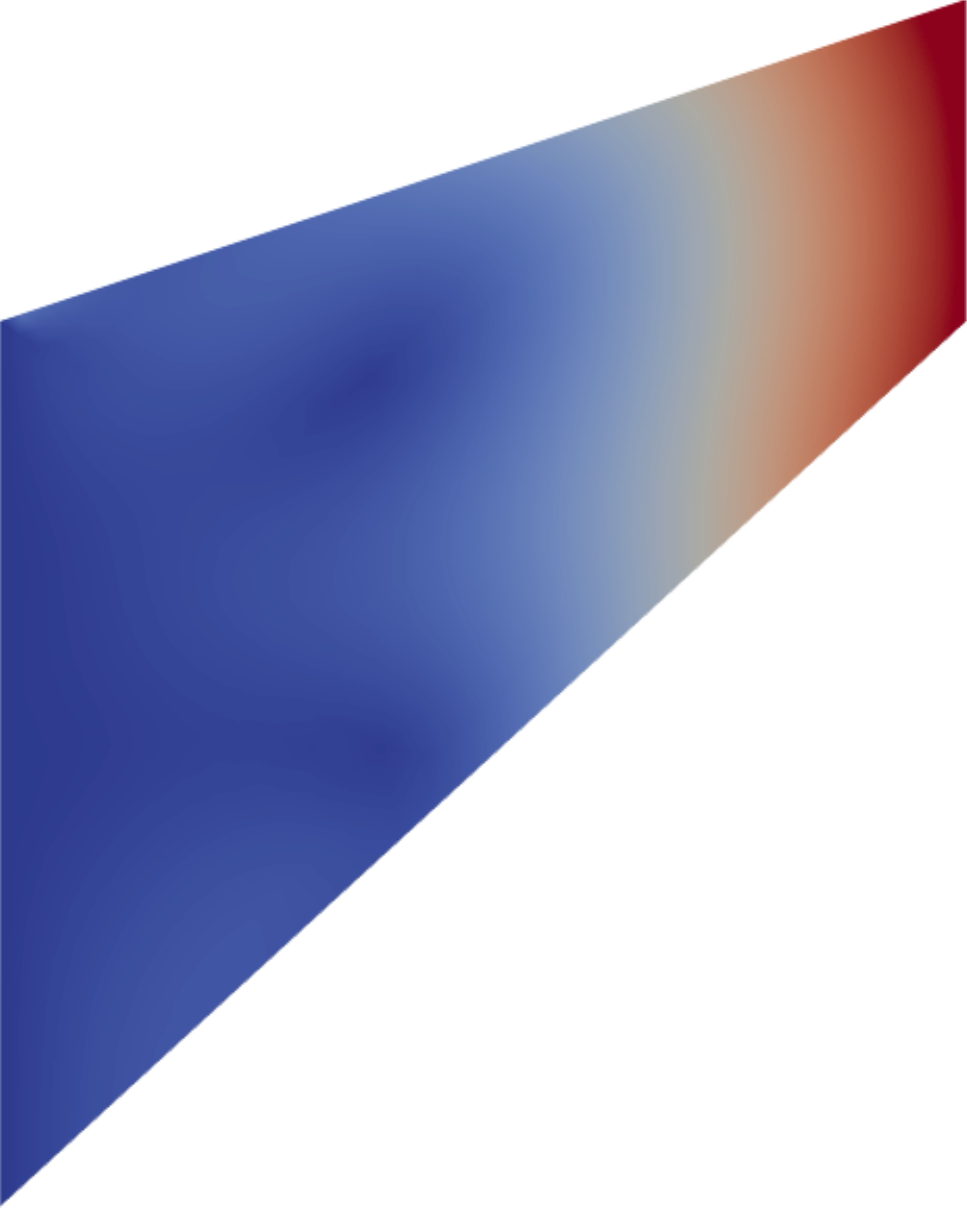}
		\caption{Stress trained NN relative displacement error, largest error is $0.32\%$.}
		\label{fig:error_NN_disp_contour_cook_C}
	\end{subfigure}
	\begin{subfigure}{\textwidth}
		\centering
		\includegraphics{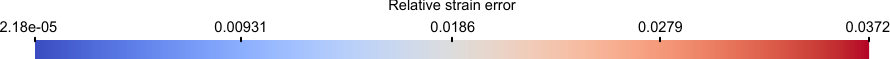}
	\end{subfigure}
	\begin{subfigure}{0.5\textwidth}
		\centering
		\includegraphics[width=0.95\textwidth]{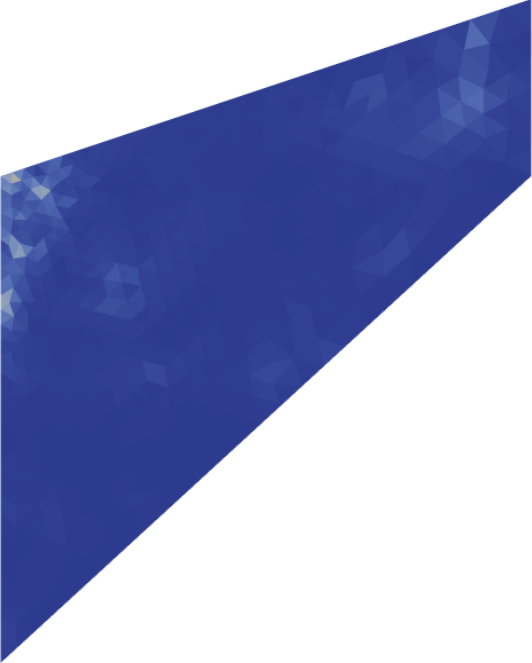}
		\caption{DDLCiso relative strain errors contour plot, largest error is $0.17\%$.}
		\label{fig:error_DDLCiso_strain_contour_cook_C}
	\end{subfigure}%
	\hfill
	\begin{subfigure}{0.5\textwidth}
		\centering
		\includegraphics[width=0.95\textwidth]{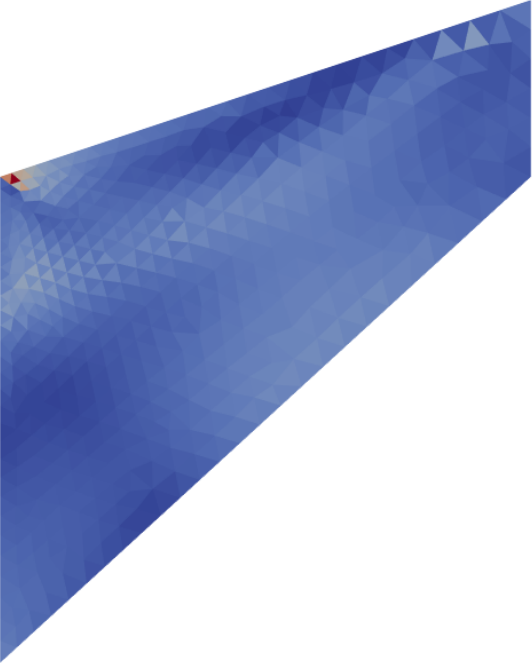}
		\caption{Stress trained NN relative strain errors contour plot, largest error is $3.72\%$.}
		\label{fig:error_NN_strain_contour_cook_C}
	\end{subfigure}
\end{figure}
~
\begin{figure}\ContinuedFloat
	\begin{subfigure}{\textwidth}
		\centering
		\includegraphics{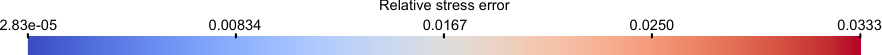}
	\end{subfigure}
	\begin{subfigure}{0.5\textwidth}
		\centering
		\includegraphics[width=0.95\textwidth]{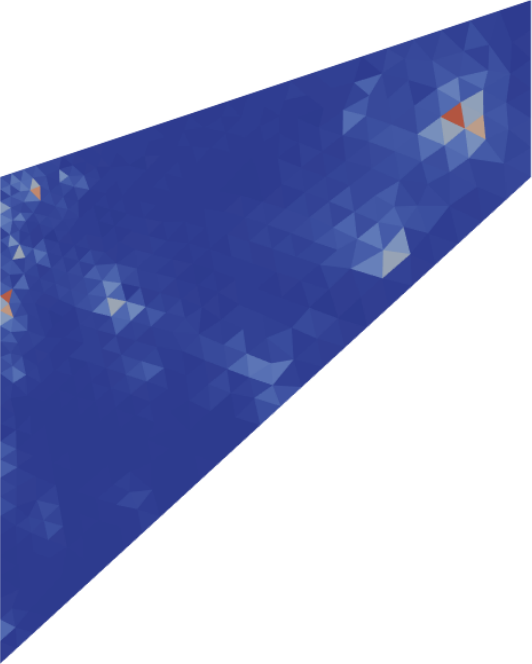}
		\caption{DDLCiso relative stress errors contour plot, largest error is $2.81\%$.}
		\label{fig:error_DDLCiso_stress_contour_cook_C}
	\end{subfigure}%
	\hfill
	\begin{subfigure}{0.5\textwidth}
		\centering
		\includegraphics[width=0.95\textwidth]{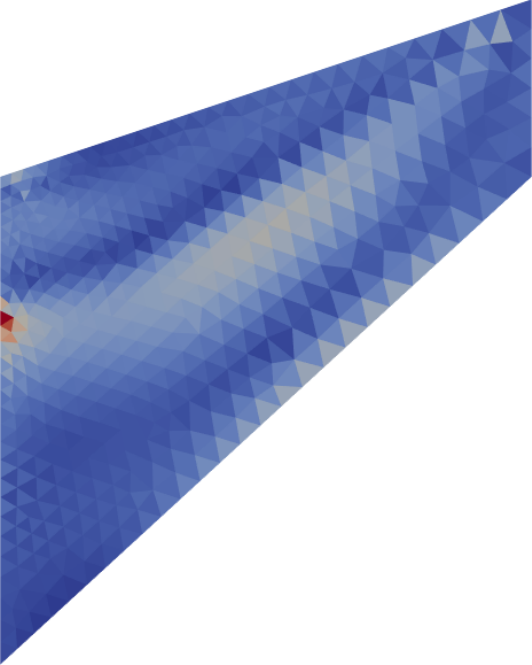}
		\caption{Stress trained NN relative stress errors contour plot, largest error is $3.33\%$.}
		\label{fig:error_NN_stress_contour_cook_C}
	\end{subfigure}
	\caption{Contour plots of the displacement, strain and stress errors of the DDLCiso and stress trained NN solutions. The results are obtained by using the source dataset without noise on a modified mesh.}
	\label{fig:error_DDLCiso_NN_contour_cook_C}
\end{figure}

In Fig.~\ref{fig:error_DDLCiso_NN_contour_cook_C_source} the relative displacement, strain and stress error plots over the undeformed mesh of the original Cook membrane from which the source dataset was sampled are shown. In Figs.~\ref{fig:error_DDLCiso_disp_contour_cook_C_source}~\&~\ref{fig:error_NN_disp_contour_cook_C_source}, the relative errors are obtained by dividing with the maximum displacement of the top-right corner. The DDCM errors are lower than the NN errors which is to be expected since the problem solved is exactly the same one from which the data is gathered. The results on the source mesh are shown in order to compare the methods when solving the same problem, but using a modified FE mesh from which the source dataset was not gathered.

In Fig.~\ref{fig:error_DDLCiso_NN_contour_cook_C} the relative displacement, strain and stress error plots over the undeformed mesh of the Cook membrane are shown using a modified mesh. The results for the NN model are smoother than for the DDLCiso solution. The error values are similar in both solutions, with the DDCM approach giving slightly better results. In Figs.~\ref{fig:error_DDLCiso_disp_contour_cook_C}~\&~\ref{fig:error_NN_disp_contour_cook_C}, the relative errors are obtained in the same manner as in Figs.~\ref{fig:error_DDLCiso_disp_contour_cook_C_source}~\&~\ref{fig:error_NN_disp_contour_cook_C_source}.

When comparing the results of the unmodified, Fig.~\ref{fig:error_DDLCiso_NN_contour_cook_C_source}, and the modified, Fig.~\ref{fig:error_DDLCiso_NN_contour_cook_C}, the main differences between the purely data-driven and the NN model approach become clear. The problem is basically the same, only the FE discretisation is different, and when observing the NN solutions in both figures, it can be seen that the maximum errors are almost identical. However, the DDLCiso solution has almost perfect solutions when it solves the mesh on which the dataset was gathered, which is to be expected since it contains the exact solutions of the problem present in the dataset.

When obtaining the solutions for Figs.~\ref{fig:error_DDLCiso_NN_contour_cook_C_source}~and~\ref{fig:error_DDLCiso_NN_contour_cook_C}the source dataset without noise was used. The presented results are in line with those shown in Fig.~\ref{fig:error_diff_DDLC_C_disp_comparison} where in that case the DDCM solution outperformed the NN one.

\begin{figure}
	\centering
	\includegraphics{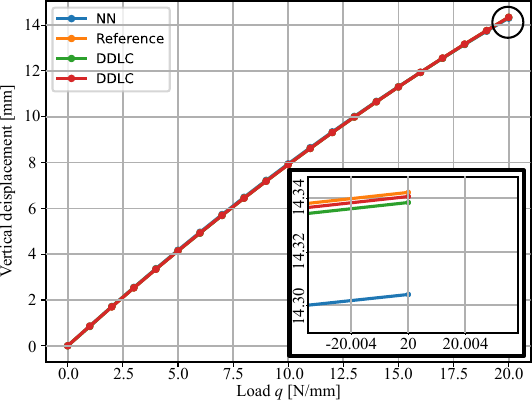}
	\caption{Total displacement of the top-right corner of Cook's membrane. The results are obtained by using the source dataset without noise and a modified mesh.}
	\label{fig:tip_displacement_cook_C}
\end{figure}

The displacement of the top-right corner is shown in Fig.~\ref{fig:tip_displacement_cook_C}, where both the DDCM and NN solutions are nearly identical to the reference solution with errors of 0.01\% and 0.27\% for the DDLCiso and NN solutions respectively.

\FloatBarrier

Finally, the online performance of the different approaches can be compared in Table~\ref{tab:execution_times}. Results are shown for the NNs trained on energy and on stress as well as the results for the DDLC and DDLCiso approaches. Reported values are relative to the original FE simulation, i.e. execution time ratios greater than 1 mean that the calculation is slower than the original one. When comparing the results, it must be taken into account that the training of a single NN took about 2 hours. Thus, if 10 NNs were trained per case, it took 20 hours to obtain the best NN for a single case. Therefore, as offline time are several orders of magnitude higher, total time comparison are pointless. On the other hand, the DDCM approaches are ready to use without the need for training (the only preparation is the creation of the nearest neighbour search trees, which is considered an online cost here for the sake of simplicity and is negligible). In this respect, the DDCM approaches are much faster to deploy. However, if many calculations are required, e.g. in an optimisation, then NNs may be the better choice. Also, NNs are a model, and as can be seen in the table, the execution times between the different cases are quite similar and can be said to be almost identical. Indeed, most the additional running time can be attributed the non-optimised management of stress and tangent tensor data at integration points. Our implementation for incorporation NN-based constitutive laws alongside FEniCS is analogous to the performed for FE$^2$ simulations as in \texttt{micmacsfenics} \cite{micmacsfenics}. A potentially more optimised implementation can be built upon the native abstraction \texttt{ExternalOperator} \cite{Bouziani2021,Latyshev2024}, recently introduced in FEniCSx. For the DDCM approaches and especially for DDLCiso, in addition to aforementioned data management issues, the execution times increase when the data is noisy and larger as more iterations are needed for DDCM convergence and nearest neighbours searches becomes costlier. As already commented, the latter cost can be further decreased using more efficient nearest neighbours algorithms.

\begin{table}
	\centering
	\begin{tabular}{ c||c|c||c|c }
		& \multicolumn{2}{c}{Energy trained NNs} & \multicolumn{2}{c}{Stress trained NNs} \\
		\hline
		\text{noise} & N100 & source & N100 & source\\
		\hline
		\hline
		0\% & 4.198 & 4.054 & 4.085 & 4.085 \\
		\hline
		1\% & 5.045 & 4.058 & 4.166 & 4.094 \\
		\hline
		5\% & 4.702 & 4.032 & 4.012 & 4.157 \\
		\hline
		10\% & 3.916 & 4.06 & 4.155 & 4.162 \\
		\hline
		\hline
		& \multicolumn{2}{c}{DDLC} & \multicolumn{2}{c}{DDLCiso} \\
		\hline
		0\% & 9.868 & 10.52 & 40.94 & 164.2 \\
		\hline
		1\% & 10.57 & 15.7 & 30.92 & 189.1 \\
		\hline
		5\% & 11.16 & 13.59 & 33.41 & 168.7 \\
		\hline
		10\% & 10.44 & 11.22 & 29.61 & 171.5
	\end{tabular}
	\caption{Online computational performance comparison for smallest dataset with 100 samples and full source dataset used in terms of execution time ratios (DDCM or NN relative the original FE solution with prescribed material model). These values show how many times slower the NN or DDCM solution is in comparison to the original FE solution.}
	\label{tab:execution_times}
\end{table}

\FloatBarrier
\subsection{Cook Membrane with Hartmann-Neff law}\label{sec:cook_hn}

To further test the applicability of the DD and NN approaches, the Cook membrane problem is again solved using the hyperelastic Hartmann-Neff law. 
It should be noted that the dataset from Section~\ref{sec:cook_ciarlet} is also used here, retaining the same strain values but recalculating the stresses to match the Hartmann-Neff law. Therefore, the database in this example does not necessarily contain stress values that are close to the original FE solution, as is the case in Sec.~\ref{sec:cook_ciarlet}. In this way, the sensitivity of the DD and the NN approach to non-ideal data-sets is tested, in addition to addressing a material with $I_2$-dependency. The results are presented in the same manner as in the previous section.

\begin{figure}
	\centering
	\begin{subfigure}{0.5\textwidth}
		\centering
		\includegraphics[width=\textwidth]{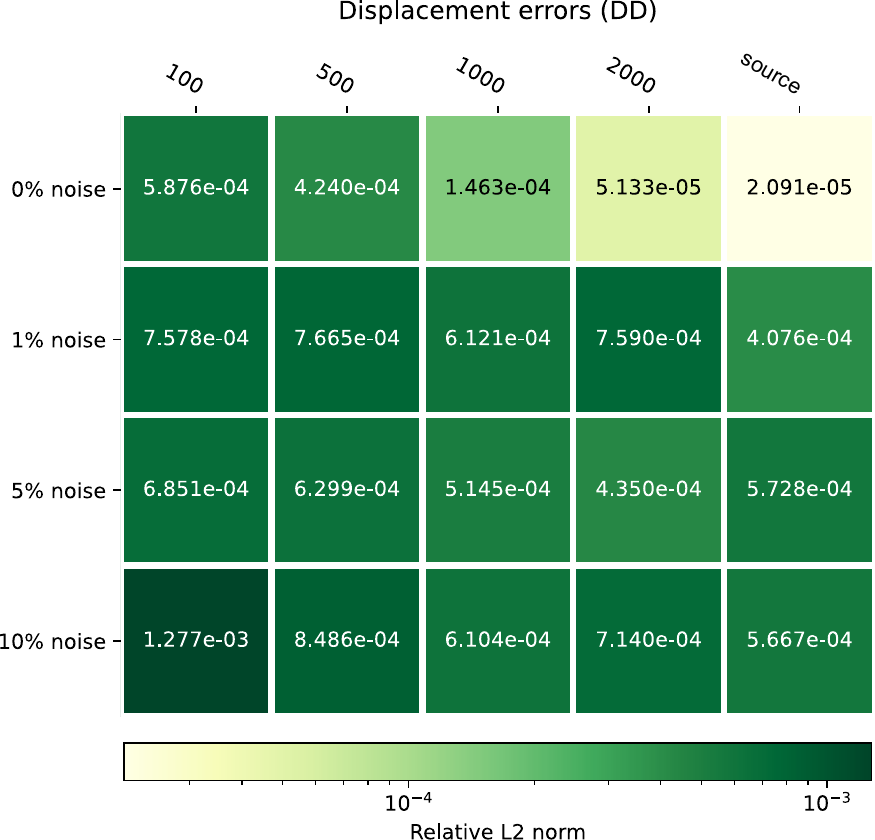}
		\caption{Standard search, original database.}
		\label{fig:errors_disp_HN_DD}
	\end{subfigure}%
	\hfill
	\begin{subfigure}{0.5\textwidth}
		\centering
		\includegraphics[scale=0.462, trim={59 0 0 0},clip=true ]{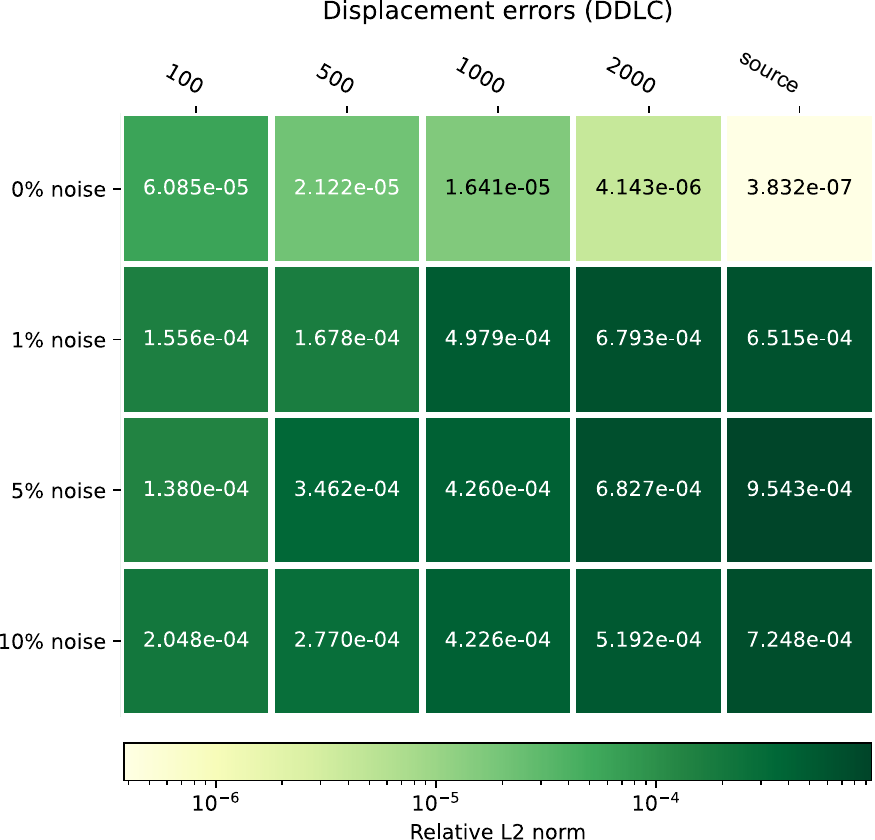}
		\caption{Locally convex search, original database.}
		\label{fig:errors_disp_HN_DDLC}
	\end{subfigure}
	
	\begin{subfigure}{0.5\textwidth}
		\centering
		\includegraphics[width=\textwidth]{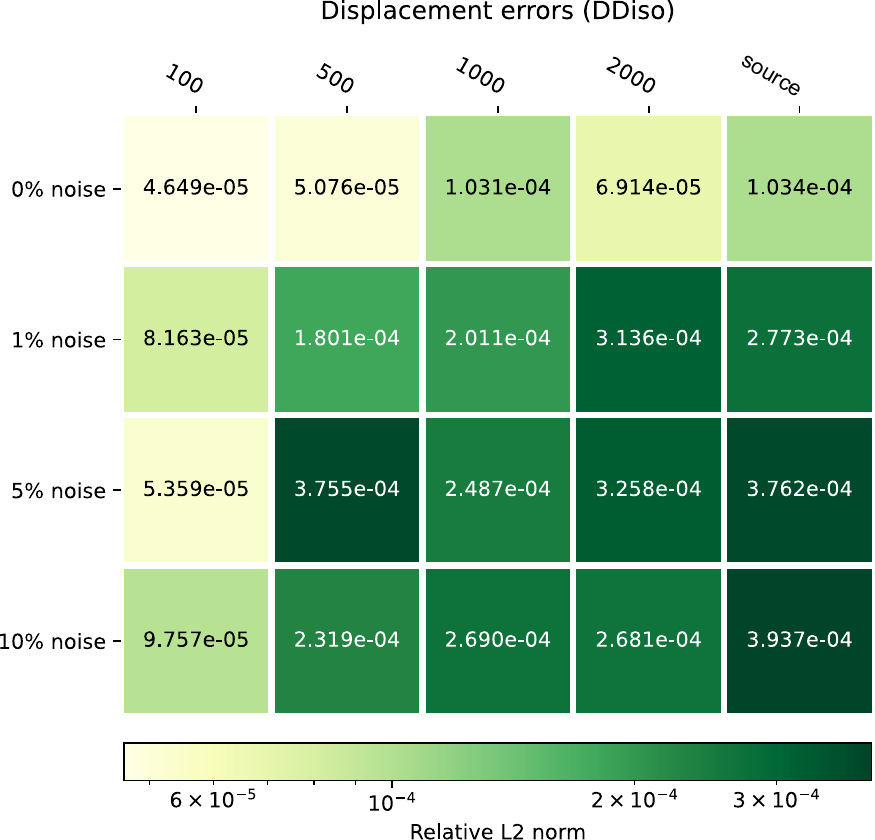}
		\caption{Standard search, enriched database with discretised orbits.}
		\label{fig:errors_disp_HN_DDiso}
	\end{subfigure}%
	\hfill
	\begin{subfigure}{0.5\textwidth}
		\centering
		\includegraphics[scale=0.462,trim={59 0 0 0}, clip=true ]{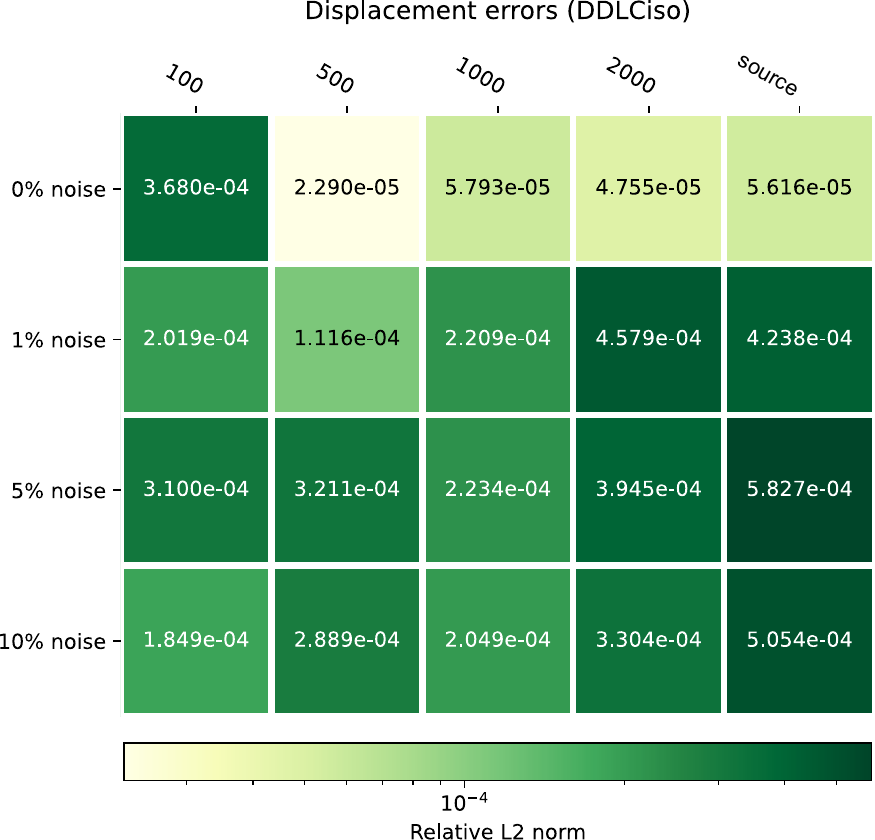}
		\caption{Locally convex, enriched database with discretised orbits.}
		\label{fig:errors_disp_HN_DDLCiso}
	\end{subfigure}
	\caption{Relative $L^2$ norm of displacements for Cook membrane with Hartmann-Neff law.}
\end{figure}

\begin{figure}
	\centering
	\begin{subfigure}{0.5\textwidth}
		\centering
		\includegraphics[width=\textwidth]{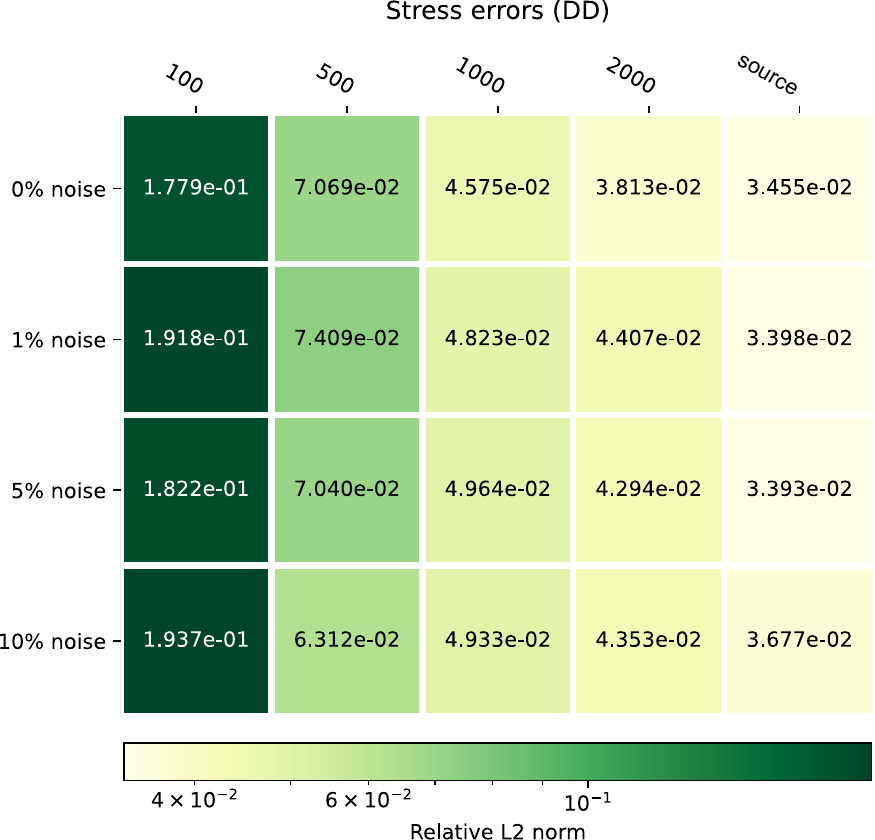}
		\caption{Standard search, original database.}
		\label{fig:errors_stress_HN_DD}
	\end{subfigure}%
	\hfill
	\begin{subfigure}{0.5\textwidth}
		\centering
		\includegraphics[scale=0.462, trim={59 0 0 0}, clip=true]{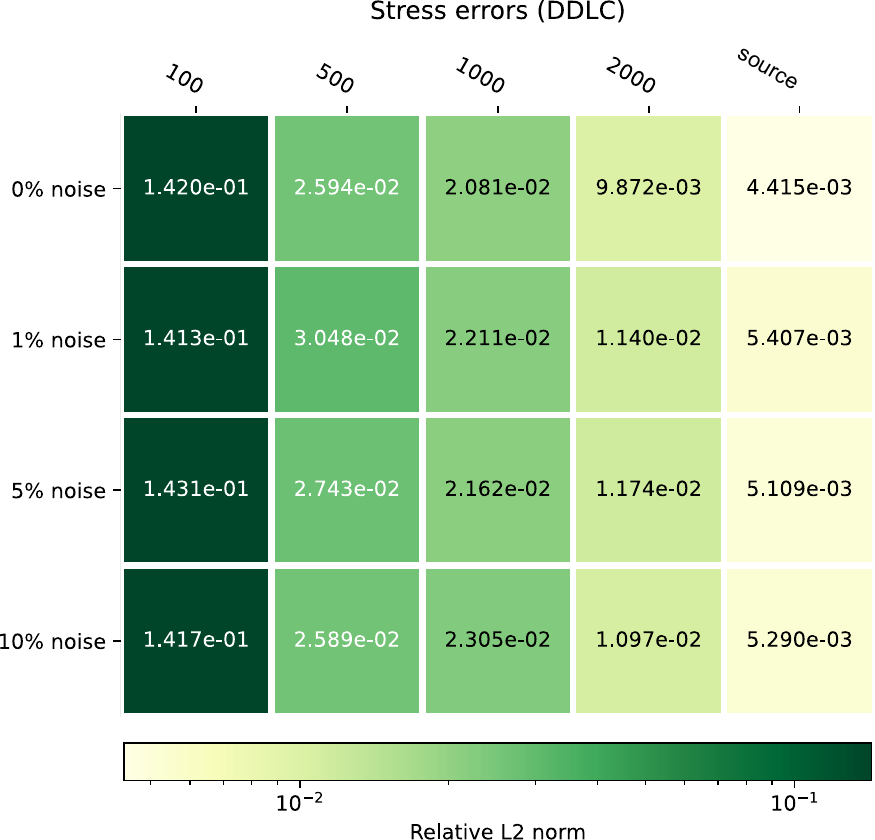}
		\caption{Locally convex search, original database.}
		\label{fig:errors_stress_HN_DDLC}
	\end{subfigure}
	
	\begin{subfigure}{0.5\textwidth}
		\centering
		\includegraphics[width=\textwidth]{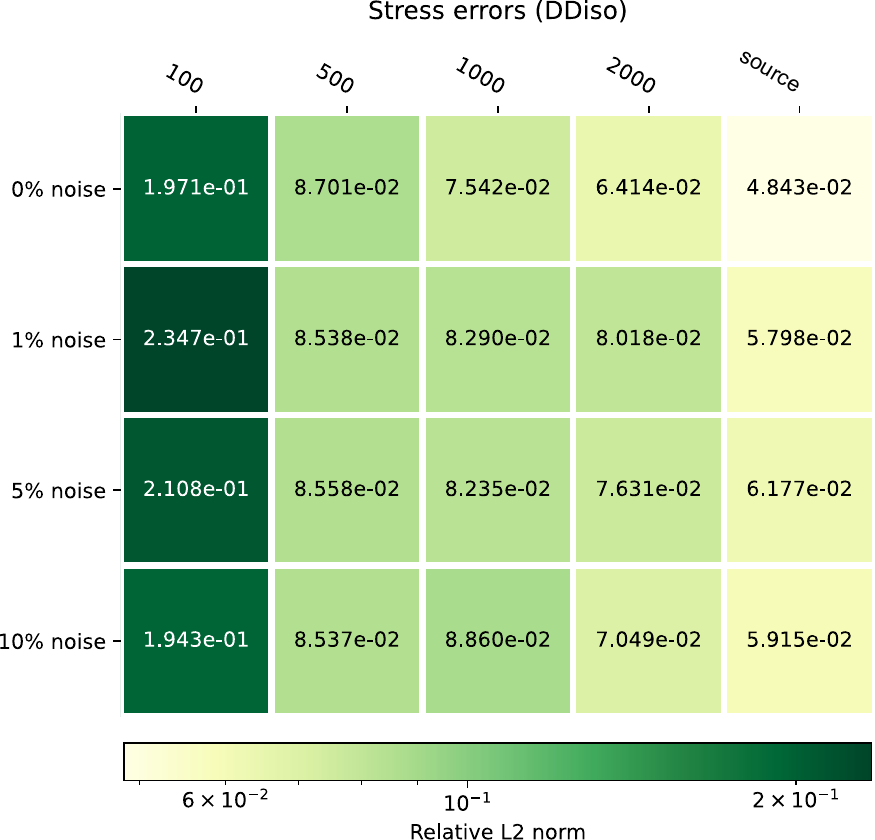}
		\caption{Standard search, enriched database with discretised orbits.}
		\label{fig:errors_stress_HN_DDiso}
	\end{subfigure}%
	\hfill
	\begin{subfigure}{0.5\textwidth}
		\centering
		\includegraphics[scale=0.462, trim={59  0 0 0}, clip=true ]{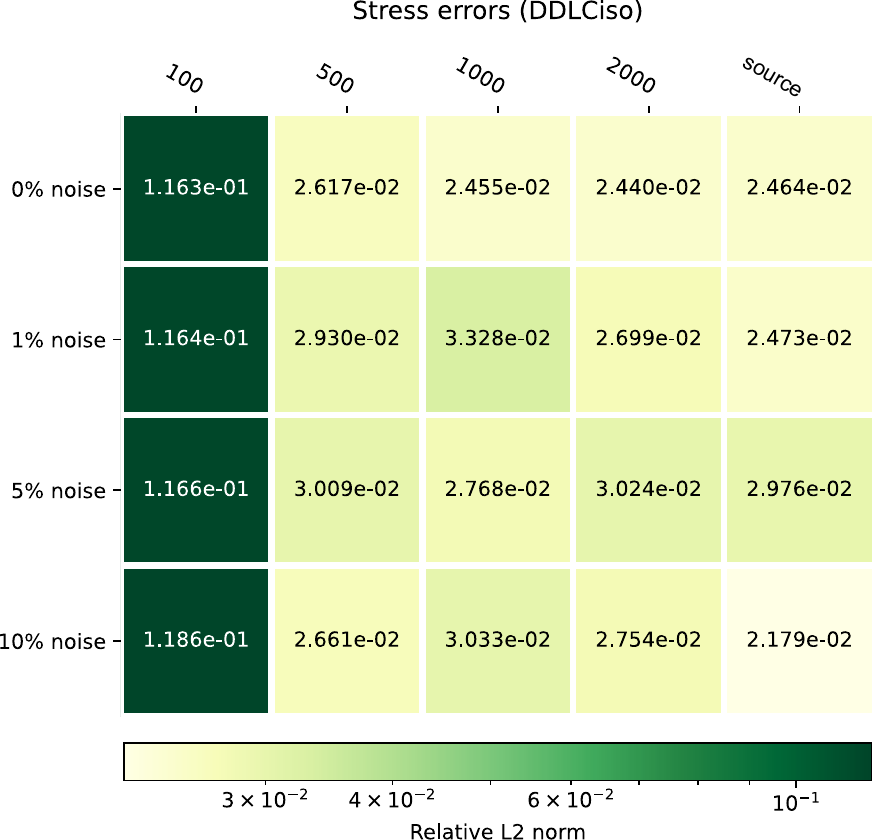}
		\caption{Locally convex, enriched database with discretised orbits.}
		\label{fig:errors_stress_HN_DDLCiso}
	\end{subfigure}
	\caption{Relative $L^2$ norm of stress for Cook membrane with Hartmann-Neff law.}
\end{figure}

\FloatBarrier

\begin{figure}
	\centering
	\begin{subfigure}{0.5\textwidth}
		\centering
		\includegraphics[width=\textwidth]{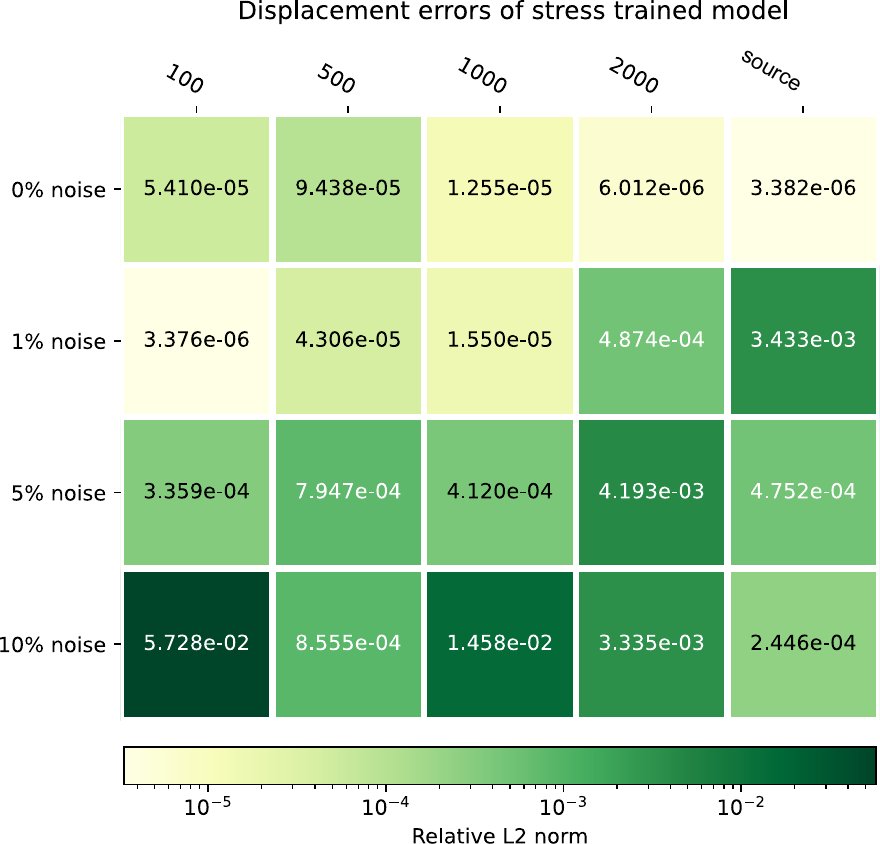}
		\caption{Displacement errors of best stress trained models.}
		\label{fig:errors_disp_stress_HN}
	\end{subfigure}%
	\hfill
	\begin{subfigure}{0.5\textwidth}
		\centering
		\includegraphics[scale=0.462, trim={59 0 0 0}, clip=true ]{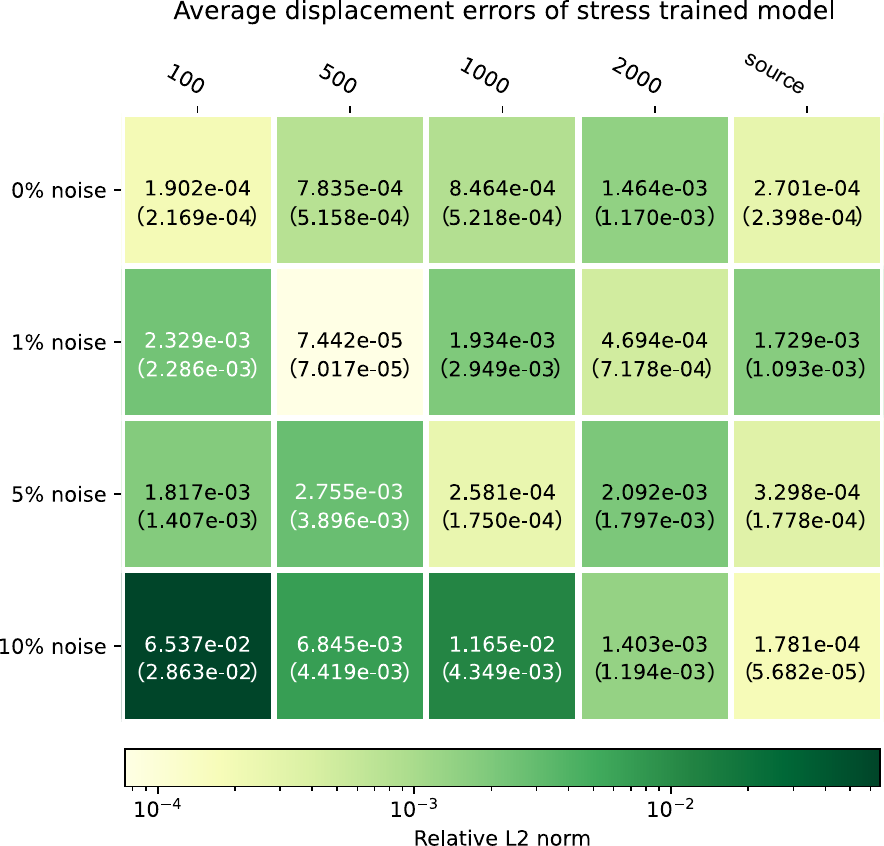}
		\caption{Average displacement errors of stress  models.}
		\label{fig:errors_disp_stress_HN_avg}
	\end{subfigure}
	\caption{Relative $L^2$ norm of displacement for neural networks trained on stress, results for Cook membrane with Hartmann-Neff law. The averaged results shown are for all training runs of neural networks. Standard deviations of the averaged errors are shown in parentheses.}
\end{figure}

\FloatBarrier

Overall, the DD approaches have larger errors. The displacement errors are comparable to those in Sec.~\ref{sec:cook_ciarlet}. However, the stress error is 4 orders of magnitude higher as seen in Fig.~\ref{fig:errors_stress_HN_DDLCiso} than in Fig.~\ref{fig:errors_stress_C_DDLCiso}, where the Ciarlet model with a database that is taken from the original FE solution is used. 

The comparison between the DD and NN approach when applied to Cook's membrane with Hartmann-Neffs law is given in the error plots on Fig.~\ref{fig:error_diff_HN_disp}. The error differences are calculated as before by subtracting $(\bullet)_{DD} - (\bullet)_{NN}$, and the relative performance of the NN models to DDLC as $\left((\bullet)_{DD} - (\bullet)_{NN}\right)/(\bullet)_{DD}$.
\FloatBarrier

\begin{figure}[h!]
	\centering
	\begin{subfigure}{0.5\textwidth}
		\centering
		\includegraphics[width=\textwidth]{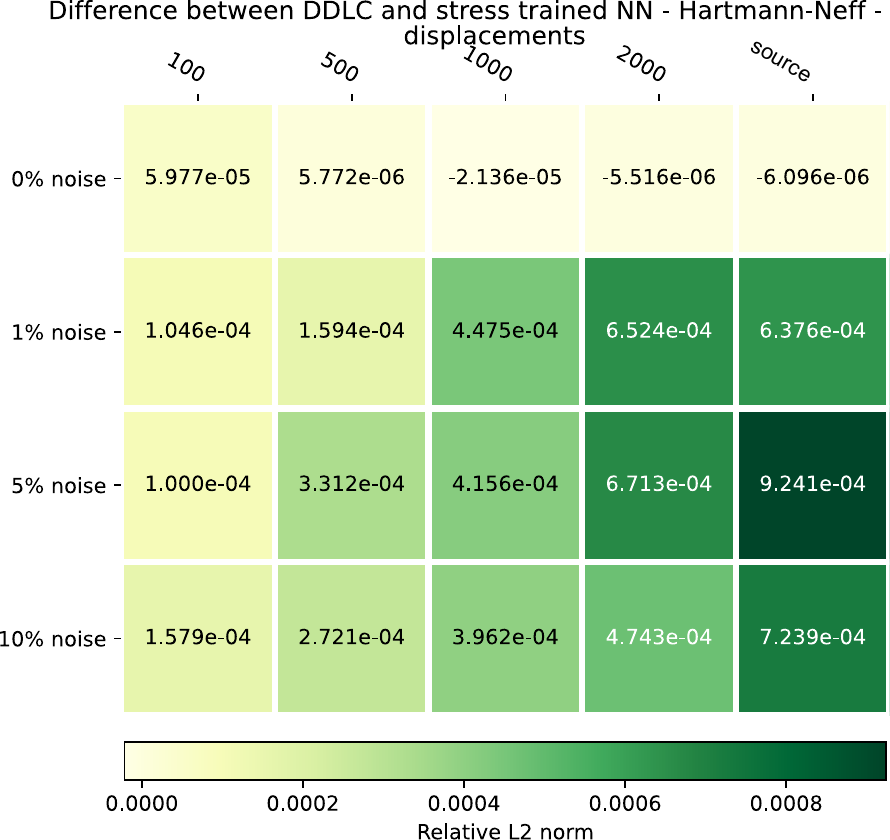}
		\caption{Absolute error difference, displacements.}
		\label{fig:error_diff_DDLC_HN_disp}
	\end{subfigure}%
	\hfill
	\begin{subfigure}{0.5\textwidth}
		\centering
		\includegraphics[width=\textwidth]{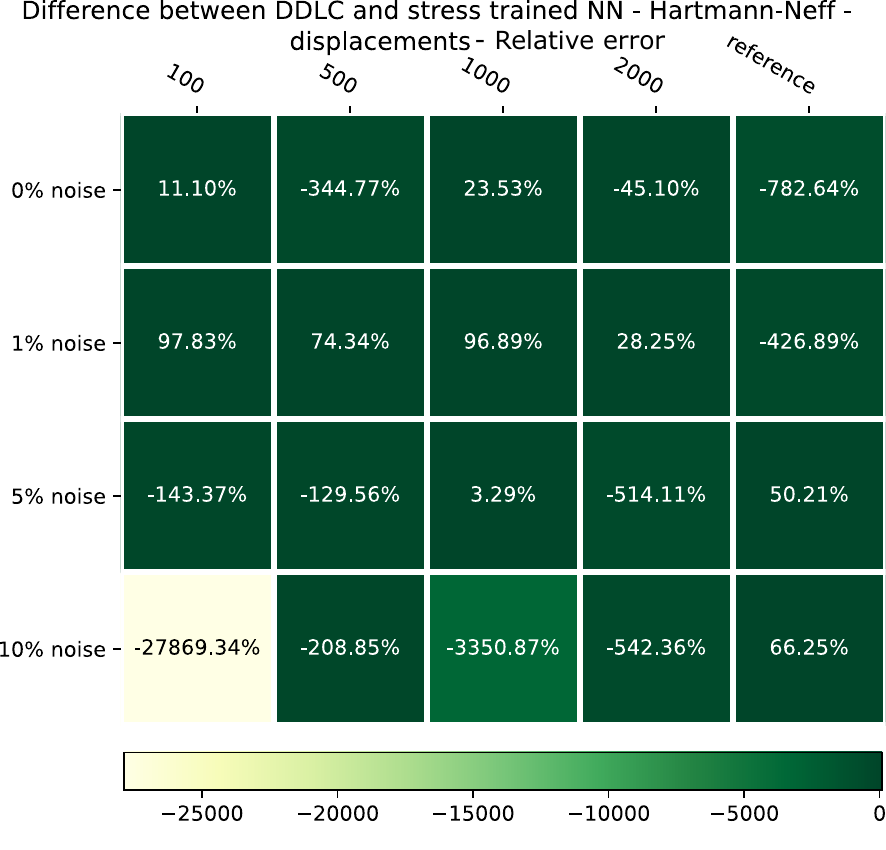}
		\caption{Relative error difference, displacements.}
		\label{fig:rel_error_diff_DDLC_HN_disp}
	\end{subfigure}
	\hfill
	\begin{subfigure}{0.5\textwidth}
		\centering
		\includegraphics[width=\textwidth]{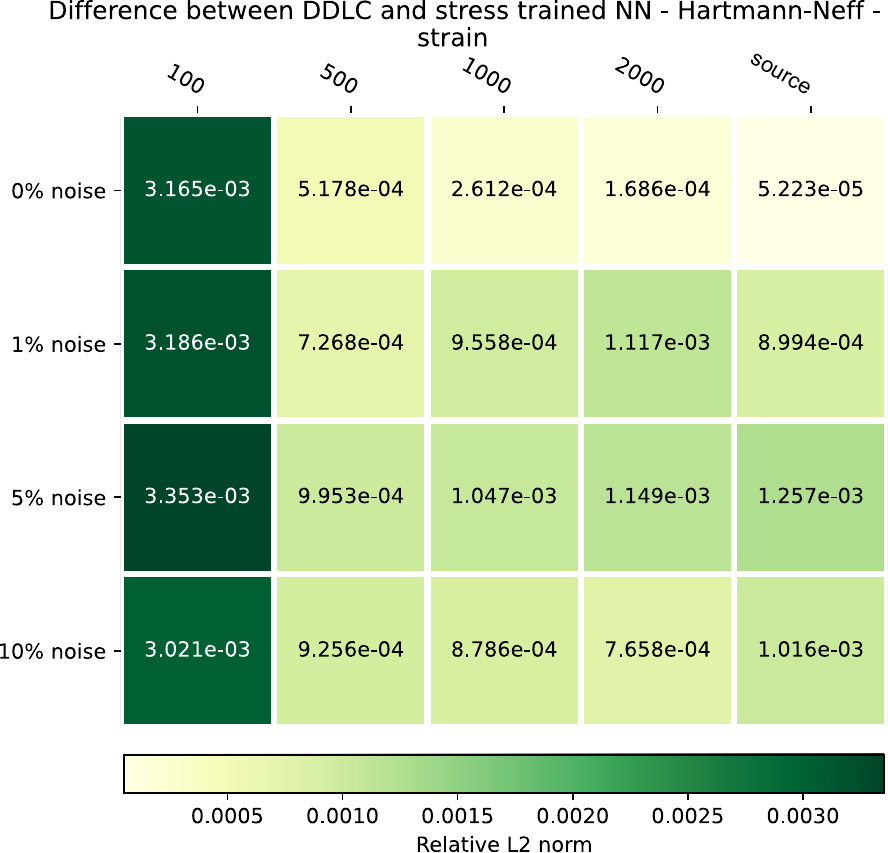}
		\caption{Absolute error difference, strain.}
		\label{fig:error_diff_DDLC_HN_strain}
	\end{subfigure}%
	\hfill
	\begin{subfigure}{0.5\textwidth}
		\centering
		\includegraphics[width=\textwidth]{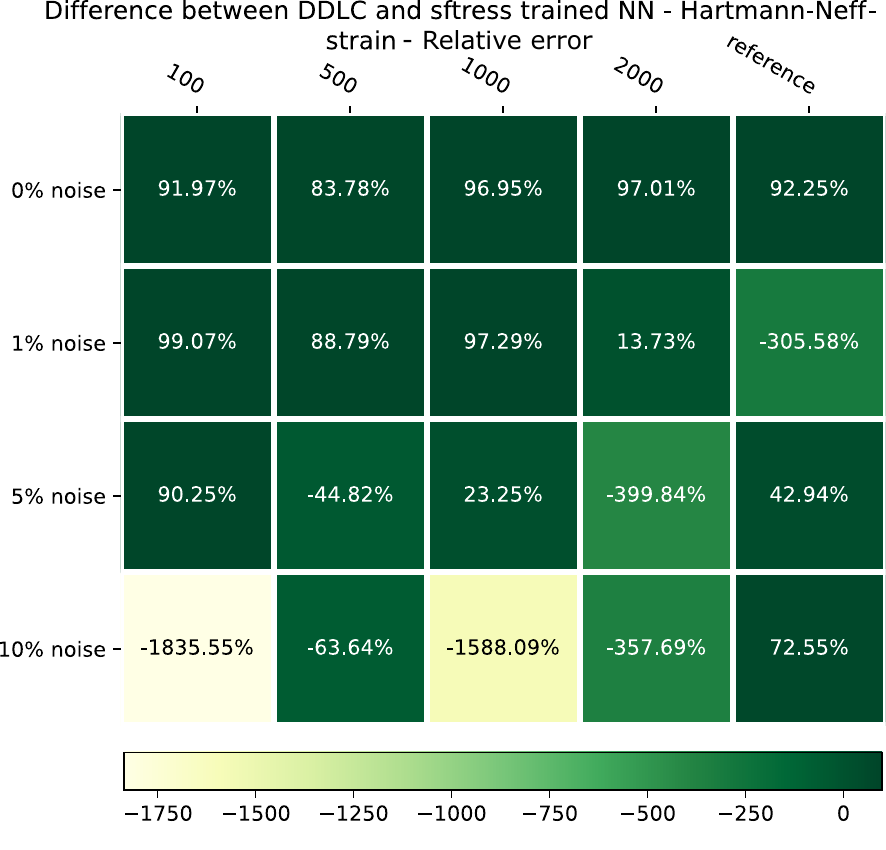}
		\caption{Relative error difference, strain.}
		\label{fig:rel_error_diff_DDLC_HN_strain}
	\end{subfigure}
\end{figure}
\begin{figure}\ContinuedFloat
	\begin{subfigure}{0.5\textwidth}
		\centering
		\includegraphics[width=\textwidth]{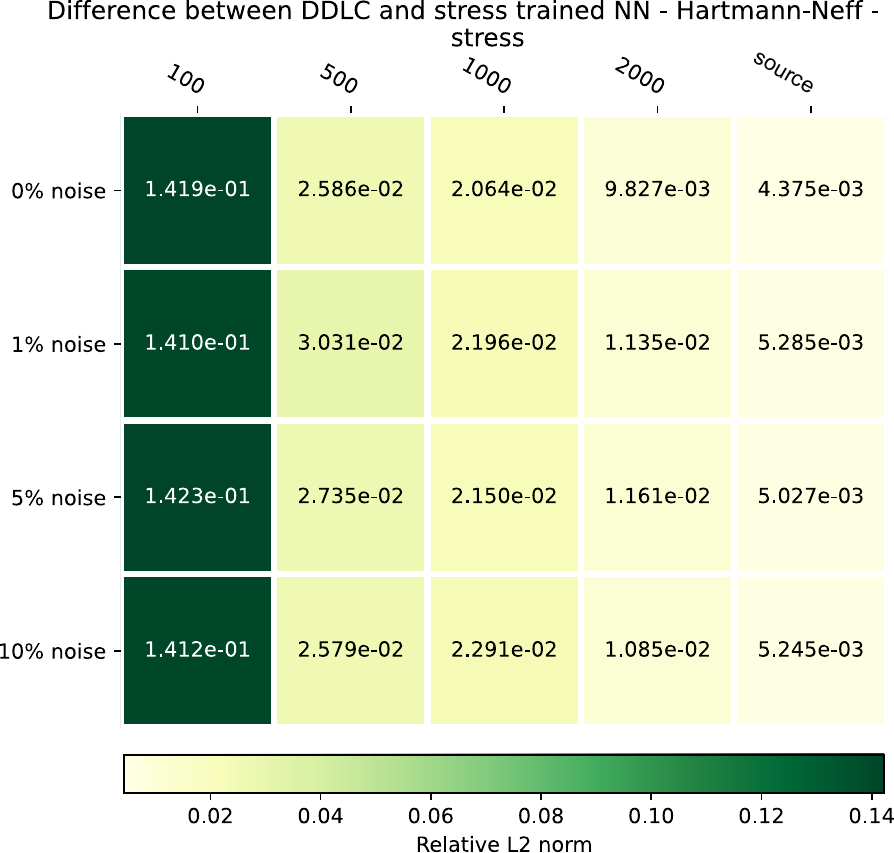}
		\caption{Absolute error difference, stress.}
		\label{fig:error_diff_DDLC_HN_stress}
	\end{subfigure}%
	\hfill
	\begin{subfigure}{0.5\textwidth}
		\centering
		\includegraphics[width=\textwidth]{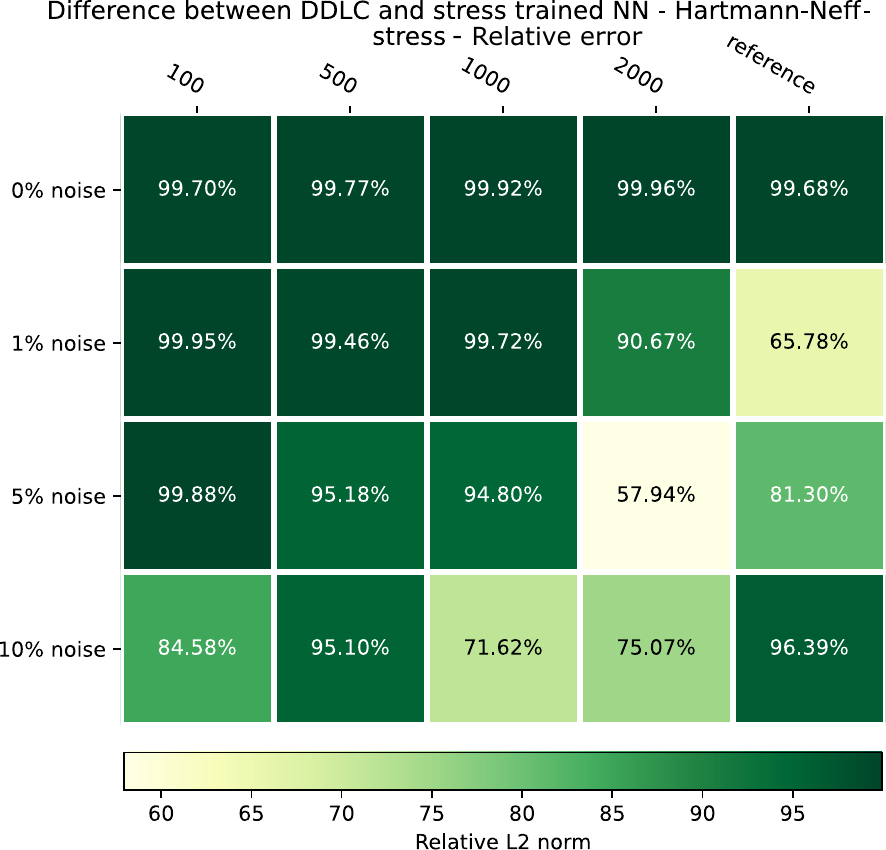}
		\caption{Relative error difference, stress.}
		\label{fig:rel_error_diff_DDLC_HN_stress}
	\end{subfigure}
	\caption{Relative $L^2$ norm error difference between DDLC and NN approaches for Cook's membrane with Hartmann-Neff law. Figures in the left column show the absolute differences between the errors, whereas those in the right column show the respective relative errors.}
	\label{fig:error_diff_HN_disp}
\end{figure}

\FloatBarrier

Again, using the Hartmann-Neff law, the same behaviour can be observed where the NNs outperform the DD variants when noise is included, for comparison see Fig.~\ref{fig:error_diff_DDLC_HN_disp}. In addition, the errors for strain and stress are much lower when using a NN model, see Figs.~\ref{fig:error_diff_DDLC_HN_strain}~\&~\ref{fig:error_diff_DDLC_HN_stress}. 
These results suggest that NNs can be used in a more general setting than DDCM, as they can model the behaviour of one example and apply it to others without loss of generality. Looking at the error differences in Fig.~\ref{fig:error_diff_DDLC_HN_disp}, the NN outperformed the DDCM approach for the smaller, non noisy datasets of 100 and 500 samples, similar to Fig.~\ref{fig:error_diff_DDLC_C_disp}, where it performed better on the smallest dataset of 100 samples. From these comparisons, it can be concluded that the NN approach is more promising for smaller data sets.

\FloatBarrier

\subsection{Punch problem}

\begin{figure}[h!]
	\centering
	\centering
	\includegraphics[width=0.7\textwidth]{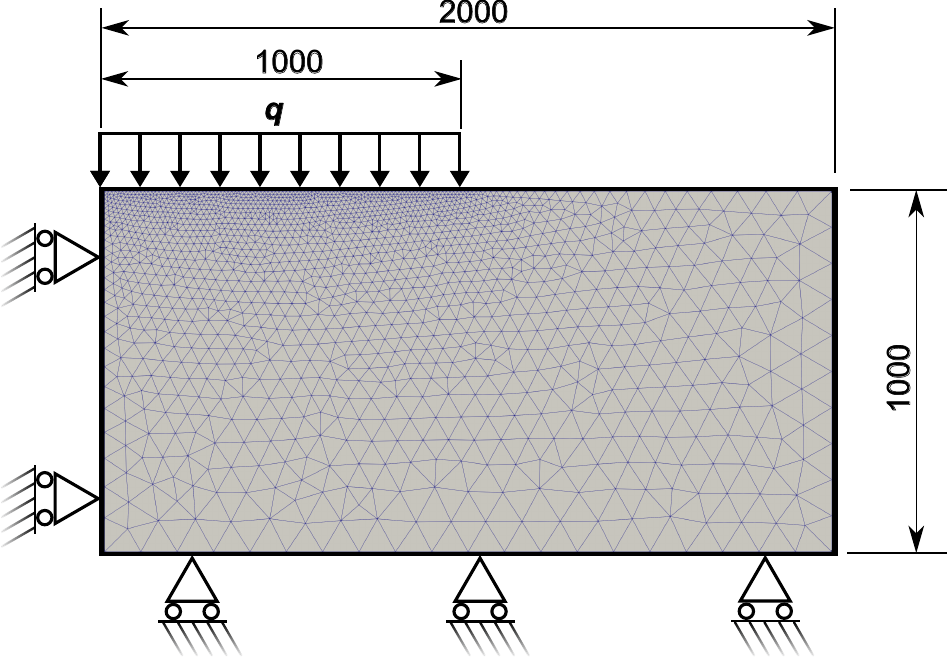}
	\caption{Geometry, mesh, boundary conditions and load of the punch problem.}
	\label{fig:punch_geometry_and_mesh}
\end{figure}

The punch problem was used as a benchmark in \cite{GammBenchmarks2021,Zlatic2023}. In this paper, the 2D problem with the Ciarlet law is considered. The motivation of using this example is to compare the viability of both the DDCM and NN approaches on different geometries and loading conditions. The geometry and boundary conditions are given in Fig.~\ref{fig:punch_geometry_and_mesh} and the mesh is refined towards the top-left corner as shown. The mesh consists of 3108 linear Lagrange triangle elements. The material parameters are given in Section~\ref{sec:cook_ciarlet}, while the load $q$ is assumed to be 100 $\textrm{N/mm}$. The NNs trained in Section~\ref{sec:cook_ciarlet} are used to solve the problem.

Given that the stress trained NNs have outperformed the energy trained ones, only the results of the former will be presented.

\begin{figure}[h!]
	\centering
	\includegraphics{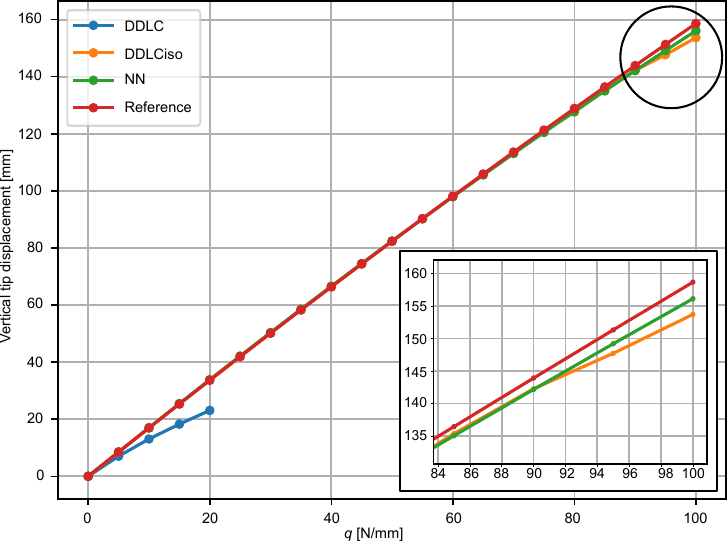}
	\caption{Top-left displacement versus applied load comparison. A close up view of the curves is shown for a clearer comparison.}
	\label{fig:plotutip}
\end{figure}

The top-left corner displacement can be seen in Fig.~\ref{fig:plotutip}. The advantage of the isotropy-enhanced DDLCiso over the standard DDLC approach is also evident in Fig.~\ref{fig:plotutip}. The DDLCiso and NN approaches converge quite well, while the DDLC approach diverges from the beginning. The errors for the DDLCiso and NN solutions are 3.14\% and 1.61\% respectively.

\begin{figure}
	\centering
	\begin{subfigure}{\textwidth}
		\centering
		\includegraphics{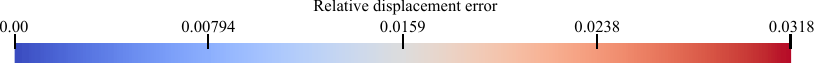}
	\end{subfigure}
	\begin{subfigure}{0.5\textwidth}
		\centering
		\includegraphics[width=0.95\textwidth]{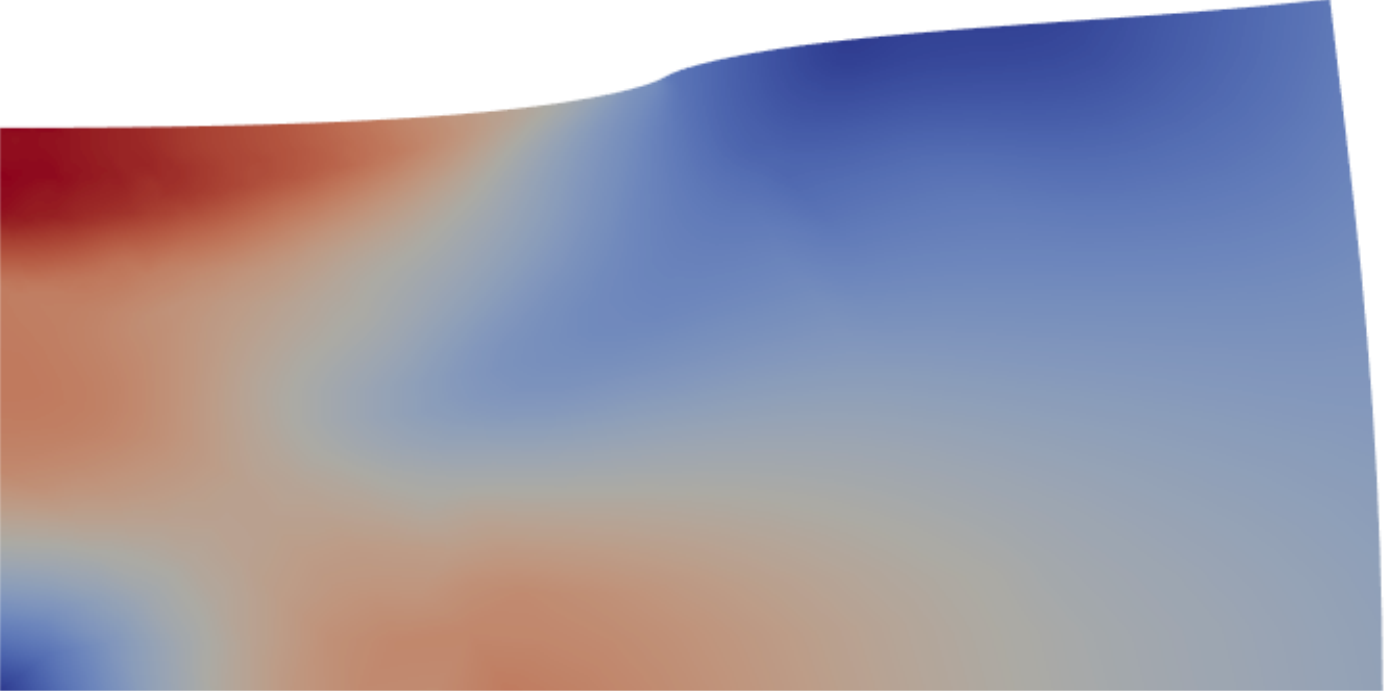}
		\caption{Relative displacement error for DDLCiso, largest error is $3.18\%$.}
		\label{fig:error_disp_rel_q100_DDLCiso_punch}
	\end{subfigure}%
	\hfill
	\begin{subfigure}{0.5\textwidth}
		\centering
		\includegraphics[width=0.95\textwidth]{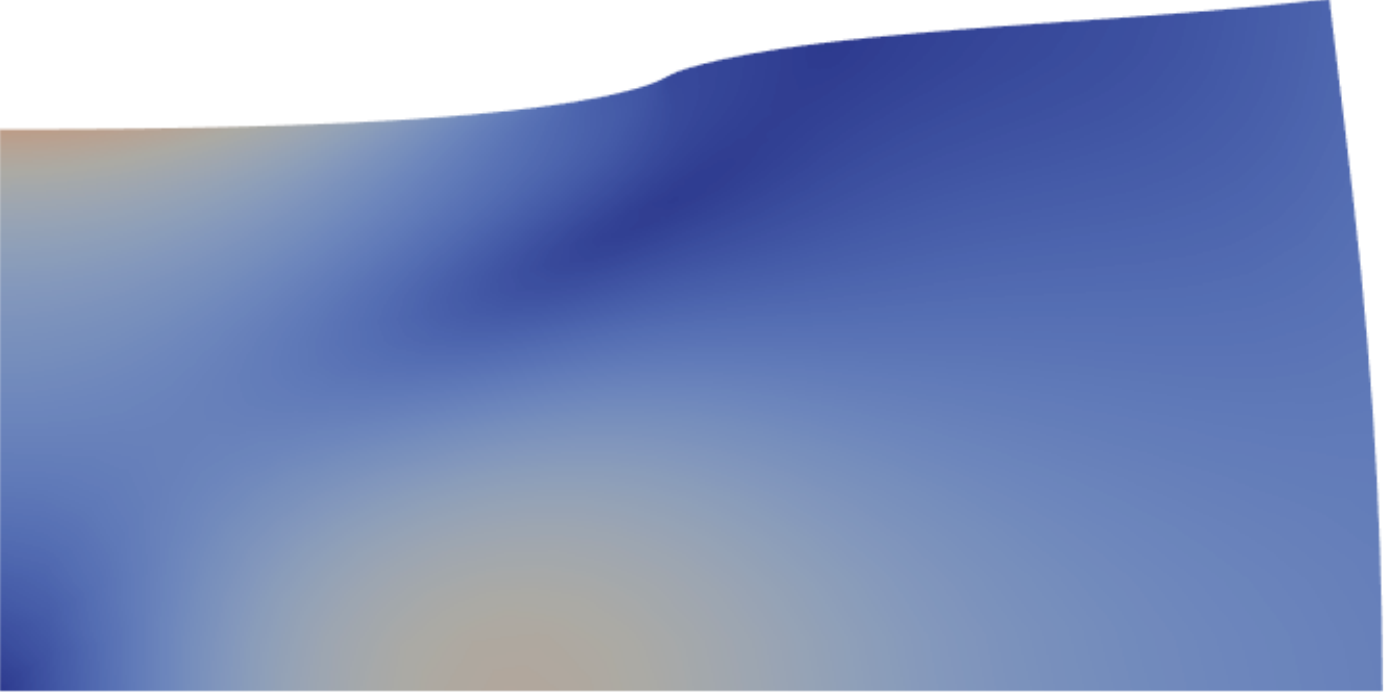}
		\caption{Relative displacement error for stress trained NN, largest error is $1.91\%$.}
		\label{fig:error_disp_rel_q100_NN_punch}
	\end{subfigure}
	\begin{subfigure}{\textwidth}
		\centering
		\includegraphics[width=\textwidth]{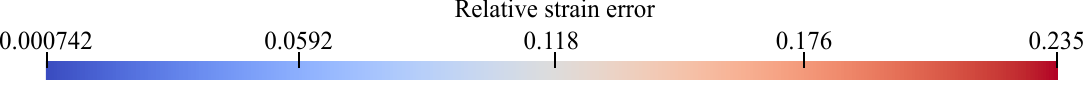}
	\end{subfigure}
	\begin{subfigure}{0.5\textwidth}
		\centering
		\includegraphics[width=0.95\textwidth]{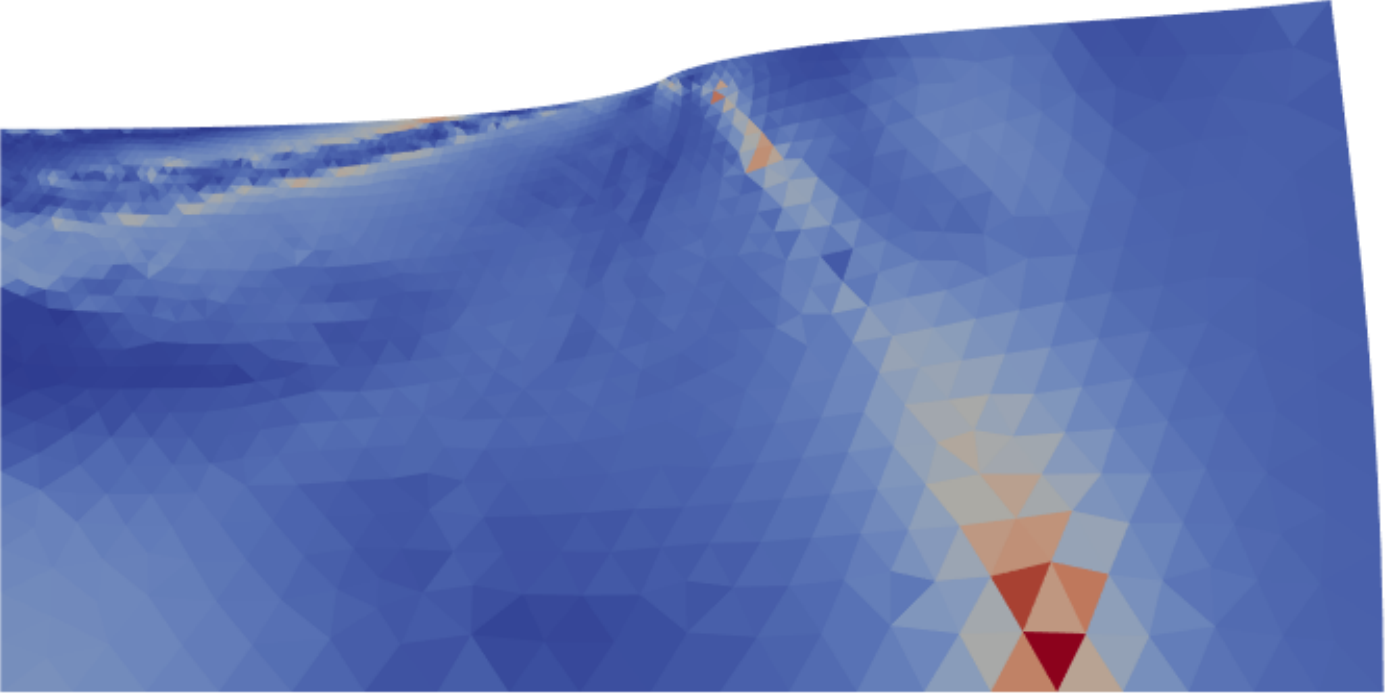}
		\caption{Relative strain error for DDLCiso, largest error is $23.5\%$.}
		\label{fig:error_E_rel_q100_DDLCiso_punch}
	\end{subfigure}%
	\hfill
	\begin{subfigure}{0.5\textwidth}
		\centering
		\includegraphics[width=0.95\textwidth]{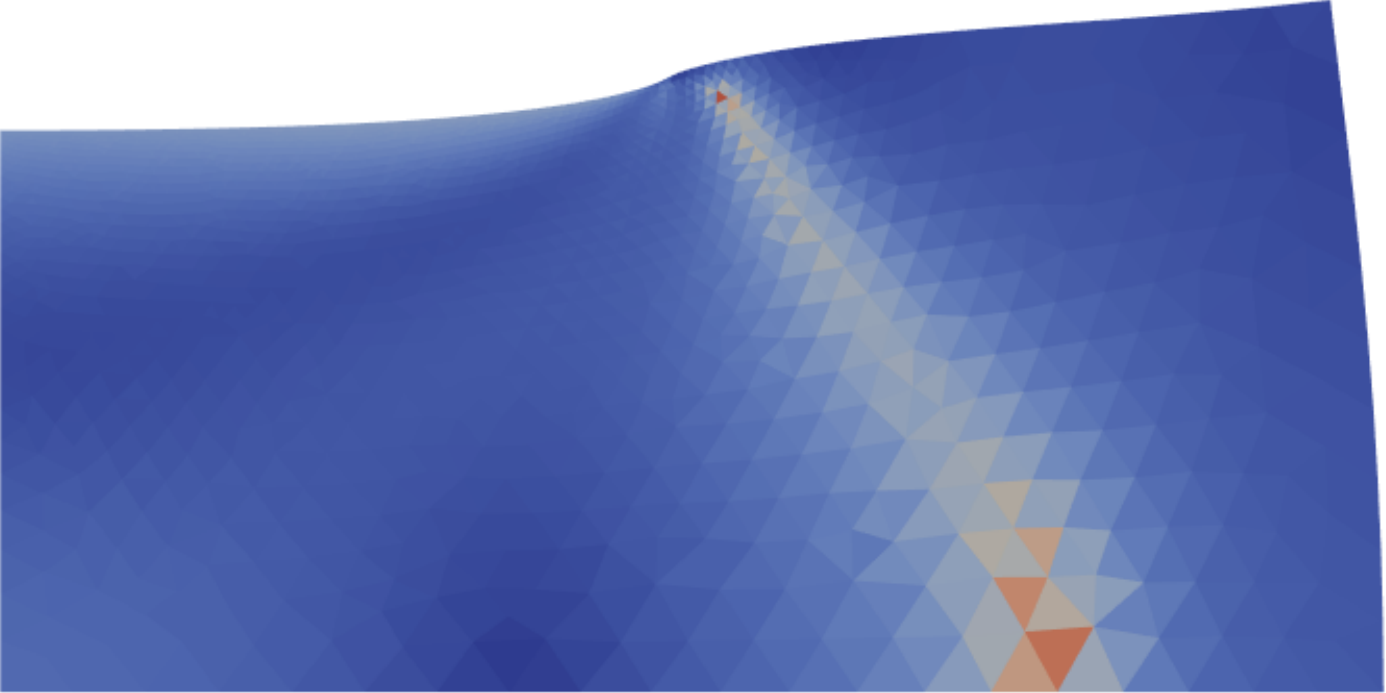}
		\caption{Relative strain error for stress trained NN, largest error is $20.1\%$.}
		\label{fig:error_E_rel_q100_NN_punch}
	\end{subfigure}
	\begin{subfigure}{\textwidth}
		\centering
		\includegraphics[width=\textwidth]{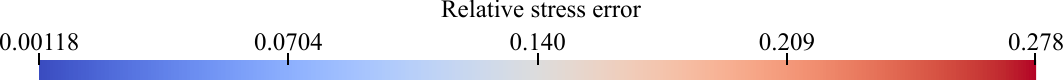}
	\end{subfigure}
	\begin{subfigure}{0.5\textwidth}
		\centering
		\includegraphics[width=0.95\textwidth]{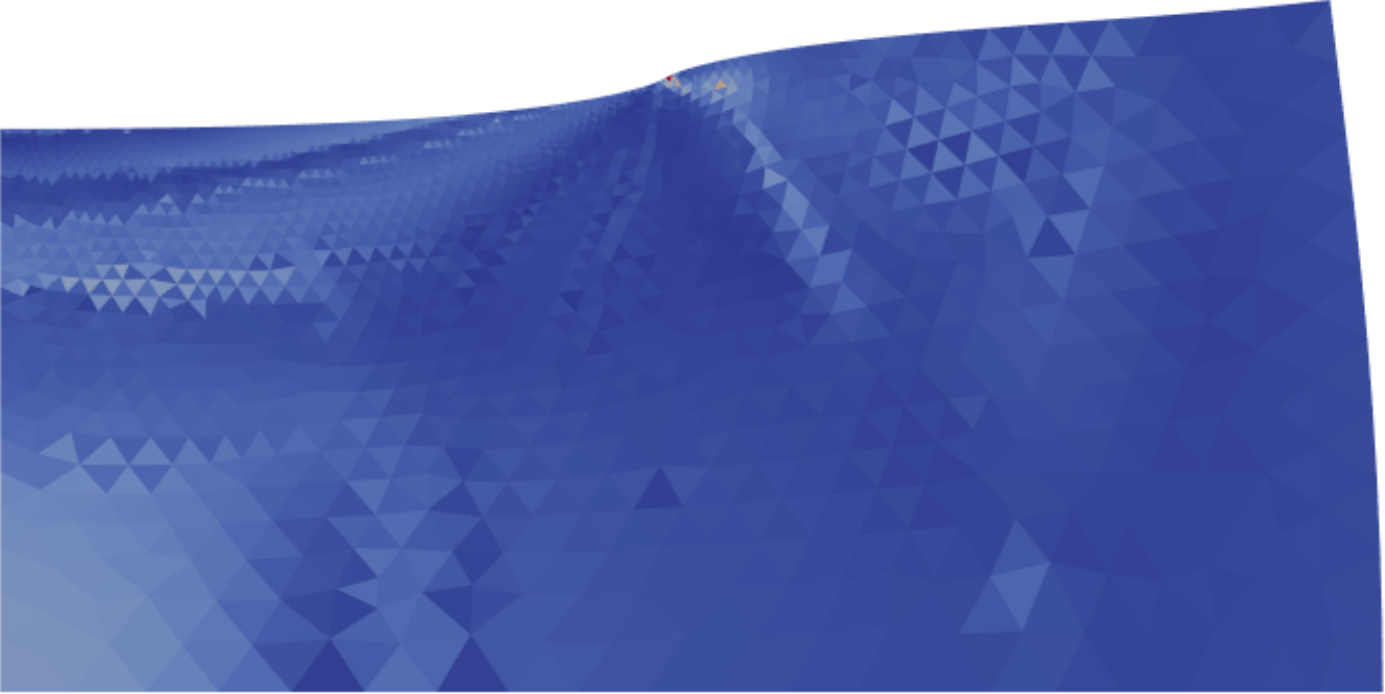}
		\caption{Relative error stress for DDLCiso, largest error is $27.8\%$.}
		\label{fig:error_S_rel_q100_DDLCiso}
	\end{subfigure}%
	\hfill
	\begin{subfigure}{0.5\textwidth}
		\includegraphics[width=0.95\textwidth]{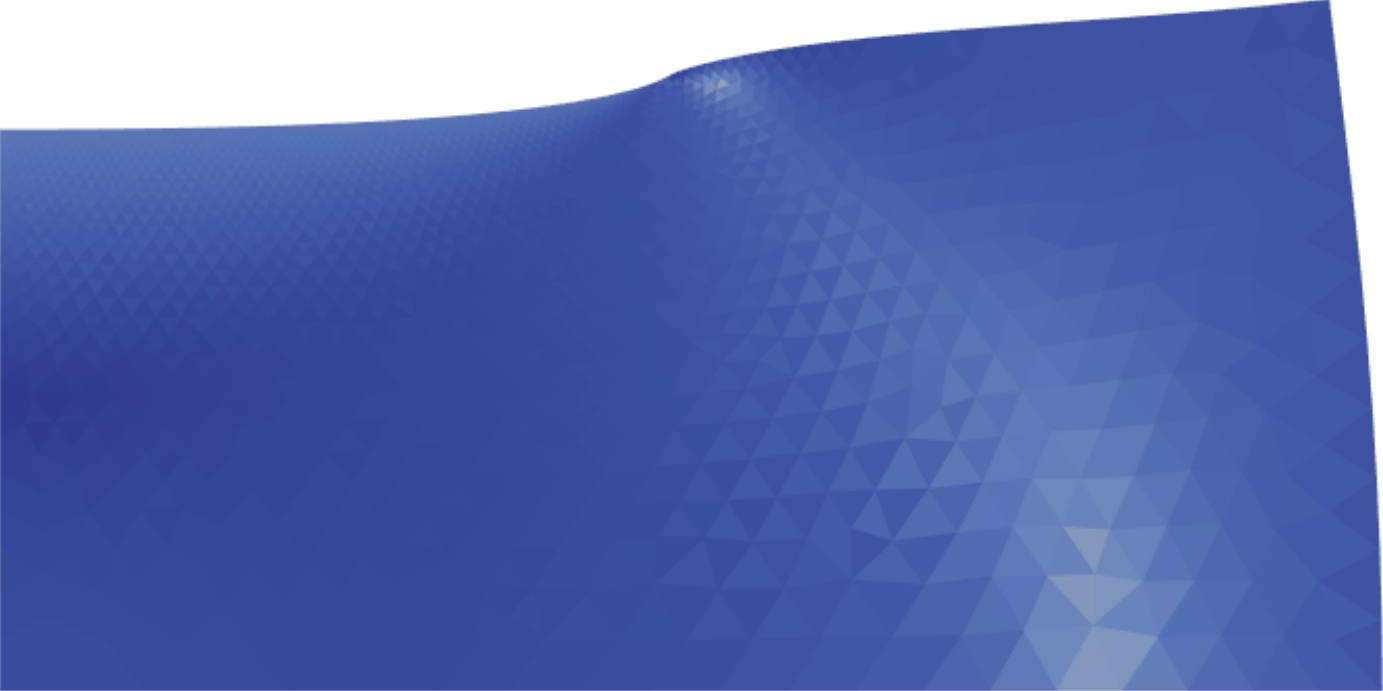}
		\caption{Relative stress error for stress trained NN, largest error is $10.9\%$.}
		\label{fig:error_S_rel_q100_NN}
	\end{subfigure}
	\caption{Relative stress errors for the DDLCiso and stress trained NN for the punch problem using the Ciarlet material model.}
	\label{fig:error_S_rel_punch_comparison}
\end{figure}

Fig.~\ref{fig:error_S_rel_punch_comparison} shows the relative errors for the displacement, strain and stress on the deformed mesh of the punch problem for DDLCiso and the stress trained NN. Note that for the relative displacement errors in Figs~\ref{fig:error_disp_rel_q100_DDLCiso_punch}~\&~\ref{fig:error_disp_rel_q100_NN_punch} the absolute errors are divided with the maximum value of displacement, while the strain and stress relative errors are obtained by dividing with the base solution, as in Section~\ref{sec:cook_ciarlet}. The errors are comparable, but the extremes appear at different locations. When comparing the stresses, for the DDLCiso solution the highest strain error occurs near the lower edge, whereas the highest stress error occurs at the endpoint of the load $q$. This is different from the NN, where the highest strain and stress errors appear close together (neighbouring elements) near the endpoint of the load.

Looking at the highest relative errors of the NN, in Fig.~\ref{fig:error_E_rel_q100_NN_punch} the highest strain error of 20.1\% occurs in the element where the lowest strain component is $2.5\cdot10^{-3}$, and in Fig.~\ref{fig:error_S_rel_q100_NN} the highest error of 10.9\% occurs in the element where the lowest stress component is 0.09 MPa. Looking at Fig.~\ref{fig:error_E_rel_q100_DDLCiso_punch} the highest strain error of 23.5\% occurs in the element where the lowest strain component is $2.5\cdot10^{-3}$ which is very similar to the NN, however the highest stress error in Fig.~\ref{fig:error_S_rel_q100_DDLCiso} occurs in the element where the lowest stress component is 14.19 MPa which is different to previous errors.

The NN solution is smoother, whereas the DDCM solution has discontinuities within the domain similar to what was noticed in Fig.~\ref{fig:error_DDLCiso_NN_contour_cook_C}.

\FloatBarrier

\section{Conclusion}

In this paper, two different approaches to modelling mechanical behaviour based on data were studied. The neural network based model approach has been developed to incorporate several physics based restrictions and two approaches to training have been presented, either directly on the strain-energy or on the stresses using the same data as the DDCM approach. The incorporated physics based restrictions include \textit{thermodynamic consistency},  \textit{convexity}, \textit{objectivity}, as well as \textit{normalization of the strain-energy}, and is in accordance to the proposition given by \cite{Linden2023}. In addition, it is advantageous to train the NNs on stresses, i.e. their derivatives, instead of directly on the energy. The model-free data-driven computational mechanics approach has been adapted to hyperelastic behaviour and several variations have been presented. 

We have implemented the DD approach for finite strains using the conjugate Lagrangian pairs ($\ten{E},\ten{S}$), i.e. the Green-Lagrange strain and the second Piola-Kirchhoff stress, as proposed by \cite{Nguyen2018}. To avoid a lack of information in coarse database scenarios, we used a local regularisation technique based on a linear local approximation hyperplane using the $k$-nearest neighbours.

We can highlight different advantages for each of the approaches in terms of accuracy. The DDCM method has shown superior performance when the underlying dataset is free of noise and sufficiently covers the needed ranges required for a given problem. On the other hand, the neural network models have shown better performance on noisy data and when the training data does not necessarily cover the exact range required for the application. Furthermore, the NN models have shown to outperform the DDCM solutions when smaller amounts of data are available. Even though one family of methods may outperform the other according to the trends mentioned above, the errors obtained are still comparable in most situations. An important difference in the methods becomes clear when comparing Figs.~\ref{fig:error_DDLCiso_NN_contour_cook_C}~and~\ref{fig:error_S_rel_punch_comparison}. In Fig.~\ref{fig:error_DDLCiso_NN_contour_cook_C} the problem is almost exactly the same as the one from which the dataset was taken, whereas in Fig.~\ref{fig:error_S_rel_punch_comparison} the problem is different. Not all the states of deformation occurring in the punch problem have been seen in the source dataset and the base DD and DDLC approaches cannot be applied. Only the computationally much more demanding DDLCiso approach provided a solution that contains discontinuities in the strain and stress fields. The NN approach, on the other hand, does not need to be adapted in any way. The choice of approach depends on whether the underlying database is sufficiently large and dense enough to cover the various states of deformation that can occur for a given problem. If it is large and dense enough with precise data then DDCM can be applied as it had higher accuracy in such a case in Section~\ref{sec:cook_ciarlet}. Otherwise, in the case of few data or possible noise in the data, the NN approach would be preferable since it naturally extrapolates the solution on regions with fewer data and filters out noise. It is worth mentioning that there are further, although more intrusive, alternatives to extend DDCM to tackle similar difficulties in addition to the Locally Convex Embedding \cite{He2020a} already considered in this work, e.g., the so-called max-ent DDCM based on the principle of maximum-entropy regularisation \cite{Kirchdoerfer2017}, and also distance-preserving nonlinear manifold denoising approach \cite{Bahmani2023} through the incorporation of a low-dimensional geometric autoenconder to fill scarce data regions, just to name a few alternatives.

In terms of computational cost, DDCM is better than NN for one-shot simulations as no training phase is required. Once the NN model is trained, its deployment adds up negligible computational cost compared to the evaluation of a simple constitutive model. Therefore, the NN approach can become favourable compared to DDCM for problems with multiple queries. It is also worth mentioning that while very dense databases can have an impact on the computational cost, this impact can be largely reduced by adopting data structures such as tree-based search with reasonable computational complexity. DDCM computational costs increase as the database becomes denser due to nearest neighbour search. That being said, the isotropic enrichment for DDCM by enriching the data set with rotated strain-stress couples increases the accuracy, but also reduces computational efficiency. The NN approach does not require such enrichment as it is based on invariants.

In summary, both the DDCM and NN approaches have proven to be viable choices. The addition of isotropy to the DDCM methods has proved invaluable for extending to examples when using data that does not come from an example itself, see Fig.~\ref{fig:plotutip}. The errors for the DDLCiso approach and stress trained NNs are comparable and the choice between DDCM and NNs would largely depend on the quality of data in the dataset and the quantity of data. With less data of lower quality, NNs seem to be the better choice, but as seen in the Cook membrane problem, where the data set was sampled from a very similar case, DDCM gives more accurate results.

\section*{Acknowledgements}

This work is supported by the Croatian Science Foundation under the project IP-2019-04-4703, and by the Erasmus+ staff mobility programme. This support is gratefully acknowledged. Felipe Rocha and Laurent Stainier gratefully acknowledge financial support from ANR through project D3MecA (ANR-19-CE46-0012).


\section*{CRediT authorship contribution statement}

\noindent\textbf{Martin Zlati\'{c}:} Conceptualization, Methodology, Software, Validation, Formal analysis, Visualization, Writing - Original Draft, Writing - Review \& Editing. 
\textbf{Felipe Rocha:} Conceptualization, Methodology, Software, Validation, Formal analysis, Visualization, Writing - Original Draft, Writing - Review \& Editing. 
\textbf{Laurent Stainier:}  Resources, Conceptualization, Supervision, Writing - Review \& Editing, Funding Acquisition.
\textbf{Marko \v{C}ana\dj ija:} Resources, Supervision, Writing - Review \& Editing, Funding Acquisition.

\section*{Data availability}

Datasets used to train the NNs and run the DDCM algorithms are available on Zenodo doi:10.5281/zenodo.12725683.

\appendix

\section{Convexity of the activation function}\label{sec:convexity_proof}

To ensure the convexity of the activation function the Hessian must be positive semi-definite. In this proof the activation of the topmost neuron in the hidden layer is considered. The Hessian of the activation function $h$ with respect to the invariants $I_1$, $I_2$, and $I_3$ is as follows:

\begin{equation}
	\ten{H} = 
	\begin{bmatrix}
		\frac{\partial^2{h}}{\partial{I_1}^2} & \frac{\partial^2{h}}{\partial{I_1}\partial{I_2}} & \frac{\partial^2{h}}{\partial{I_1}\partial{I_3}} \\[5pt]
		\frac{\partial^2{h}}{\partial{I_2}\partial{I_1}}&
		\frac{\partial^2{h}}{\partial{I_2}^2} & \frac{\partial^2{h}}{\partial{I_2}\partial{I_3}} \\[5pt]
		\frac{\partial^2{h}}{\partial{I_3}\partial{I_1}} &
		\frac{\partial^2{h}}{\partial{I_3}\partial{I_2}} &
		\frac{\partial^2{h}}{\partial{I_3}^2}
	\end{bmatrix},
	\label{eq:hessian_of_activation_function}
\end{equation}
where the partial derivatives of the activation function from Eq.~(\ref{eq:output_one_neuron}) w.r.t. the invariants are defined as follows:
\begin{equation}
	\frac{\partial^2{h_i}}{\partial{I_k}\partial{I_l}} = w_{k,i}^{[1]}w_{l,i}^{[1]}\underbrace{\alpha_i^2\exp\big[\alpha[w_{1,i}^{[1]}(I_1-3) + w_{2,i}^{[1]}(I_2 - 3) + w_{3,i}^{[1]}(I_3 - 1)]\big]}_{:=A},
	\label{eq:activation_2nd_derivative_generic} 
\end{equation}
where $k, l = 1,2,3$, and $i = 1,...,n_h$. Note that the Hessian is symmetric. The positive semi-definitiveness requires:
\begin{equation}
	\ten{x}^T\ten{H}\ten{x} = A \underbrace{(x_1 w_{1,i}^{[1]} + x_2 w_{2,i}^{[1]} + x_3 w_{3,i}^{[1]})^2}_{:=B} \geq 0, \quad \forall \ten{x} \in \mathbb{R}^3.
	\label{eq:pos_sem_def_cond}
\end{equation}

The inequality holds since $A$ in \ref{eq:activation_2nd_derivative_generic} and $B$ in \ref{eq:pos_sem_def_cond} are always non-negative. Since each activation function is convex, it holds that the neural network as a whole is convex.

\section{DDCM equilibrium projection subproblem} \label{sec:DDCM_PE}

Given $\bf{z}^* \in \mathcal{Z}_D$, let us find  $\bf{z} = \text{arg min}_{\bf{z}' \in \mathcal{Z}_E} d(\bf{z}',\bf{z}^*)$. First, we should recall that the kinematical admissibility is obtained by deriving $\ten{E}$ from a regular displacement field $\bf{u}$ through Eq.~\eqref{eq:compatibility}, eventually satisfying some Dirichlet boundary condition. Such space of admissible displacements is denoted $\mathcal{U}$. As commented, in order to relax the linear momentum balance constraint, we introduce a displacement-like Lagrange Multiplier $\bs{\eta}$ living in $\mathcal{V}$, which is analogous to $\mathcal{U}$ but with homogeneous Dirichlet conditions. The space of symmetric second-order tensor fields is denoted $\mathcal{S}$. The constrained problem is then reformulated in a unconstrained fashion as follows
\begin{align}
	(\bf{u}, \ten{S}, \bs{\eta}) = \arg \min_{\bf{u}' \in \mathcal{U}} \min_{\ten{S}' \in \mathcal{S}} \max_{\bs{\eta}' \in \mathcal{V}} 
	\mathcal{L}(\bf{u}', \ten{S}', \bs{\eta}'), 
\end{align}
with the Lagrangian functional
\begin{equation} \label{eq:lagragian}
	\mathcal{L} (\bf{u}, \ten{S}, \bs{\eta} ) = d^2\left((\ten{E}(\bf{u}), \ten{S}), (\ten{E}^*, \ten{S}^*\right) - \int_{\Omega_0} \ten{F}(\bf{u}) \ten{S}: \nabla \bs{\eta} + \int_{\Omega_0} \bf{b}_0 \cdot \bs{\eta} + \int_{\partial \Omega^N_0} \bar{\bf{t}}_0 \cdot \bs{\eta},
\end{equation}
where for the sake of simplicity, we consider the case where $\bf{b}_0$ and $\bar{\bf{t}}_0$ do not depend on the displacements. By the stationary condition along the direction $\delta \ten{S}$ we obtain $\ten{S} = \ten{S}^* + \Ten4{C} \delta \ten{E}(\bf{u},\bs{\eta})$. Replacing the former expression into the Lagrangian, we obtain $\tilde{\mathcal{L}}(\bf{u}, \bs{\eta}) = \mathcal{L}(\bf{u}, \ten{S}(\bf{u},\bs{\eta}),  \bs{\eta})$. In terms of minimization, the extremality condition leads to the following nonlinear residuals expressions 

\begin{equation}
	\begin{split}
		\bf{R}_{\bf{u}}(\bf{u}, \bs{\eta}; \delta \bf{u}) = (\Ten4{C} (\ten{E}(\bf{u}) - \hat{\ten{E}}), \ten{F}^{T} \nabla \delta \bf{u}) - (\Ten4{C} \ten{F}^{T} \nabla \bs{\eta}, \nabla^T \delta \bf{u} \nabla \bs{\eta}) \\ - (\hat{\ten{S}}, \nabla^T \delta \bf{u} \nabla \bs{\eta}),
	\end{split}
\end{equation}
\begin{equation}
	\begin{split}
		\bf{R}_{\bs{\eta}}(\bf{u}, \bs{\eta}; \delta \bs{\eta} ) = - (\Ten4{C} \ten{F}^{T}(\bf{u}) \nabla \bs{\eta}, \ten{F}^{T} \nabla \delta \bs{\eta}) - (\hat{\ten{S}}, \ten{F}^{T} \nabla \delta \bs{\eta}) + \int_{\Omega_0} \bf{b}_0 \cdot \bs{\delta \eta} \\ + \int_{\partial \Omega^N_0} \bar{\bf{t}}_0 \cdot \delta\bs{\eta},
	\end{split}
\end{equation}
obtained by the directional derivative of the Lagrangian for each direction $\{\delta \bf{u}, \delta \bs{\eta}\}$. We proceed iteratively by the Newton-Raphson method, with the tangent operator given by:
\begin{equation}
	\begin{split}
		\bf{D}_{\bf{u}\bf{u}}(\bf{u}, \bs{\eta}; d\bf{u}, \delta \bf{u}) = (\Ten4{C} \ten{F}^T \nabla d\bf{u}, \ten{F}^{T} \nabla \delta \bf{u}) + 
		(\Ten4{C} (\ten{E}(\bf{u}) - \hat{\ten{E}}), \nabla^T d\bf{u} \nabla \delta \bf{u}) \\
		-(\Ten4{C} \nabla^T d\bf{u} \nabla \bs{\eta} , \nabla^T \delta d\bf{u} \nabla \delta \bs{\eta}),
	\end{split}
\end{equation}
\begin{equation}
	\begin{split}
		\bf{D}_{\bf{u}\bs{\eta}}(\bf{u}, \bs{\eta}; d\bf{u}, \delta \bs{\eta}) = -(\Ten4{C} \ten{F}^T \nabla d\bs{\eta}, \nabla^T \delta \bf{u} \nabla \bs{\eta})  
		-(\Ten4{C} \ten{F}^T \nabla \bs{\eta}, \nabla^T \delta \bf{u} \nabla d  \bs{\eta} ) \\
		-(\hat{\ten{S}},  \nabla^T \delta \bf{u} \nabla d \bs{\eta}),
	\end{split}
\end{equation}
\begin{flalign}
	&\;\;\bf{D}_{\bs{\eta}\bs{\eta}}(\bf{u}, \bs{\eta}; d\bs{\eta}, \delta \bs{\eta}) = -(\Ten4{C} \ten{F}^T \nabla d \bs{\eta}, \ten{F}^{T} \nabla \delta \bs{\eta}),& &
\end{flalign}
where the directional derivatives are computed along $\{ d \bf{u}, d \bs{\eta} \}$. Note that $\ten{D}_{\bs{u} \bf{\eta}}$ is symmetric such that$\ten{D}_{\bf{\eta} \bs{u}}$ is not written explicitly. A discretized version of the notation shown here can be found in other works such as \cite{Kirchdoefer2016,Platzer2021,Nguyen2018}.

\section{Additional figures}\label{sec:additional_figs}

This appendix contains additional error plots that were obtained when comparing the DDCM and NN approaches. For clarity, they were put in this appendix so as to not clutter the main body of the paper.

\begin{figure}
	\centering
	\begin{subfigure}{0.5\textwidth}
		\centering
		\includegraphics[width=\textwidth]{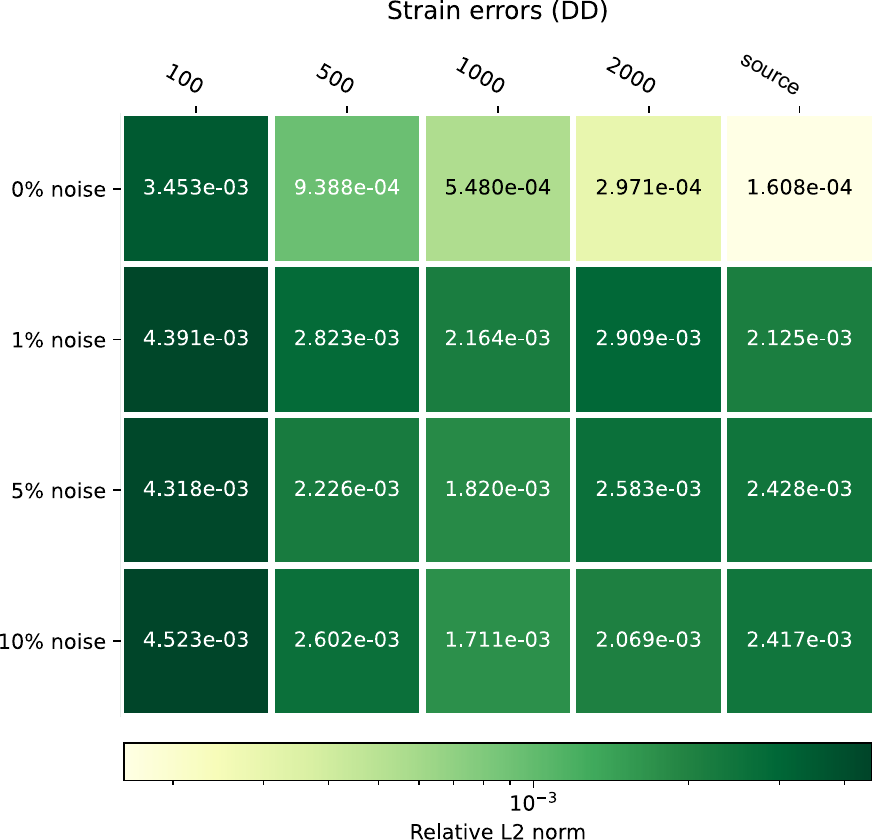}
		\caption{Standard search, original database.}
		\label{fig:errors_strain_C_DD}
	\end{subfigure}%
	\hfill
	\begin{subfigure}{0.5\textwidth}
		\centering
		\includegraphics[width=\textwidth]{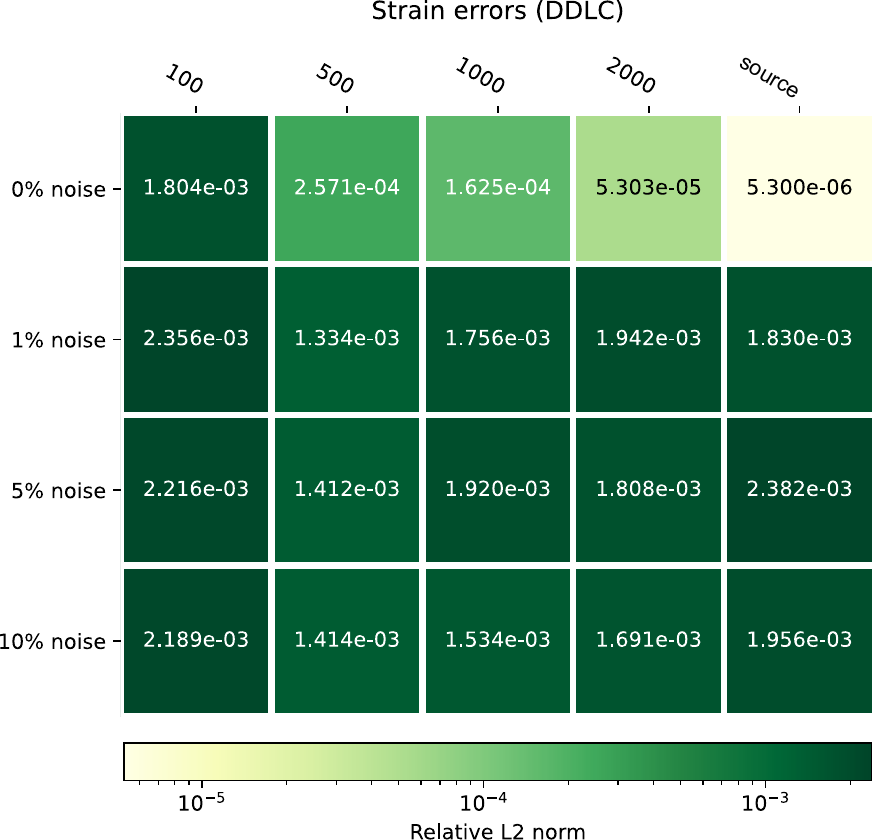}
		\caption{Locally convex search, original database.}
		\label{fig:errors_strain_C_DDLC}
	\end{subfigure}
	
	\begin{subfigure}{0.5\textwidth}
		\centering
		\includegraphics[width=\textwidth]{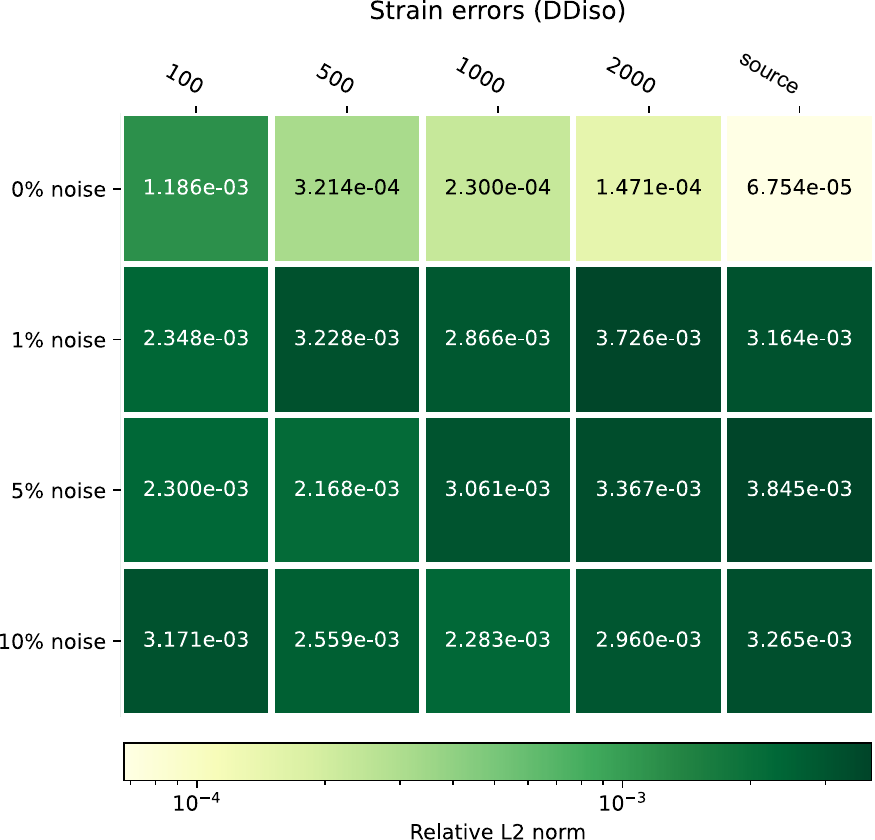}
		\caption{Standard search, enriched database with discretised orbits.}
		\label{fig:errors_strain_C_DDiso}
	\end{subfigure}%
	\hfill
	\begin{subfigure}{0.5\textwidth}
		\centering
		\includegraphics[width=\textwidth]{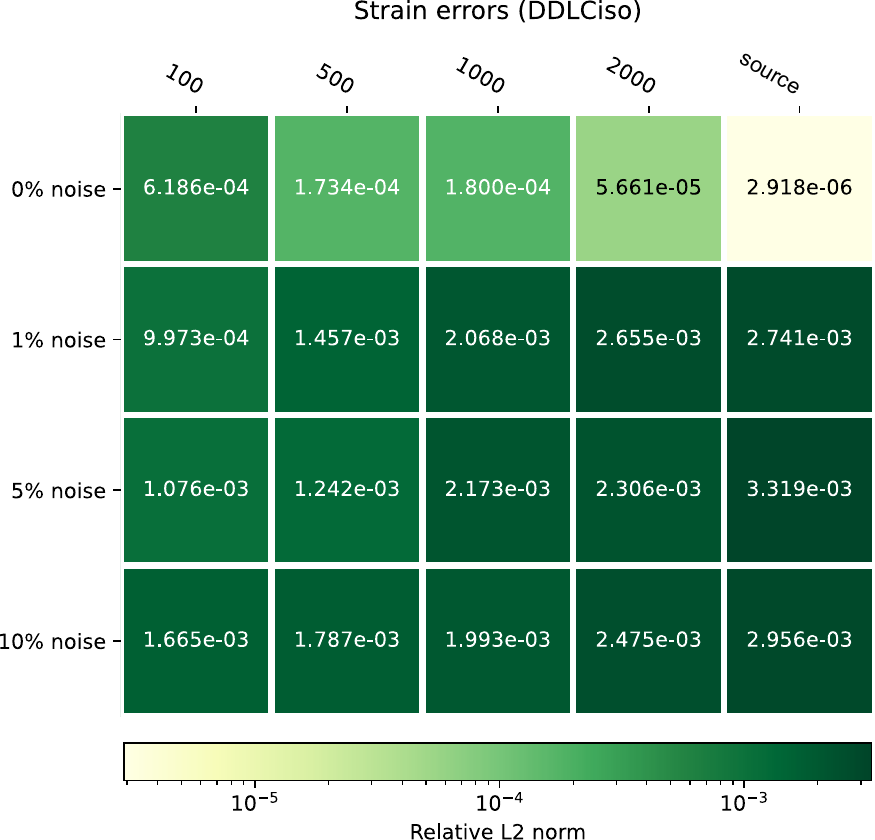}
		\caption{Locally convex, enriched database with discretised orbits.}
		\label{fig:errors_strain_C_DDLCiso}
	\end{subfigure}
	\caption{Relative $L^2$ norm of strains for Cook membrane with Ciarlet law.}
	\label{fig:errors_strain_C_data_driven}
\end{figure}

\begin{figure}
	\centering
	\begin{subfigure}{0.5\textwidth}
		\centering
		\includegraphics[width=\textwidth]{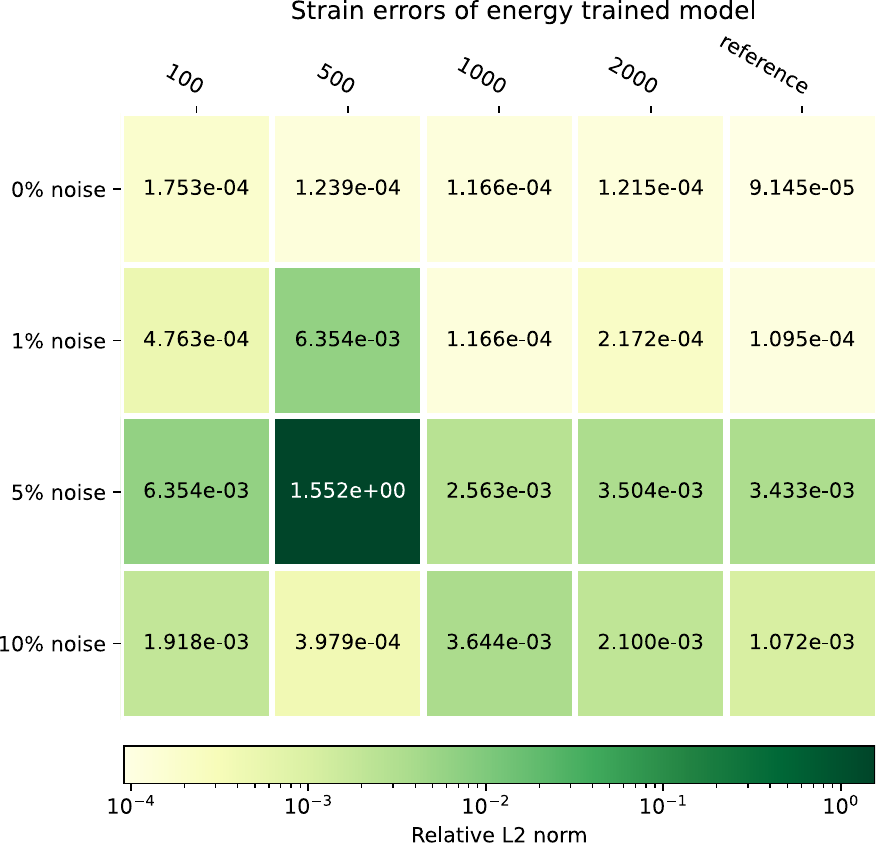}
		\caption{Strain errors of best energy trained models.}
		\label{fig:errors_strain_energy_Ciarlet}
	\end{subfigure}%
	\hfill
	\begin{subfigure}{0.5\textwidth}
		\centering
		\includegraphics[width=\textwidth]{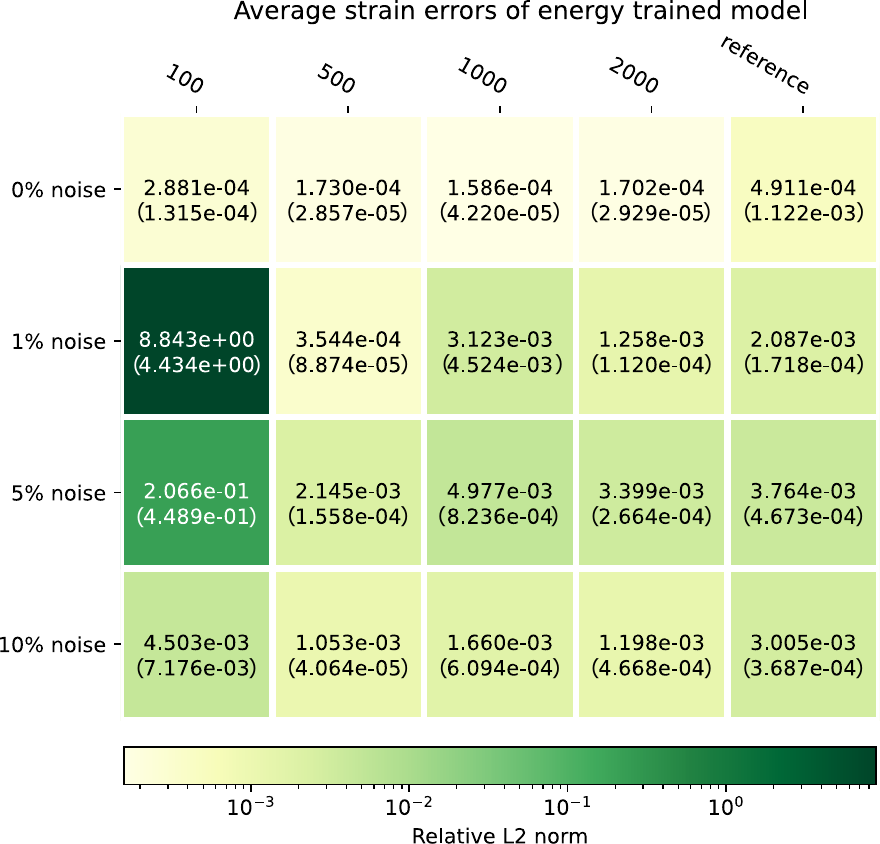}
		\caption{Average strain errors of energy trained models.}
		\label{fig:errors_strain_energy_Ciarlet_avg}
	\end{subfigure}
	\caption{Relative $L^2$ norm of strain for neural networks trained on energy, results for Cook membrane with Ciarlet law. The results shown are averaged for all training runs of neural networks. Standard deviations of the averaged errors are shown in parentheses.}
\end{figure}

\begin{figure}
	\centering
	\begin{subfigure}{0.5\textwidth}
		\centering
		\includegraphics[width=\textwidth]{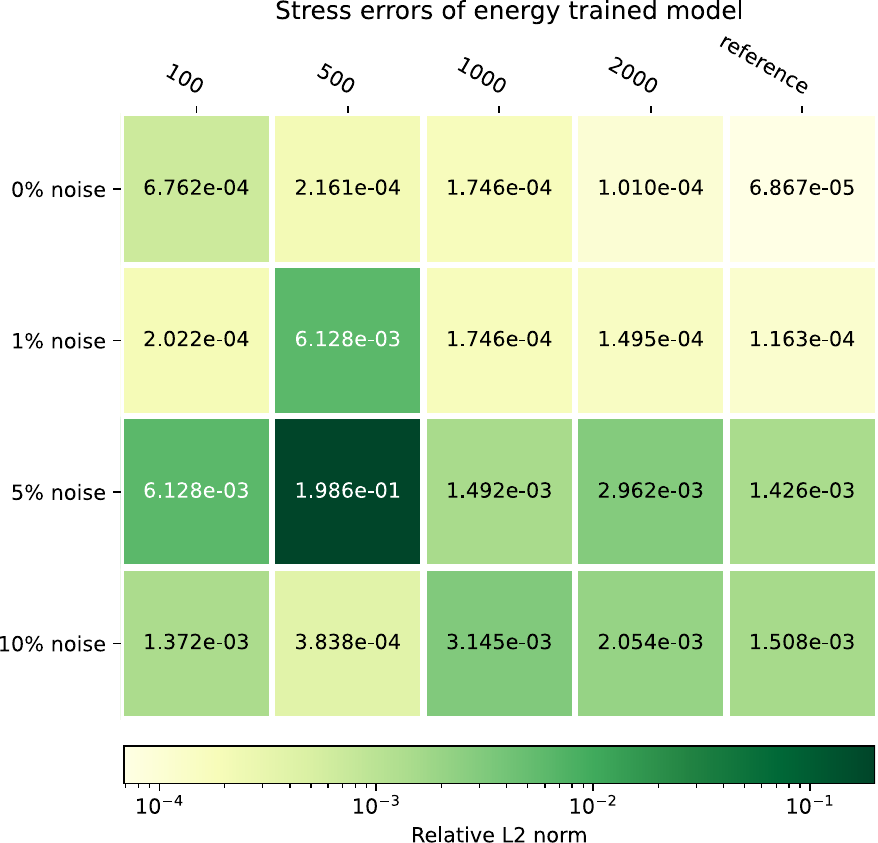}
		\caption{Stress errors of best energy trained models.}
		\label{fig:errors_stress_energy_Ciarlet}
	\end{subfigure}%
	\hfill
	\begin{subfigure}{0.5\textwidth}
		\centering
		\includegraphics[width=\textwidth]{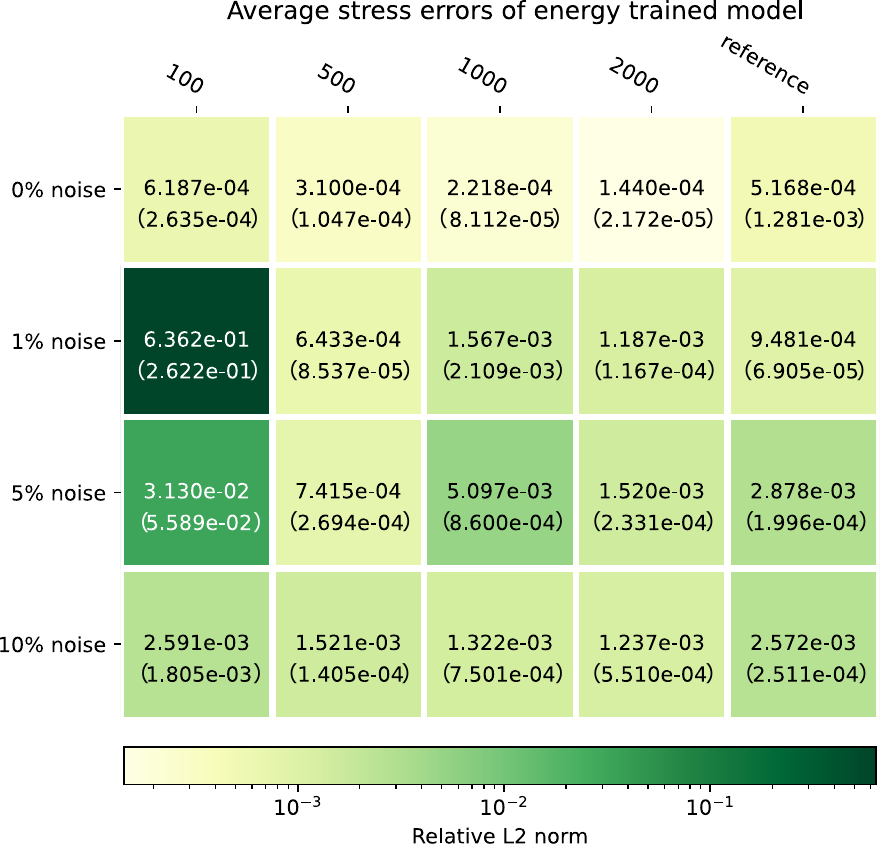}
		\caption{Average stress errors of energy trained models.}
		\label{fig:errors_stress_energy_Ciarlet_avg}
	\end{subfigure}
	\caption{Relative $L^2$ norm of stress for neural networks trained on energy, results for Cook membrane with Ciarlet law. The results shown are averaged for all training runs of neural networks. Standard deviations of the averaged errors are shown in parentheses.}
\end{figure}

\begin{figure}
	\centering
	\begin{subfigure}{0.5\textwidth}
		\centering
		\includegraphics[width=\textwidth]{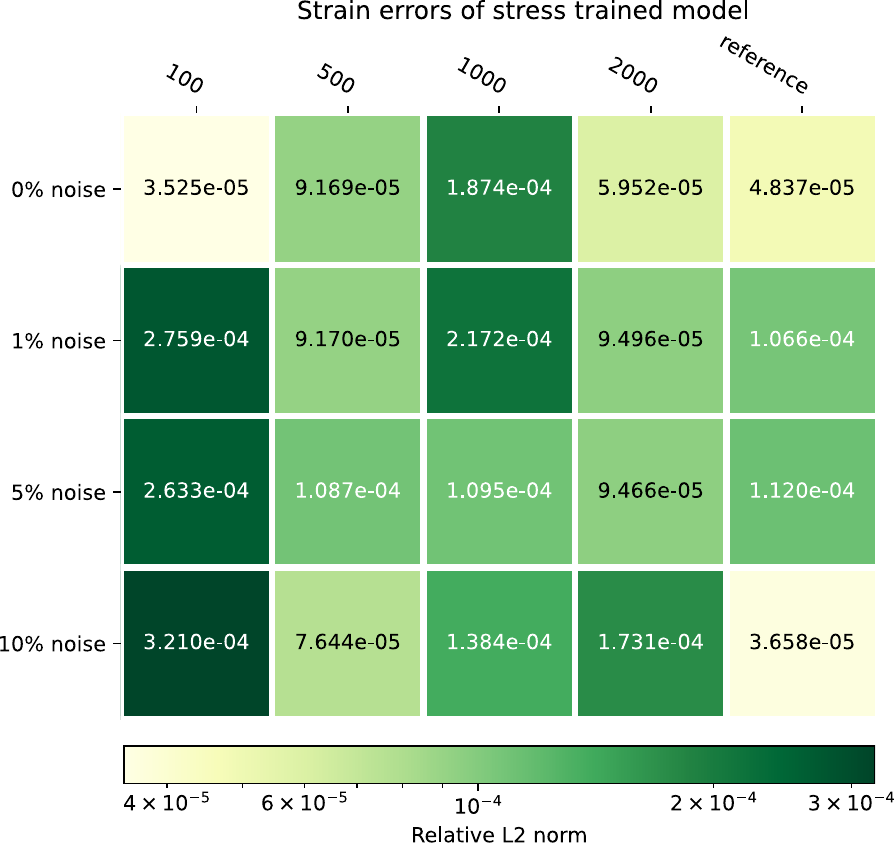}
		\caption{Strain errors of best stress trained models.}
		\label{fig:errors_strain_stress_Ciarlet}
	\end{subfigure}%
	\hfill
	\begin{subfigure}{0.5\textwidth}
		\centering
		\includegraphics[width=\textwidth]{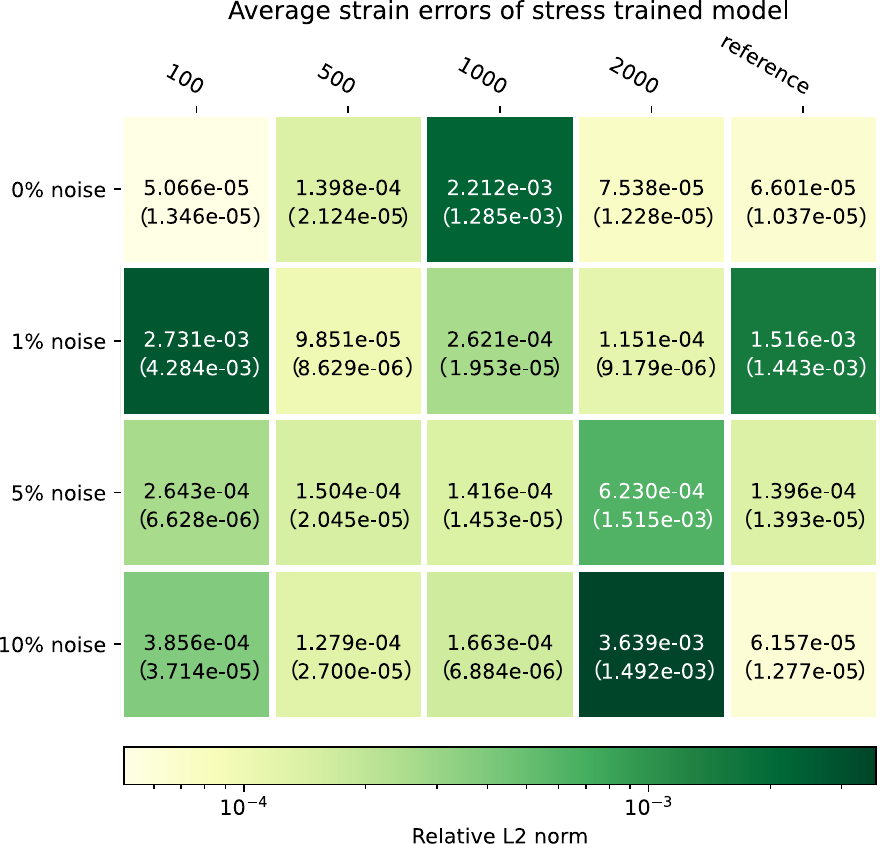}
		\caption{Average strain errors of stress trained models.}
		\label{fig:errors_strain_stress_Ciarlet_avg}
	\end{subfigure}
	\caption{Relative $L^2$ norm of strain for neural networks trained on stress, results for Cook membrane with Ciarlet law. The results shown are averaged for all training runs of neural networks. Standard deviations of the averaged errors are shown in parentheses.}
\end{figure}
\begin{figure}
	\centering
	\begin{subfigure}{0.5\textwidth}
		\centering
		\includegraphics[width=\textwidth]{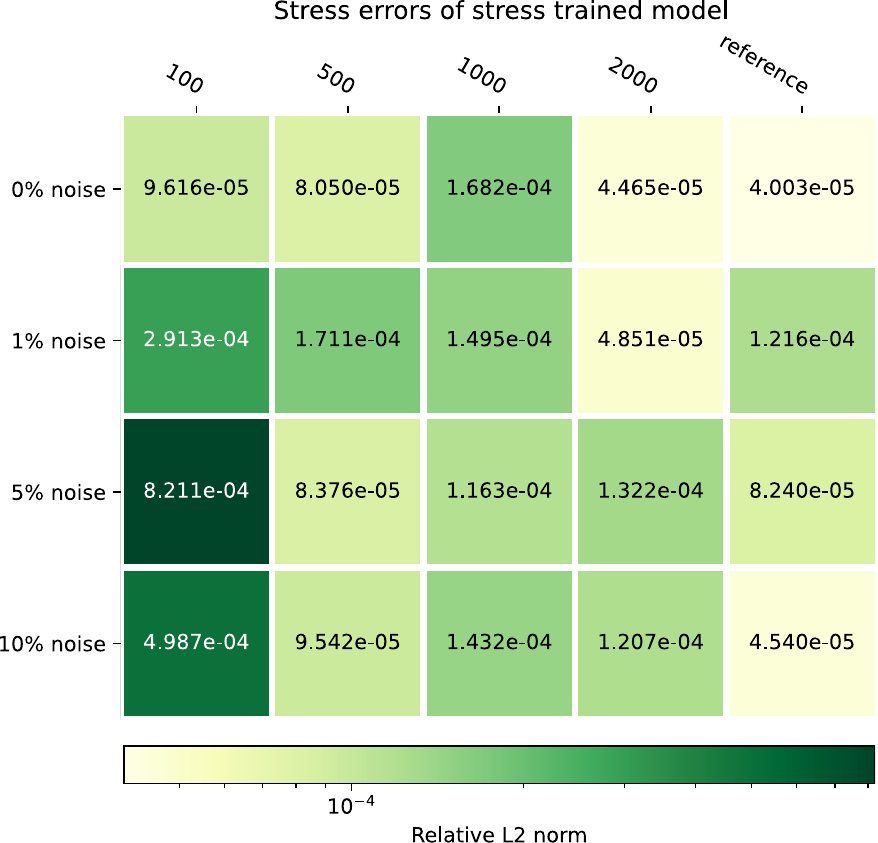}
		\caption{Stress errors of best stress trained models.}
		\label{fig:errors_stress_stress_Ciarlet}
	\end{subfigure}%
	\hfill
	\begin{subfigure}{0.5\textwidth}
		\centering
		\includegraphics[width=\textwidth]{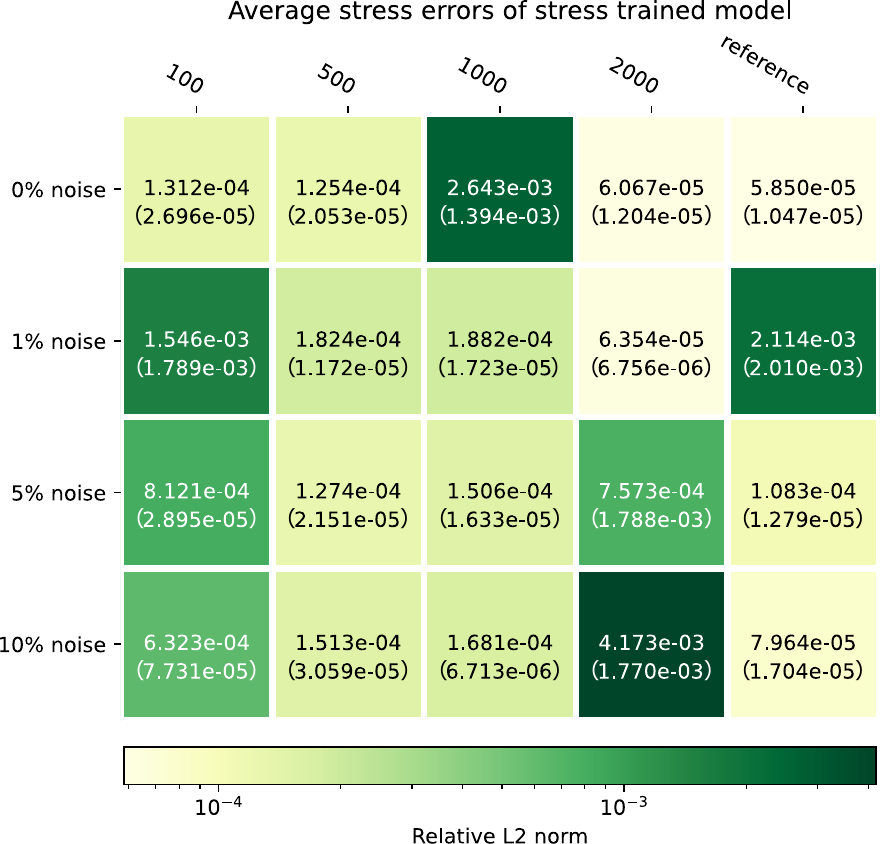}
		\caption{Average stress errors of stress trained models.}
		\label{fig:errors_stress_stress_Ciarlet_avg}
	\end{subfigure}
	\caption{Relative $L^2$ norm of stress for neural networks trained on stress, results for Cook membrane with Ciarlet law. The results shown are averaged for all training runs of neural networks. Standard deviations of the averaged errors are shown in parentheses.}
\end{figure}

\begin{figure}
	\centering
	\begin{subfigure}{0.5\textwidth}
		\centering
		\includegraphics[width=\textwidth]{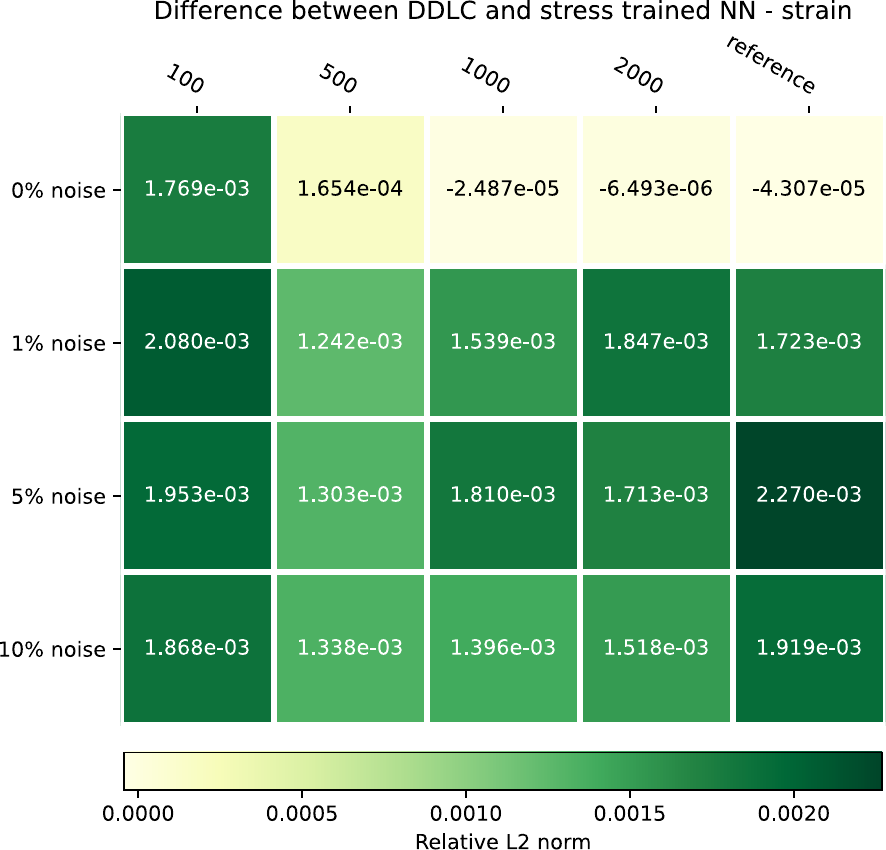}
		\caption{Absolute error difference, strain.}
		\label{fig:error_diff_DDLC_C_strain}
	\end{subfigure}%
	\hfill
	\begin{subfigure}{0.5\textwidth}
		\centering
		\includegraphics[width=\textwidth]{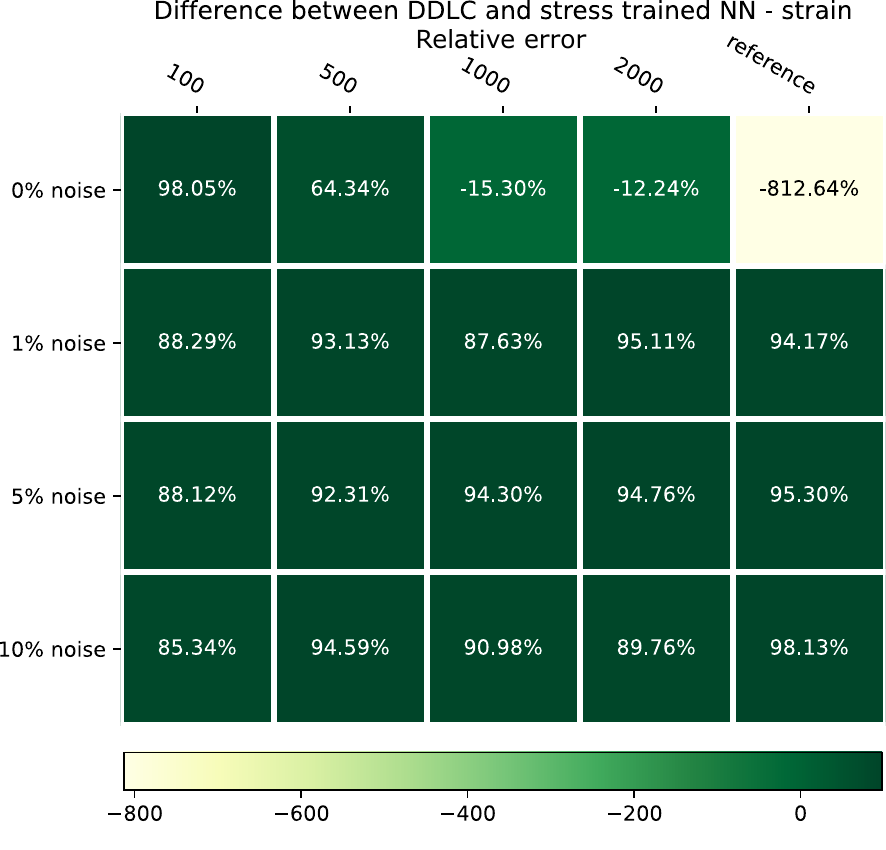}
		\caption{Relative error difference, strain.}
		\label{fig:rel_error_diff_DDLC_C_strain}
	\end{subfigure}
	\hfill
	\begin{subfigure}{0.5\textwidth}
		\centering
		\includegraphics[width=\textwidth]{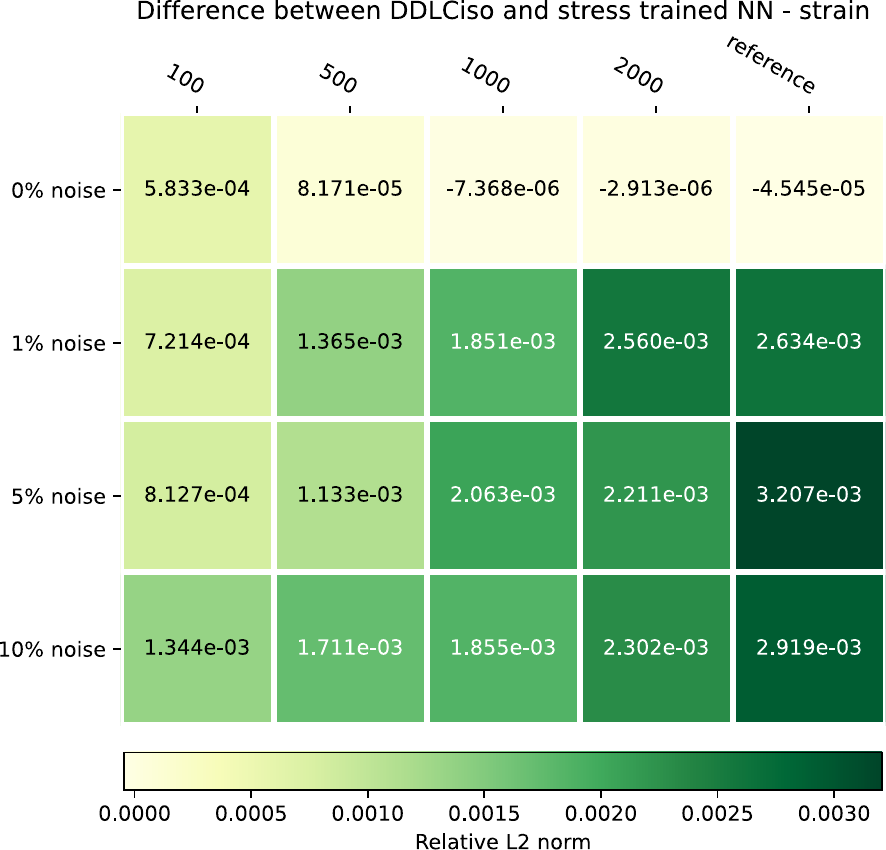}
		\caption{Absolute error difference, strain.}
		\label{fig:error_diff_DDLCiso_C_strain}
	\end{subfigure}%
	\hfill
	\begin{subfigure}{0.5\textwidth}
		\centering
		\includegraphics[width=\textwidth]{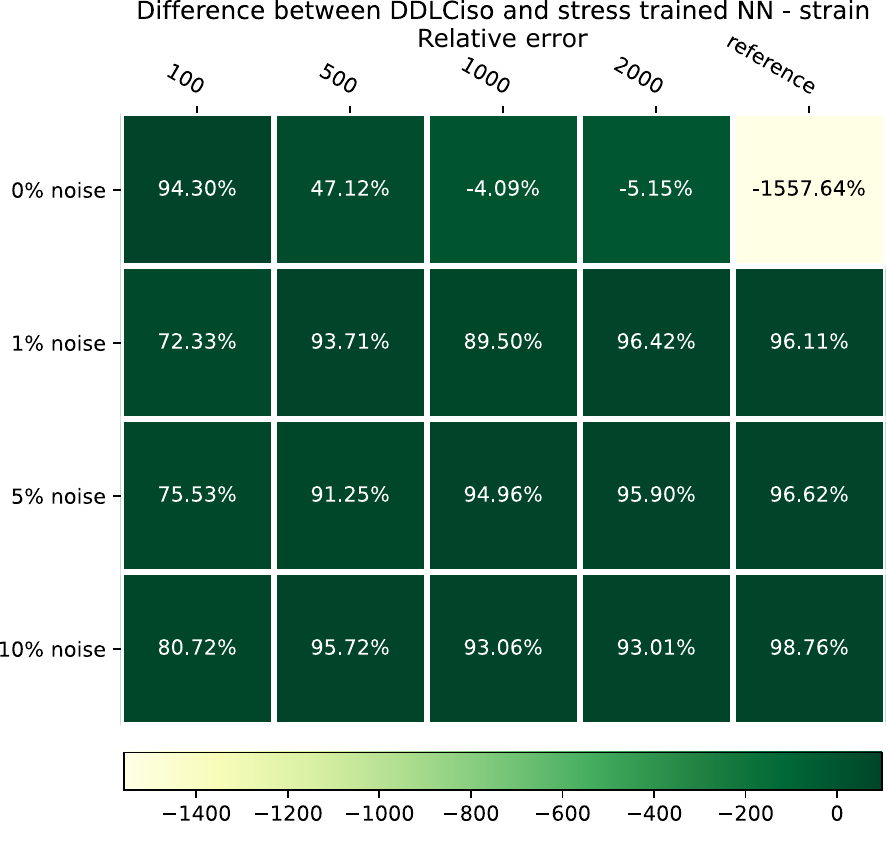}
		\caption{Relative error difference, strain.}
		\label{fig:rel_error_diff_DDLCiso_C_strain}
	\end{subfigure}
	\caption{Relative $L^2$ norm error differences between DD and NN approaches for Cook's membrane with Ciarlet law, strain. The upper row shows the absolute error difference and relative error when comparing DDLC to stress trained NNs, while the bottom row shows the errors when comparing DDLCiso to stress trained NNs.}
\end{figure}

\begin{figure}
	\centering
	\begin{subfigure}{0.5\textwidth}
		\centering
		\includegraphics[width=\textwidth]{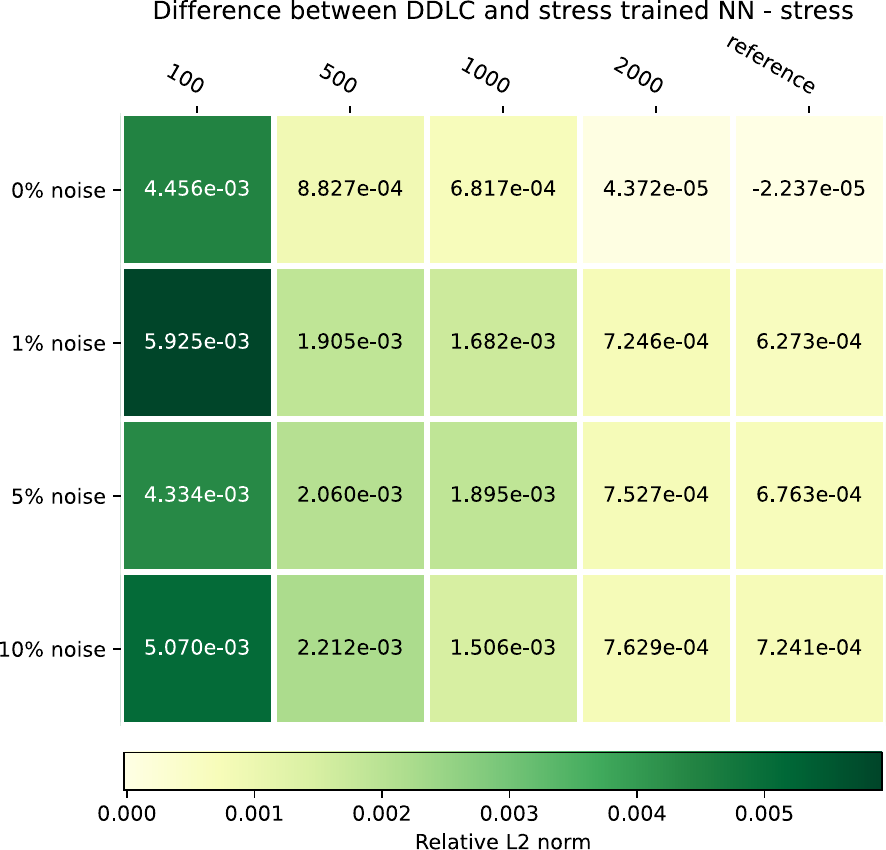}
		\caption{Absolute error difference, stress.}
		\label{fig:error_diff_DDLC_C_stress}
	\end{subfigure}%
	\hfill
	\begin{subfigure}{0.5\textwidth}
		\centering
		\includegraphics[width=\textwidth]{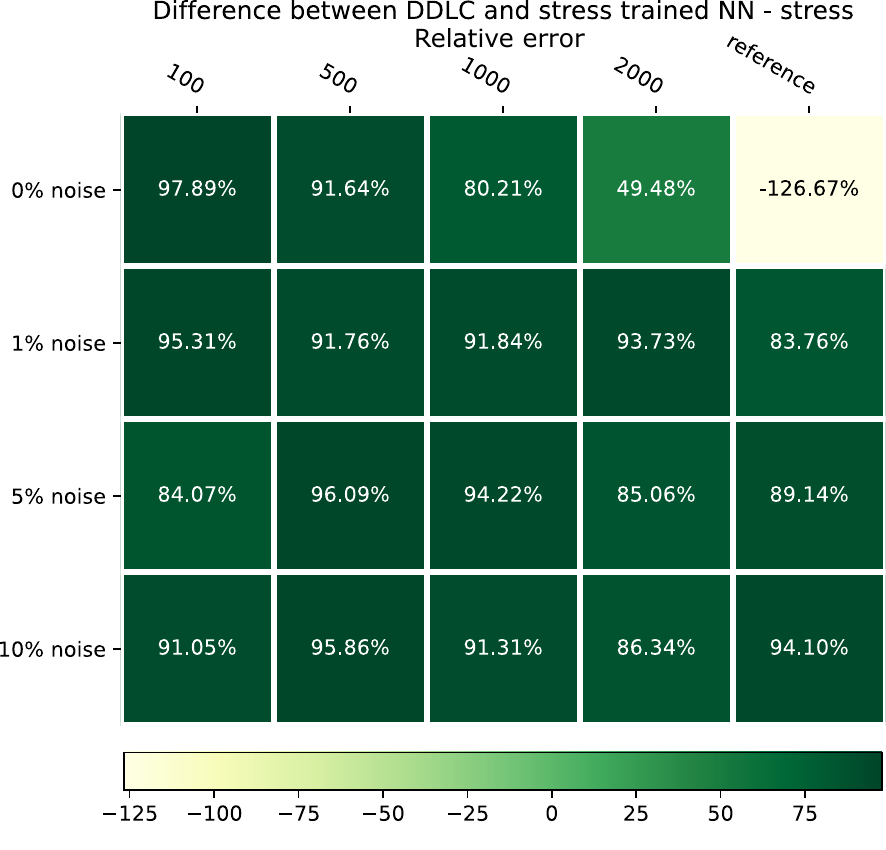}
		\caption{Relative error difference, stress.}
		\label{fig:rel_error_diff_DDLC_C_stress}
	\end{subfigure}
	\begin{subfigure}{0.5\textwidth}
		\centering
		\includegraphics[width=\textwidth]{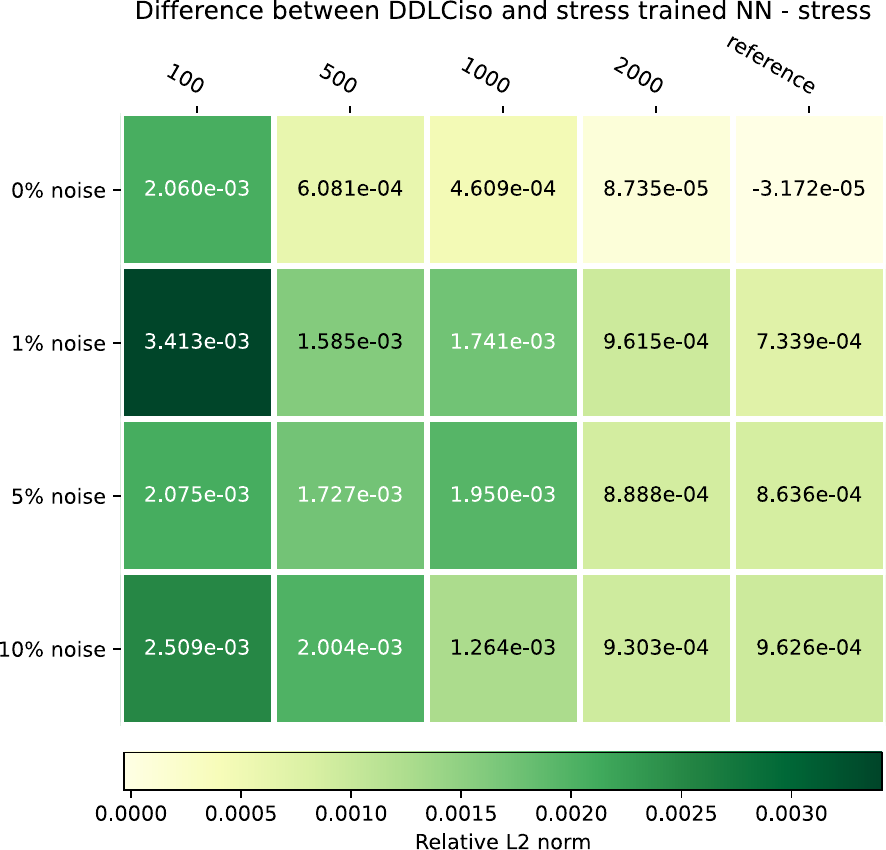}
		\caption{Absolute error difference, stress.}
		\label{fig:error_diff_DDLCiso_C_stress}
	\end{subfigure}%
	\hfill
	\begin{subfigure}{0.5\textwidth}
		\centering
		\includegraphics[width=\textwidth]{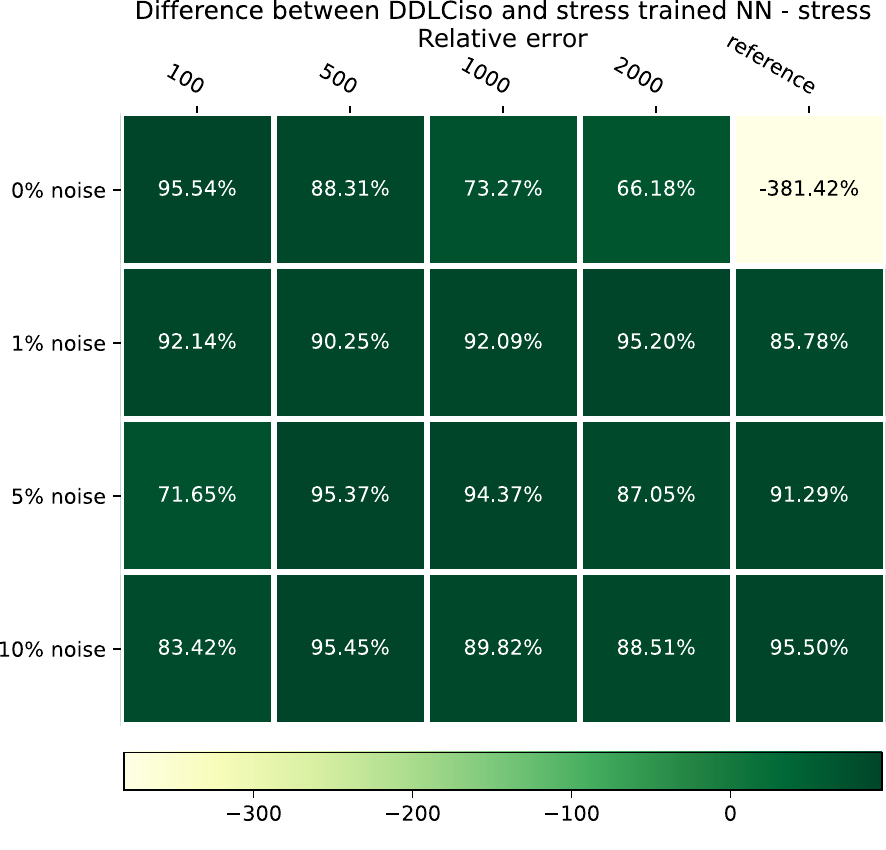}
		\caption{Relative error difference, stress.}
		\label{fig:rel_error_diff_DDLCiso_C_stress}
	\end{subfigure}
	\caption{Relative $L^2$ norm error difference between DD and NN approaches for Cook's membrane with Ciarlet law, stress. The upper row shows the absolute error difference and relative error when comparing DDLC to stress trained NNs, while the bottom row shows the errors when comparing DDLCiso to stress trained NNs.}
\end{figure}

\begin{figure}
	\centering
	\begin{subfigure}{0.5\textwidth}
		\centering
		\includegraphics[width=\textwidth]{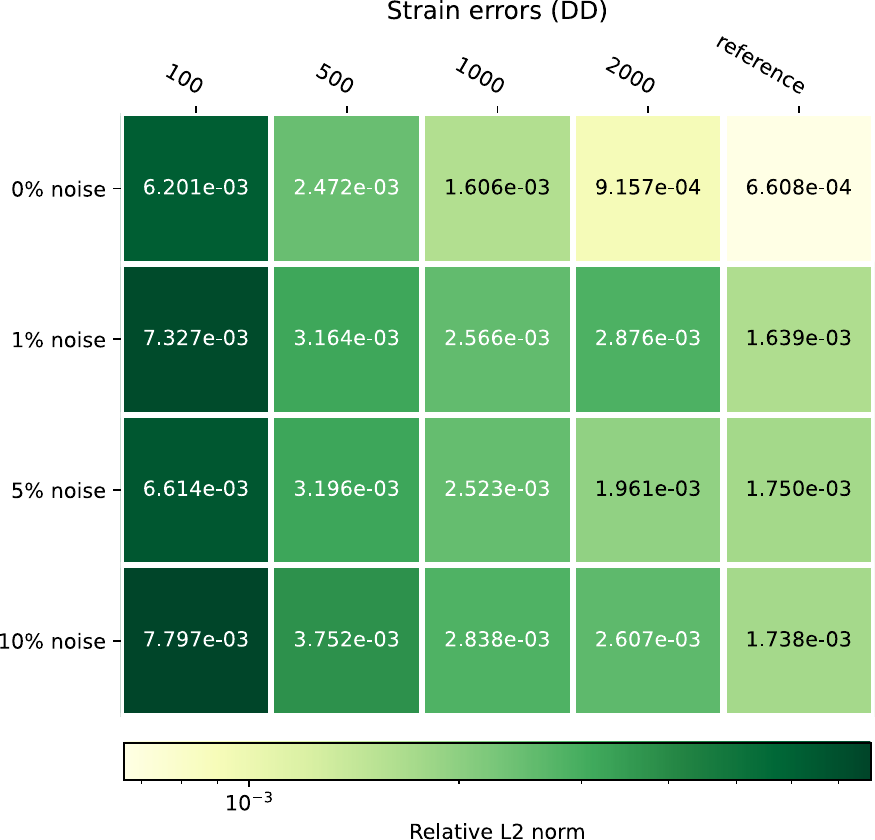}
		\caption{Standard search, original database.}
		\label{fig:errors_strain_HN_DD}
	\end{subfigure}%
	\hfill
	\begin{subfigure}{0.5\textwidth}
		\centering
		\includegraphics[width=\textwidth]{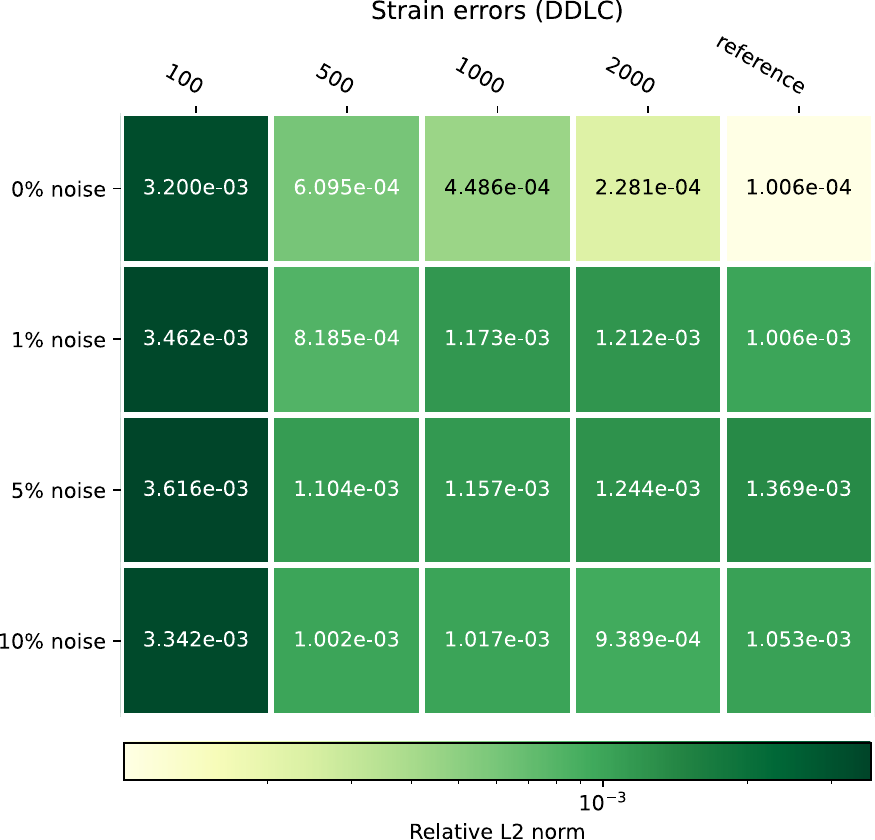}
		\caption{Locally convex search, original database.}
		\label{fig:errors_strain_HN_DDLC}
	\end{subfigure}
	
	\begin{subfigure}{0.5\textwidth}
		\centering
		\includegraphics[width=\textwidth]{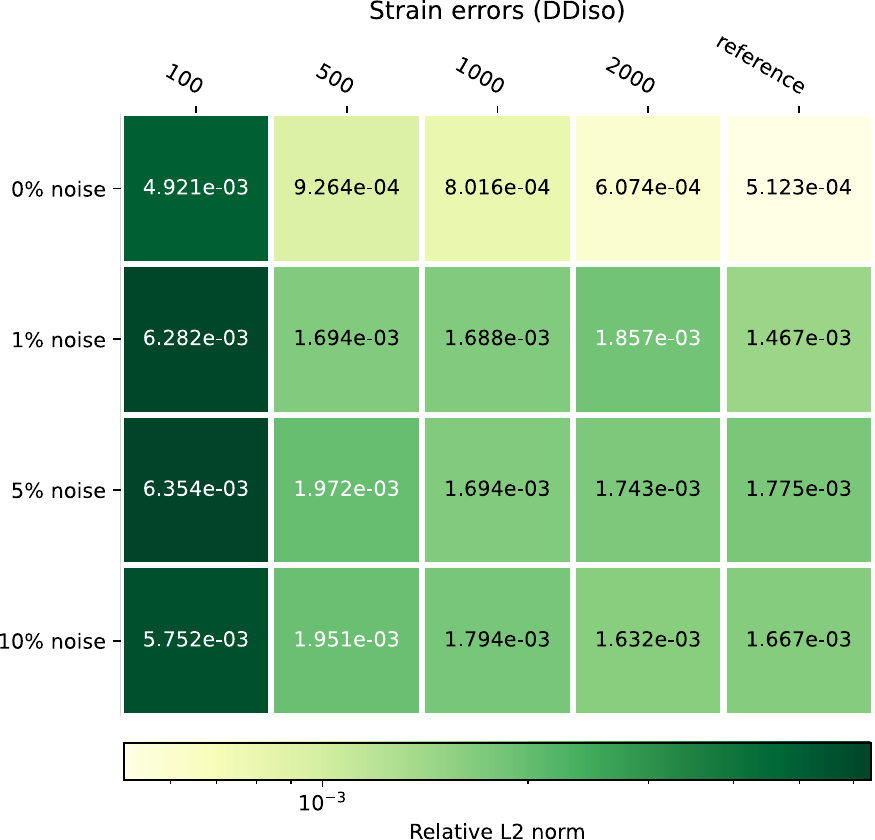}
		\caption{Standard search, enriched database with discretised orbits.}
		\label{fig:errors_strain_HN_DDiso}
	\end{subfigure}%
	\hfill
	\begin{subfigure}{0.5\textwidth}
		\centering
		\includegraphics[width=\textwidth]{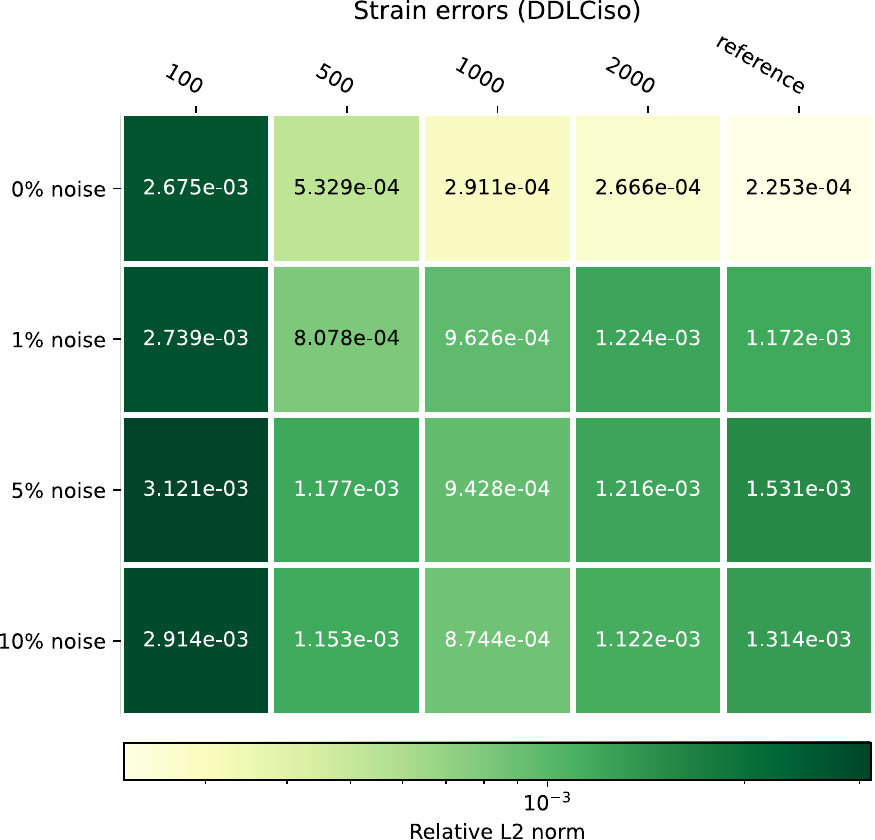}
		\caption{Locally convex, enriched database with discretised orbits.}
		\label{fig:errors_strain_HN_DDLCiso}
	\end{subfigure}
	\caption{Relative $L^2$ norm of strains for Cook membrane with Hartmann-Neff law.}
\end{figure}

\begin{figure}
	\centering
	\begin{subfigure}{0.5\textwidth}
		\centering
		\includegraphics[width=\textwidth]{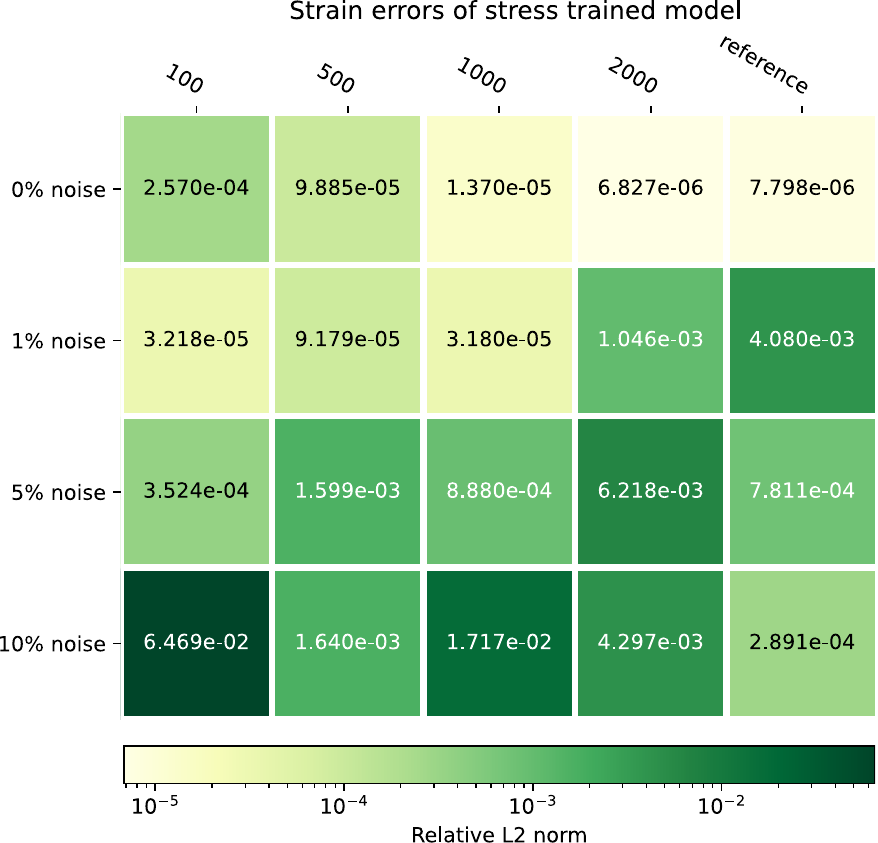}
		\caption{Strain errors of best stress trained models.}
		\label{fig:errors_strain_stress_HN}
	\end{subfigure}%
	\hfill
	\begin{subfigure}{0.5\textwidth}
		\centering
		\includegraphics[width=\textwidth]{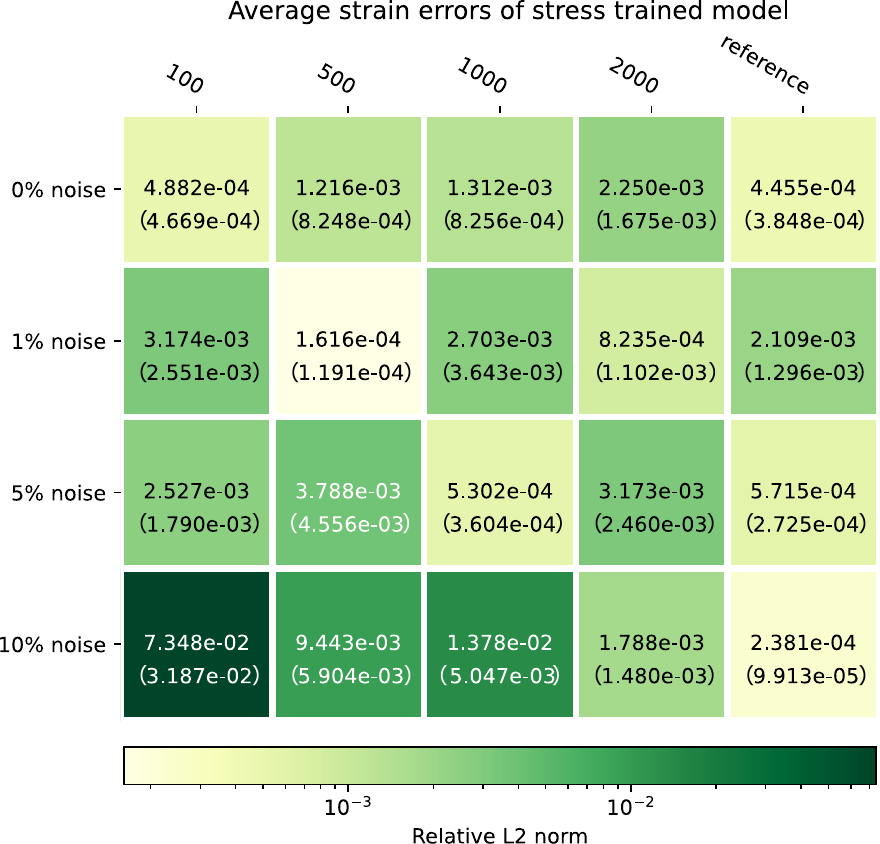}
		\caption{Average strain errors of stress trained models.}
		\label{fig:errors_strain_stress_HN_avg}
	\end{subfigure}
	\caption{Relative $L^2$ norm of strain for neural networks trained on stress, results for Cook membrane with Hartmann-Neff law. The results shown are averaged for all training runs of neural networks. Standard deviations of the averaged errors are shown in parentheses.}
\end{figure}

\begin{figure}
	\centering
	\begin{subfigure}{0.5\textwidth}
		\centering
		\includegraphics[width=\textwidth]{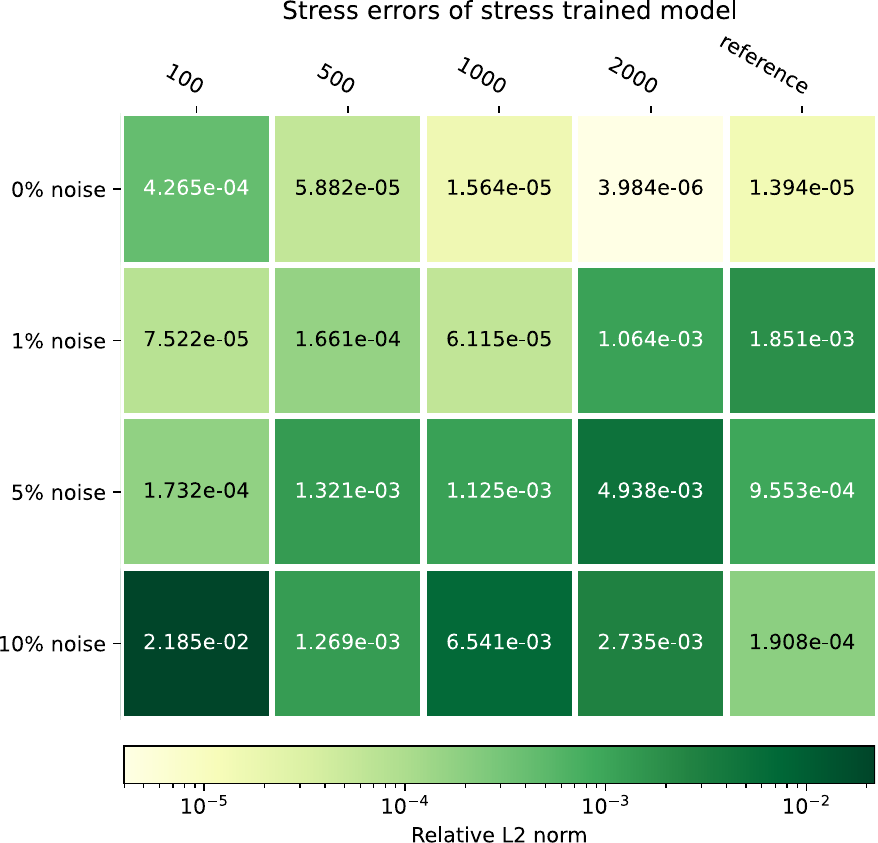}
		\caption{Stress errors of best stress trained models.}
		\label{fig:errors_stress_stress_HN}
	\end{subfigure}%
	\hfill
	\begin{subfigure}{0.5\textwidth}
		\centering
		\includegraphics[width=\textwidth]{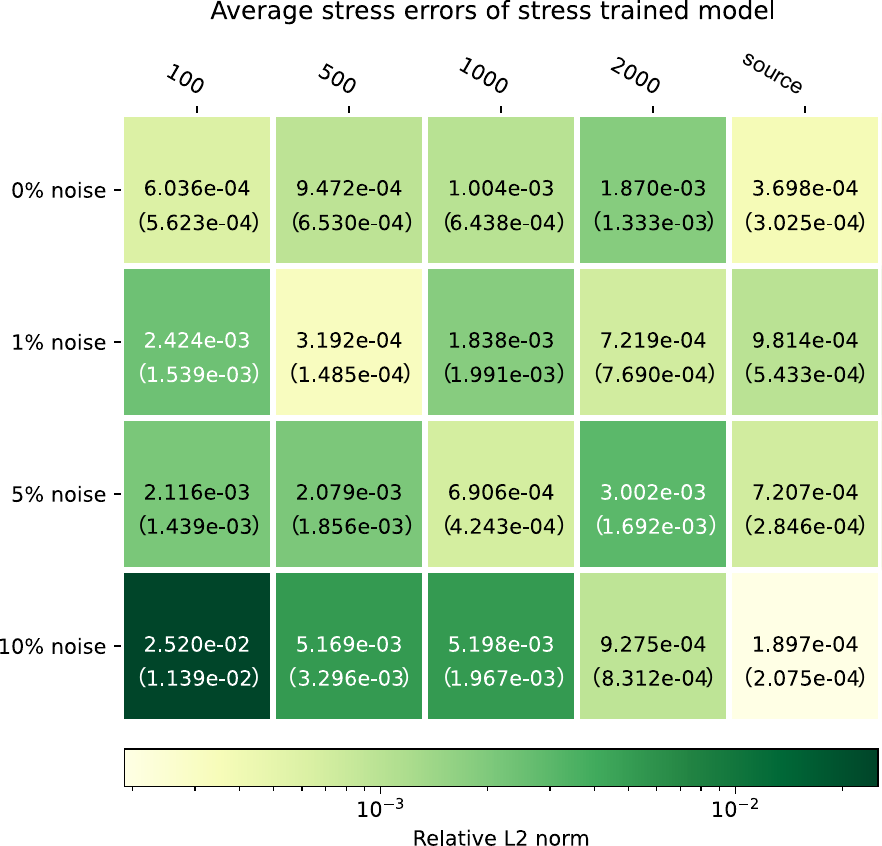}
		\caption{Average stress errors of stress trained models.}
		\label{fig:errors_stress_stress_HN_avg}
	\end{subfigure}
	\caption{Relative $L^2$ norm of stress for neural networks trained on stress, results for Cook membrane with Hartmann-Neff law. The results shown are averaged for all training runs of neural networks. Standard deviations of the averaged errors are shown in parentheses.}
\end{figure}

\FloatBarrier




%


\end{document}